\documentclass[referee]{aa}
\usepackage{graphicx}
\usepackage{epigraph}

\begin{document}

\title{Are explanations of the
Poynting-Robertson effect correct?}
\author{J. Kla\v{c}ka, J. Petr\v{z}ala, P. P\'{a}stor, L. K\'{o}mar}
\institute{Faculty of Mathematics,
Physics and Informatics, Comenius University \\
Mlynsk\'{a} dolina, 842 48 Bratislava, Slovak Republic \\
e-mail: klacka@fmph.uniba.sk, petrzala@fmph.uniba.sk \\
Pavol.Pastor@fmph.uniba.sk, komar@fmph.uniba.sk}

\date{}


\abstract{
Physics of the Poynting-Robertson (P-R) effect is discussed and
compared with the statements published in the past thirty years.
Relativistically covariant formulation reveals the essence of the
P-R effect and points out to nonphysical explanations
in scientific papers and monographs.
Although the final equation of motion \\
$m$ $d\vec{v} / dt$ $=$ ( $S A'\bar{Q'}_{pr}$ $/$ $c$ ) 
$\left \{ \left ( 1 ~-~ \vec{v} \cdot \vec{e} / c \right ) \vec{e} ~-~
	\vec{v} / c \right \}$ \\
has been usually correctly presented and used, its derivation
and explanation of its essence is frequently incorrect.

The relativistically covariant form of the equation of motion yields the
P-R effect as an action of the radiation pressure force on a moving 
spherical body. No "P-R drag", as a particular relativistically 
covariant equation of motion, exists. 
Omission of the nonphysical term "P-R drag"
excludes any confusion in the published definitions.

The difference between the effects of solar electromagnetic
and corpuscular (solar wind) radiation is stressed. The force acting on
the particle due to the solar wind
(the simple case of radial solar wind velocity is considered) is \\
$\vec{F}_{sw}$ $=$ $F_{sw}$ [ ( 1 $-$
$\vec{v} \cdot \vec{e} /  v_{sw}$ ) $\vec{e}$ $-$
$x'$ $\vec{v} / v_{sw}$ ]~, \\
where $F_{sw}$ is the force on the stationary particle,
$v_{sw}$ is the heliocentric solar-wind speed, and,
the value of $x'$ depends on material
properties of the particle (1 $<$ $x'$ $<$ 3).

Secular evolution of orbital elements is presented. 
Initial conditions are included.

}

\authorrunning{Kla\v{c}ka et al.}
\titlerunning{Poynting-Robertson effect}

\maketitle

\newpage

\section{Introduction}

\epigraph{Motto:\\
What is popular is not always right; what is right is not always popular.}

The action of electromagnetic radiation on a moving spherical body is
called the Poynting-Robertson (P-R) effect. It is used in astrophysical modelling of orbital evolution of dust grains for many decades
(e. g., Poynting 1903, Robertson 1937, Wyatt and Whipple 1950,
Robertson and Noonan 1968, Dohnanyi 1978, Burns {\it et al.} 1979, Kapi\v{s}insk\'{y} 1984, Jackson and Zook 1989, Leinert and Gr\"{u}n 1990, Gustafson 1994, Dermott {\it et al.} 1994, Reach {\it et al.} 1995,
Murray and Dermott 1999, Woolfson 2000, Danby 2003, Quinn 2005,
Harwit 2006, Gr\"{u}n 2007, Sykes 2007, Kr\"{u}gel 2008, Wyatt 2009).
The P-R effect holds when material of the particle is distributed in a 
spherically symmetric way. It is assumed that the spherical particle 
can be used as an approximation to
real, arbitrarily shaped, particle. Relativistically covariant
equation of motion for arbitrarily shaped dust particle can be found in
Kla\v{c}ka (2004, 2008b), where also thermal emission of the particle is
taken into account (Mishchenko 2001: the radiation pressure on arbitrarily
shaped particles arises also from an anisotropy of thermal emission).
Krauss and Wurm (2004) have published the first experimental evidence that
nonspherical dust grains behave in a different way than the spherical ones.

We will deal with the P-R effect in this paper. It has elapsed thirty
years since the time of Dohnanyi's statement "Much confusion in the
literature has existed on this subject and to some extent still does,
unfortunately." (Dohnanyi 1978, p. 563). The situation has not changed 
and many "explanations" of the P-R effect have been
offered during the past thirty years. It is important to have correct explanations of the effect, in disposal.

The paper presents many statements on the Poynting-Robertson
effect, which exist in the literature. We have tried to show physically
incorrect points of the statements and to put them into
a correct way. The enormous amount of
incorrect presentations probably confirms the ideas of Weisskopf (1973):
"The pressure of fast publication is so great that people rush into print
with hurriedly written papers and books that show little concern for
careful formulation of ideas. ....
Essential ideas of physics are often lost ..."

We have already tried to explain the physics of the P-R effect (see, e. g., Kla\v{c}ka 1992a, 1993b, 2000b, 2001b). However, our comments
and explanations have not met, in general, with a positive reaction and
the incorrect statements and derivations are repeatedly cited and
reproduced in many other scientific papers and textbooks. This avalanche propagation is the reason why we have decided to write a paper presenting comments on various derivations and explanations.
This will lead to a deeper and more complete understanding of the effect.

Our paper is based on the results of Kla\v{c}ka (2008a, 2008b).
At first, we summarize the most important results on the P-R effect.
Then, we discuss many "explanations" of the P-R effect, which have been
presented in the literature during the past thirty years.
The scientists should concentrate on Sec. 6.1, where comments on the
most frequently cited paper by Burns {\it et al.} (1979) are given:
not only the important parts of Kla\v{c}ka (1992a -- Sec. 2.5) are repeated,
but the relativistic results of Kla\v{c}ka (2008a, 2008b) are used. It is
clear that the statement "the interpretation of the low-order velocity terms
in classical (nonrelativistic) physics are still debated" (Gustafson 1994,
p. 565) is already not relevant, now. Really, even the simplest case of
Poynting (1903) and Robertson (1937) is correctly understood on the basis
of relativistic approach. Moreover, the dimensionless efficiency
factor $\bar{Q}'_{pr}$, $\bar{Q}'_{ext}$ are obtained from Mie's solution
of Maxwell's equations and it cannot be said that Maxwell's equations
belong to classical (nonrelativistic) physics. These relativistic results
are used also in other parts of the paper; e.g., the case of the
aberration of light is discussed in detail in Sec. 6.2.

To be sure that the physical results are correct, one has to consider
relativistically covariant equation of motion. Kla\v{c}ka (2008a, 2008b) explicitly showed the advantage of relativistically covariant formulations: the application of this approach enabled to found that even the statement of Poynting (1903) and Robertson (1937) (and other explanations up to the 
year 2008) is incorrect: as for the outgoing radiation, the particle's 
"own radiation outwards being equal in all directions has zero resultant pressure" (Poynting 1903) in the particle's own frame of reference, or, 
"the process of absorption and re-emission produces no net force 
on a particle when one chooses to work with a stationary frame referred 
to the particle" (Wyatt and Whipple 1950).

The importance of relativistically covariant equations of motion
is stressed, e. g., by Feynman and Mould. 
Feynman states: "The laws of physics must be such that after a Lorentz transformation, the new form of the laws looks just like the old form."
(Feynman {\it et al.} 2006, volume II, p. 25-1), or,
"There is, however, another reason for writing our equations this way.
It has been discovered -- after Einstein guessed that it might be so --
that {\it all} of the laws of physics are invariant under the Lorentz
transformation. That is the principle of relativity. Therefore, if we
invent a notation which shows immediately when a law is written down
whether it is invariant or not, we can be sure that in trying to make
new theories we will write only equations which are consistent with the
principle of relativity. ... the interesting physical thing is that
{\it every law} of physics must have this same invariance under the same
transformation.  ... It is because the principle of relativity is a fact
of nature that in the notation of four-dimensional vectors the equations
of the world will look simple."
(Feynman {\it et al.} 2006, volume II, p. 25-11).
Feynman has motivated a reader to the importance of using tensors (or, as
a special case, four-vectors) in formulation of equation of motion. In order to
be the reader more familiar with the relevance of this access, we reproduce the
corresponding explanation from Mould (2002, p. 123): "It is a goal of relativity
to formulate the fundamental laws of nature in {\it covariant} form; that is,
in a form that is independent of coordinates. A four-vector equation, for
instance, is covariant in this sense. The particular values of a four-vector
may very well depend on coordinates, but a four-vector {\it equation} will
have the same form in all coordinate systems. Tensor equations have this same
property. They have the same form in all coordinate systems even though their
specific component values differ from one coordinate system to another. Scalar
functions have a {\it form} invariance as well as {\it value} invariance."

Sections 2-5 present the most relevant results on the physics
of the P-R effect. Sec. 6 discusses various "explanations"
of the P-R effect, published during the past 30 years. Sec. 7
offers two simple correct explanations of the P-R effect. Finally,
Sec. 8 summarizes the relevant equations for the secular evolution
of orbital elements of a spherical body under the action of 
central star's gravity and electromagnetic radiation.
Discussions presented in the paper can help in understanding 
the P-R effect. The paper should ensure that
only correct explanations will be published in future.

\section{Covariant formulations}
Relativistically covariant formulations are
useful in understanding of the interaction between a particle
and an incident electromagnetic radiation. Spherically
symmetric mass distribution within the particle is tacitly assumed.

Interaction of the spherical particle with the incident electromagnetic
radiation  is known as the Poynting-Robertson effect. The emission
component of the radiation force, due to thermal emission, is zero,
i. e. $\vec{F}'_{e}$ $=$ 0. The conditions for the incoming and outgoing
radiation are (Kla\v{c}ka 2008b):
\begin{eqnarray}\label{1}
\frac{d p^{\mu}_{in}}{d \tau} &=&
	\frac{w^{2} S A' \bar{Q'}_{ext}}{c}~b^{\mu} ~,
\nonumber\\
\frac{d p^{\mu}_{out}}{d \tau} &=&
     \frac{w^{2} S A' \bar{Q'}_{ext}}{c} ~ \left \{
     \bar{Q}' ~ \frac{u^{\mu}}{c} ~+~
     \left ( 1 - \bar{Q'} \right ) ~ b^{\mu} \right \} ~,
\nonumber \\
\bar{Q}' &\equiv& \frac{\bar{Q'}_{pr}}{\bar{Q'}_{ext}} ~,
\end{eqnarray}
where the left-hand sides are four-momenta, e.g.
$p^{\mu}_{out}$ $=$ ($E_{out} / c$ ; $\vec{p}_{out}$ ),
$c$ is the speed of light in vacuum,
$S$ is the flux density of radiation energy (energy flow through
unit area perpendicular to the ray per unit time), $A'$ is geometrical
cross-section of the spherical particle, $\bar{Q'}_{ext}$ is dimensionless
extinction efficiency factor given by optical properties of the particle and
averaged over the stellar spectrum, $\bar{Q'}_{pr}$ is dimensionless efficiency
factor for radiation pressure integrated over stellar spectrum and calculated
for radial direction (as for the dimensionless factors of effectivity of
radiation pressure and extinction, see Mie 1908, or, e. g.,
Kla\v{c}ka and Kocifaj 2007, or, Sec. 4.5 in Bohren and
Huffman 1983) and 4-vectors $u^{\mu}$ (four-velocity),
$b^{\mu}$ and the quantity $w$	are given by the following equations:
\begin{eqnarray}\label{2}
u^{\mu} &=& \left( \gamma~ c~;~\gamma~ \vec{v} \right) ~,
\nonumber \\
b^{\mu} &=& \left( \frac{1}{w}~;~ \frac{\vec{e}}{w}  \right) ~,
\nonumber \\
w &=& \gamma \left( 1 ~-~ \frac{\vec{v} \cdot \vec{e}}{c} \right) ~,
\end{eqnarray}
where $\vec{v}$ is particle's velocity in the reference frame of the star
and $\vec{e}$ is unit position vector of the particle with respect to
the star; $\gamma$ is the Lorentz factor. As was pointed out by
Kla\v{c}ka (2008b), Eq. (1) differs from the geometrical approach
presented in the literature since the time of Poynting and Robertson.

Equations for $dp^{\mu}_{out}/d\tau$ and $dp^{\mu}_{in}/d\tau$ yield the following covariant relation between the outgoing and the incoming
radiation:
\begin{eqnarray}\label{3}
\frac{d p^{\mu}_{out}}{d \tau} &=& \left ( 1 ~-~ \bar{Q'} \right )~
				 \frac{d p^{\mu}_{in}}{d \tau} ~+~
				 \frac{w^{2} S A' \bar{Q'}_{ext}}{c}~\bar{Q'}~
				 \frac{u^{\mu}}{c}~.
\nonumber \\
\bar{Q'} &\equiv& \frac{\bar{Q'}_{pr}}{\bar{Q'}_{ext}} ~.
\end{eqnarray}
Eq. (3) can be considered as a fundamental equation for the validity
of the P-R effect.

\subsection{Incoming radiation and particle's acceleration}
The action of the incoming radiation influences particle's motion
according to the following equation of motion:
\begin{equation}\label{4}
\frac{d p^{\mu}}{d \tau} = \frac{d p^{\mu}_{in}}{d \tau} ~,
\end{equation}
where the right-hand side is given by Eq. (1) and
\begin{equation}\label{5}
p^{\mu}~=~m u^{\mu}
\end{equation}
is a four-momentum of the particle.
Inserting Eqs. (1) and (5) into Eq. (4), we obtain
\begin{equation}\label{6}
m ~ \frac{d u^{\mu}}{d \tau} +
\frac{d m}{d \tau} ~ u^{\mu} = \frac{w^{2} S A' \bar{Q'}_{ext}}{c}~b^{\mu} ~.
\end{equation}
Multiplication of Eq. (6) by $u_{\mu}$ yields, on the basis
of $u_{\mu}$ $d u^{\mu}/d \tau$ $=$ 0,
$u_{\mu}$ $u^{\mu}$ $=$ $c^{2}$,
$u_{\mu}$ $b^{\mu}$ $=$ $c$:
\begin{equation}\label{7}
\left ( \frac{d m}{d \tau} \right )_{in} =
\frac{w^{2} S A' \bar{Q'}_{ext}}{c^{2}} ~.
\end{equation}

Inserting Eq. (7) into Eq. (6), we obtain four-acceleration
\begin{equation}\label{8}
\left ( \frac{d u^{\mu}}{d \tau} \right ) _{in}  =
\frac{w^{2} S A' \bar{Q'}_{ext}}{mc} ~
\left ( b^{\mu} - \frac{u^{\mu}}{c} \right ) ~.
\end{equation}

The result represented by Eq. (7) can be understood in the sense that
$dE_{in}/d \tau$ $=$ $w^{2} S A' \bar{Q'}_{ext}$ corresponds
to the energy of the incoming radiation interacting with the particle
[$dE_{in}/d \tau$ $=$ $c^{2}$ ($dm /d \tau$)$_{in}$].
The real system consists of radiation and the particle and the
system cannot be divided into the individual components. Thus,
the result presented by Eq. (7) does not correspond to the real
change of the particle mass, but only to a virtual change.
No direct real experiment can verify the validity of Eq. (7).
Indirect evidence of Eq. (7) is the validity of the result
0 $<$ $\bar{Q'}_{pr} / \bar{Q'}_{ext}$ $<$ 2 and experimental
confirmation of Maxwell's equations and relativity theory.

It may seem surprising that $\left ( d m /d \tau \right )_{in}$
depends on the particle velocity and $A' \bar{Q'}_{ext}$, only.
E.g., perfectly reflecting and absorbing spherical surfaces,
within geometrical optics approximation, yield the same
result for $\left ( d m /d \tau \right )_{in}$.

\subsection{Outgoing radiation and particle's acceleration}
The action of the outgoing radiation influences the motion of the
particle according to the following equation of motion:
\begin{equation}\label{9}
\frac{d p^{\mu}}{d \tau} = -~ \frac{d p^{\mu}_{out}}{d \tau} ~,
\end{equation}
where the right-hand side is given by Eq. (1). Inserting Eqs. (1) and
(5) into Eq. (9), we obtain
\begin{eqnarray}\label{10}
m ~ \frac{d u^{\mu}}{d \tau} ~+~
\frac{d m}{d \tau} ~ u^{\mu} &=& - ~ \frac{w^{2} S A' \bar{Q'}_{ext}}{c}~
     \left \{ \bar{Q'} ~ \frac{u^{\mu}}{c} ~+~
     \left ( 1 - \bar{Q'} \right ) ~ b^{\mu} \right \} ~,
\nonumber \\
\bar{Q'} &\equiv& \frac{\bar{Q'}_{pr}}{\bar{Q'}_{ext}} ~.
\end{eqnarray}
Multiplication of Eq. (10) by $u_{\mu}$ yields, on the basis of
$u_{\mu}$ $u^{\mu}$ $=$ $c^{2}$,
$u_{\mu}$ $d u^{\mu}/d \tau$ $=$ 0, and,
$u_{\mu}$ $b^{\mu}$ $=$ $c$:
\begin{equation}\label{11}
\left ( \frac{d m}{d \tau} \right )_{out} = -~ \frac{w^{2} S A'
					       \bar{Q'}_{ext}}{c^{2}} ~.
\end{equation}
Inserting of Eq. (11) into Eq. (10), we obtain four-acceleration
\begin{eqnarray}\label{12}
\left ( \frac{d u^{\mu}}{d \tau} \right ) _{out}  &=& -~
\frac{w^{2} S A' \bar{Q'}_{ext}}{m c} ~ \left ( 1 - \bar{Q'} \right )
\left ( b^{\mu} - \frac{u^{\mu}}{c} \right ) ~,
\nonumber\\
\bar{Q'} &\equiv& \frac{\bar{Q'}_{pr}}{\bar{Q'}_{ext}}~.
\end{eqnarray}

\subsection{Equation of motion}
Equation of motion of the particle is given by the condition
\begin{equation}\label{13}
\frac{d p^{\mu}}{d \tau} = \frac{d p^{\mu}_{in}}{d \tau} ~-~
			   \frac{d p^{\mu}_{out}}{d \tau} ~.
\end{equation}
Eqs. (1) and (13) yield
\begin{eqnarray}\label{14}
\frac{d p^{\mu}}{d \tau} &=&
     \frac{w^{2} S A' \bar{Q'}_{ext}}{c}~\bar{Q'} ~ \left(
     b^{\mu}~-~\frac{u^{\mu}}{c}\right) ~,
\nonumber \\
\bar{Q'} &\equiv& \frac{\bar{Q'}_{pr}}{\bar{Q'}_{ext}} ~.
\end{eqnarray}

Insertion of Eq. (5) into the left-hand side of Eq. (14) yields
$dp^{\mu} / d \tau$ $=$ $m ( du^{\mu} / d \tau )$  $+$
$( dm / d \tau ) u^{\mu}$. Using this relation and subsequent
multiplication of Eq. (14) by $u_{\mu}$, one obtains $dm/d\tau$ $=$ 0,
if also $u_{\mu}$ $(du^{\mu}/d\tau)$ $=$ 0 and
$u_{\mu}$ $u^{\mu}$ $=$ $c^{2}$,
$u_{\mu}$ $b^{\mu}$ $=$ $c$ are used. The relation $dm/d\tau$ $=$ 0
states that the mass of the particle is conserved. Thus, Eq. (14) may be
rewritten to the form
\begin{eqnarray}\label{15}
\frac{d u^{\mu}}{d \tau} &=&
     \frac{w^{2} S~ A' ~\bar{Q'}_{ext}}{m~c} ~\bar{Q'} ~ \left (
     b^{\mu} ~-~ \frac{u^{\mu}}{c} \right ) ~,
\nonumber \\
\bar{Q}' &\equiv& \frac{\bar{Q'}_{pr}}{\bar{Q'}_{ext}} ~.
\end{eqnarray}

\section{The fundamental condition}
In order to be able to correctly understand physics of the P-R effect,
we have to realize that the fundamental condition for the validity
of the P-R effect exists (Eq. 3).
The essence of the P-R effect is that nonradial components of the particle's
radiation pressure and components of the thermal force are zero (in particle's
frame of reference). Thus, the P-R effect holds for spherical particles.

The fundamental condition states: the total momentum
(per unit time) of the outgoing radiation is nonzero, in general, and
\begin{eqnarray}\label{16}
E'_{outgoing} &=& E'_{incoming} ~,
\nonumber\\
\vec{p'}_{outgoing} &=& ( 1 - \bar{Q'} ) ~ \vec{p'}_{incoming} ~,
\nonumber \\
\bar{Q'} &\equiv& \frac{\bar{Q'}_{pr}}{\bar{Q'}_{ext}} ~,
\nonumber \\
\vec{F}'_{e} &=& 0~,
\end{eqnarray}
in the proper frame of reference of the particle. The condition is
obtained from its covariant form presented by Eq. (3). The equation states
that the P-R effect holds if the total momentum (per unit time)
of the outgoing radiation, integrated over the whole space angle, is in the
direction of the incident radiation, in the particle's frame of reference.

It is generally believed that the special case $\bar{Q}'$ $=$ $1$
corresponds to the perfectly absorbing sphere, in geometrical optics
approximation. The equation of motion for this special case was discussed
by Poynting (1903) and finally written by Robertson (1937).
It is believed that this special case yields i)
$\vec{p'}_{outgoing}$ $\equiv$ $d\vec{p'}_{out}/d\tau$ $=$ 0
(see Eqs. 1 or 3) and this corresponds
to the hypothesis of "isotropic reemission of the radiation", and,
ii), the case $\bar{Q'}$ $=$ 1 is also produced by specular reflection
on the spherical particle (within the geometrical optics approximation).
However, the physical condition for the two cases, perfect absorption and
specular reflection, is $\bar{Q'}$ $=$ 1/2, in reality (Kla\v{c}ka 2008a,
2008b).

\section{Approximation of the equation of motion}
It is sufficient to approximate space component of the covariant
equation of motion (and also acceleration of the particle) to the first
order in $\vec{v} / c$. Eq. (14) yields the following relation between its
time and space components:
\begin{equation}\label{17}
\frac{dp^{0}}{d \tau}~=~\frac{\vec{v}}{c} \cdot \frac{d\vec{p}}{d \tau}~.
\end{equation}
For this reason, we have to do an approximation of the time component to
the second order in $\vec{v} / c$.

At first we will present approximations for the incident and outgoing
radiation.
Eqs. (1) yield the following approximation (the results are exact,
accurate to all orders in $v/c$):
\begin{eqnarray}\label{18}
\frac{dE_{in}}{dt} &=& S~ A' ~\bar{Q'}_{ext}~
\left( 1 ~-~ \frac{\vec{v} \cdot \vec{e}}{c} \right) ~,
\nonumber \\
\frac{d\vec{p}_{in}}{dt} &=& \frac{S A'\bar{Q'}_{ext}}{c}~
\left ( 1 ~-~ \frac{\vec{v} \cdot \vec{e}}{c} \right ) \vec{e} ~.
\end{eqnarray}
The results for the outgoing radiation (to the second order in $v/c$
in energy) are:
\begin{eqnarray}\label{19}
\frac{dE_{out}}{dt} &=& S ~A' ~\bar{Q'}_{ext}~
\left\{ 1 ~-~ \frac{\vec{v} \cdot \vec{e}}{c} ~
-~\bar{Q'}~\left[
\frac{\vec{v}\cdot \vec{e}}{c}~-~
\left(\frac{\vec{v}\cdot \vec{e}}{c}\right)^{2}~-~
\left(\frac{\vec{v}}{c}\right)^{2} \right] \right\} ~,
\nonumber \\
\frac{d \vec{p}_{out}}{dt} &=& \frac{S A'\bar{Q'}_{ext}}{c}~
\left \{ \bar{Q'} \frac{\vec{v}}{c} ~+~ \left ( 1 ~-~ \bar{Q'} \right )
\left ( 1 ~-~ \frac{\vec{v}\cdot\vec{e}}{c} \right ) ~\vec{e} \right \} ~,
\nonumber \\
\bar{Q'} &\equiv& \frac{\bar{Q'}_{pr}}{\bar{Q'}_{ext}} ~,
\end{eqnarray}
where $t$ is time measured in the stationary reference frame, i.e.
the reference frame in which the particle moves with the velocity $\vec{v}$.
Acceleration of the particle, due to the incoming and outgoing radiation,
we get from Eqs. (8) and (12)
\begin{eqnarray}\label{20}
\left ( \frac{d~\vec{v}}{d~t} \right )_{in} &=&
\frac{S A' \bar{Q'}_{ext}}{m c} ~\left [ \left ( 1 -
\frac{\vec{v}\cdot\vec{e}}{c} \right ) \vec{e} ~-~ \frac{\vec{v}}{c} \right ] ~,
\nonumber \\
\left ( \frac{d~\vec{v}}{d~t} \right)_{out} &=& -~
\frac{S A'\bar{Q'}_{ext}}{m c} \left ( 1 - \bar{Q'} \right )
\left [ \left ( 1 -
\frac{\vec{v} \cdot \vec{e}}{c} \right ) \vec{e} ~-~ \frac{\vec{v}}{c} \right ] ~,
\nonumber \\
\bar{Q'} &\equiv& \frac{\bar{Q'}_{pr}}{\bar{Q'}_{ext}} ~.
\end{eqnarray}

Covariant equation of motion of the particle yields
\begin{eqnarray}\label{21}
\frac{dE}{dt} &=& S~ A' ~\bar{Q'}_{ext} ~ \bar{Q'}~\left\{
\frac{\vec{v} \cdot \vec{e}}{c}~-~
\left(\frac{\vec{v} \cdot \vec{e}}{c}\right)^{2}~-~
\left(\frac{\vec{v}}{c}\right)^{2} \right\}~,
\nonumber \\
\frac{d\vec{p}}{dt} &=& \frac{S A'\bar{Q'}_{ext}}{c} ~\bar{Q'}~
\left \{ \left ( 1 ~-~ \frac{\vec{v} \cdot \vec{e}}{c} \right ) \vec{e} ~-~
\frac{\vec{v}}{c} \right \} ~,
\nonumber\\
\bar{Q'} &\equiv& \frac{\bar{Q'}_{pr}}{\bar{Q'}_{ext}}
\end{eqnarray}
and the acceleration of the particle is
\begin{eqnarray}\label{22}
\frac{d\vec{v}}{dt} &=& \frac{S A'\bar{Q'}_{ext}}{m c} ~\bar{Q'}~
\left \{ \left ( 1 ~-~ \frac{\vec{v} \cdot \vec{e}}{c} \right ) \vec{e} ~-~
\frac{\vec{v}}{c} \right\} ~,
\nonumber\\
\bar{Q'} &\equiv& \frac{\bar{Q'}_{pr}}{\bar{Q'}_{ext}} ~.
\end{eqnarray}

\section{Radiation pressure and the Poynting-Robertson drag}
The P-R effect is characterized by the following equation of motion
to the first order in $v/c$:
\begin{eqnarray}\label{23}
\frac{d \vec{v}}{dt} &=& \beta~\frac{GM}{r^{2}}~
\left \{ \left ( 1 ~-~ \frac{\vec{v} \cdot \vec{e}}{c} \right ) \vec{e} ~-~
\frac{\vec{v}}{c} \right \} ~,
\nonumber\\
\beta &\equiv& \frac{\pi R^{2} L}{4 \pi G M m c} ~\bar{Q'}_{pr} ~,
\end{eqnarray}
where $G$ is a gravitational constant, $L$ and $M$ are luminosity and mass
of the central star around which the particle moves, $r$ is a distance of the
particle from the star and $R$ is a radius of the spherical particle.

Gustafson (1994, p. 566) writes: "The velocity-independent radial term
representing force [acceleration] due to radiation pressure... -- although
not a pressure, it is often referred to simply as 'radiation pressure' ":
\begin{equation}\label{24}
\left(\frac{d\vec{v}}{dt}\right)_{pressure}~=~
\beta~\frac{GM}{r^{2}}~\vec{e}~.
\end{equation}
According to a definition (see, e.g. Gustafson 1994, p. 566, or, Dohnanyi
1978, pp. 562-565), the velocity-dependent part of Eq. (23) is the
P-R drag:
\begin{equation}\label{25}
\left(\frac{d\vec{v}}{dt}\right)_{P-R~ drag}~=~
-\beta~\frac{GM}{r^{2}}~\left\{\left( \frac{\vec{v}\cdot\vec{e}}{c}
\right ) ~\vec{e} ~+~ \frac{\vec{v}}{c} \right \}~.
\end{equation}

We can decompose the velocity vector $\vec{v}$ of the particle into the
radial and transversal parts
\begin{equation}\label{26}
\vec{v} ~=~ v_{R}~\vec{e}~+~v_{T}~\vec{e}_{T}~\equiv~
\dot{r}~\vec{e}~+~r \dot{\theta}~\vec{e}_{T}~,
\end{equation}
where the unit vector $\vec{e}_{T}$ is normal to $\vec{e}$ in the orbital
plane. Substituting Eq. (26) into Eq. (23) yields
\begin{equation}\label{27}
\frac{d\vec{v}}{dt} ~=~ \beta ~ \frac{GM}{r^{2}} ~\left \{ \left ( 1 ~-~
	  2 ~\frac{\dot{r}}{c} \right ) ~\vec{e} ~-~ \frac{r \dot{\theta}}{c}
	  ~\vec{e}_{T} \right \}~.
\end{equation}
On the basis of Eqs. (25) and (26) we can write for the P-R drag
\begin{equation}\label{28}
\left ( \frac{d\vec{v}}{dt} \right )_{P-R~drag} ~= -~\beta ~\frac{GM}{r^{2}}~
	\left ( 2~ \frac{\dot{r}}{c} ~\vec{e} ~+~
	\frac{r ~\dot{\theta}}{c} ~\vec{e}_{T} \right )~.
\end{equation}

We have to stress that any decomposition to the "radiation pressure" (Eq. 24)
and the "P-R drag" (Eqs. 25 and 28) is nonphysical. As follows from
covariant formulation, the physical equation is Eq. (23).

Moreover, avoiding the usage of the "P-R drag" not only represents
correct physics, but any confusion about the "P-R drag" is removed.
This is also important since, besides definition represented by Eq. (25),
also another "P-R drag" is used. It contains only
the term proportional to $-$ $\vec{v}$/$c$ in the equation of motion --
the first term in brackets on the right-hand side of Eq. (25) is omitted
(see, e.g., Wyatt 2009). One must bear in mind
that the two different definitions of the P-R drag lead to two
different results for secular changes of orbital elements
[$da_{\beta} / dt$ $=$ $-$ $\beta$ ($GM/c$) $A_{P-R~drag}$ / ($a_{\beta}$
(1 $-$ $e_{\beta})^{3/2}$),
$de_{\beta} / dt$ $=$ $-$ $\beta$ ($GM/c$) $E_{P-R~drag}$ / ($a^{2}_{\beta}$
$(1 - e_{\beta})^{1/2}$), where
$A_{P-R~drag}$ $=$ 2 $+$ 3 $e^{2}_{\beta}$,
$E_{P-R~drag}$ $=$ (5/2) $e_{\beta}$ for the first type of the
P-R drag and
$A_{P-R~drag}$ $=$ 2 $+$ 2 $e^{2}_{\beta}$,
$E_{P-R~drag}$ $=$ 2 $e_{\beta}$ for the second type of the
P-R drag;  $a_{\beta}$ and $e_{\beta}$ denote semimajor axis and eccentricity
of the particle's orbit around the central star,
see Eqs. 50-67 in Kla\v{c}ka 2004].
Unfortunately, another definition of the P-R drag exists:\\
$(d\vec{v}/dt)_{P-R~drag}$ $=$ $-~\beta~(GM/r^{2})~(r\dot{\theta}/c)~
\vec{e}_{T}$, \\
see, e.g. Minato et al. (2004). Really, confusion in the literature exists.
And physics is ignored.

The P-R effect is the effect of radiation pressure on a moving spherical
(nonrotating) body. Thus, any separation between the radiation pressure
and the P-R effect is also nonphysical (see, e.g., Leinert and Gr\"{u}n
1990; Mann 2009; see also Minato {\it et al.} 2004 -- Sec. 3.2).

\section{P-R effect and statements presented in the literature}
As we have already mentioned at the beginning of the paper, much confusion
exists in explanations of the P-R effect. As for the papers by Poynting
(1903), Robertson (1937), Wyatt and Whipple (1950) and Burns {\it et al.}
(1979), the physics was partially commented in Kla\v{c}ka (2008a, 2008b) 
and also in Kla\v{c}ka 1992a (Sec. 2.5). We will
deal mainly with newer explanations presented in the literature, in this
section.

We have to bear in mind that the results obtained by Kla\v{c}ka (2008a, 2008b)
are consistent with Maxwell's equations and relativity theory. Since
relativity theory was motivated by better understanding of Maxwell's
equations, we consider Maxwell's equations (and their consequences)
as an indispensable part of the relativity theory. Classical physics
considers that interactions spread with an infinite speed. Thus,
all the results presented in Kla\v{c}ka (2008a, 2008b) have to be
considered as a consequence of the relativity theory. Any
statement of the type "radiation forces -- including the Poynting-Robertson
drag -- are {\it fundamentally classical} forces and are not
produced by relativistic effects as commonly believed" (Burns {\it et al.}
1979, p. 8) is not physically correct.
As an example, we can mention the calculation of $\bar{Q'}_{pr}$. It is
based on Mie's solution of Maxwell's equations and it cannot be based
on classical physics inconsistent with relativity theory. Similarly,
the flux density of radiation energy transforms according to the
relation $S'$ $=$ $S$ (1 $-$ 2 $\vec{v} \cdot \vec{e} / c$) and not
according to $S'$ $=$ $S$ (1 $-$ $\vec{v} \cdot \vec{e} / c$) which
corresponds to "fundamentally classical" result of Burns {\it et al.}
(1979, p. 5). Similarly, the fundamental condition is $\vec{p'}_{outgoing}$
$=$ (1 $-$ $\bar{Q'}_{pr}$ $/$ $\bar{Q'}_{ext}$) $\vec{p'}_{incoming}$
(see Eq. 16) and not the "fundamentally classical" condition
$\vec{p'}_{outgoing}$ $=$ (1 $-$ $\bar{Q'}_{pr}$)  $\vec{p'}_{incoming}$
corresponding to ideas of Poynting (1903), Robertson (1937),
Wyatt and Whipple (1950), Burns {\it et al.} (1979, pp. 10-11) and others.
Similarly, Eqs. (7) and (11) hold. Similarly, if only thermal emission
is considered, then the decrease of particle's mass is $dm/d \tau$ $=$ $-$
$E_{o}' / c^{2}$, where $E_{o}'$ is the outgoing energy per unit time
(see Eqs. 40-41 in Kla\v{c}ka 2008b). Similarly, light does not obey the law
of addition of velocities, according to Einstein's second postulate
["Principle of the invariance (constancy) of the speed of light:
The speed of light in a vacuum has the same value in each inertial reference
frame, irrespective of the velocities of the light source or the light receiver."
(Freund 2008, p. 9)] Simply, the processes participating in the P-R effect
are correctly explained using ideas of relativity theory
and not ideas of classical nonrelativistic physics.

\subsection{Burns {\it et al.} (1979)}
Since the paper by Burns {\it et al.} (1979) is generally considered
to be the most relevant contribution on the P-R effect,
we will present the most important comments on the paper.
Some information can be found also in
papers by Kla\v{c}ka (1992a -- Sec. 2.5, 2008a, 2008b). \\
\noindent\\
1. \\
Burns {\it et al.} (1979, p. 5) write: \\
"The force on a perfectly absorbing particle due to solar (photon)
radiation can be viewed as composed of two parts: (a) a {\it radiation
pressure} term, which is due to the initial interception by the
particle of the incident momentum in the beam, and (b) a {\it mass-loading
drag}, which is due to the effective rate of mass loss from the moving
particle as it continuously reradiates the incident energy." \\

Physics: \\
\nopagebreak

We have $\bar{Q}'_{ext}$ $=$ 2, $\bar{Q}'_{pr}$ $=$ 1 for the case
(see Secs. 7.3.3 and 7.3.4 in Kla\v{c}ka 2008b). As for the incoming
radiation, we can write, on the basis of Eq. (1)
(see also Eq. 75 in Kla\v{c}ka 2008b), \\
$d p^{\mu}_{in} / d \tau$ $=$ (2 $w^{2} S A' / c$) $b^{\mu}$ ~, \\
$(d m / d \tau)_{in}$ $=$  2 $w^{2} S A' /c^{2}$ ~, \\
or, to the first order in $\vec{v} / c$, \\
$d \vec{p}_{in} / d t$ $=$ (2 $S A' / c$) (1 $-$ $\vec{v} \cdot \vec{e}
/ c$) $\vec{e}$~. \\
The outgoing radiation effects on the particle by the following force
(Eq. 76 in Kla\v{c}ka 2008b): \\
$(d p^{\mu}_{out} / d \tau)_{particle}$ $=$ $-$ ($w^{2} S A' / c$)
($b^{\mu}$ $+$ $u^{\mu} / c$)~, \\
$(d m / d \tau)_{out}$ $=$ $-$ 2 $w^{2} S A' / c^{2}$ ~, \\
or, to the first order in $\vec{v} / c$, \\
$(d \vec{p}_{out} / d t)_{particle}$ $=$ $-$ ($S A' / c$) [
(1 $-$ $\vec{v} \cdot \vec{e} / c$)
$\vec{e}$ $+$ $\vec{v} / c$ ] . \\
If (2 $S A' / c$) (1 $-$ $\vec{v} \cdot \vec{e} / c$) $\vec{e}$ would be
the "radiation pressure term", then $-$ ($S A' / c$) [
(1 $-$ $\vec{v} \cdot \vec{e} / c$) $\vec{e}$ $+$ $\vec{v} / c$ ] should be
the "P-R drag" ("mass loading drag")!
The readers should be utterly confused by the "physics"
(see also Sec. 6.5; probably, the authors Burns {\it et al.} had not in mind
these equations, when they formulated the above presented text). \\
\noindent\\
2.  \\
Burns {\it et al.} (1979, p. 5) write: \\
"$S'$ $=$ $S$ (1 $-$ $\dot{r}/c$)~, ~~~ (1-B) \\
... The momentum removed per second from the incident beam as seen
by the particle is then ($S' A / c$) $\hat{\vec{S}}$." \\

Physics: \\
\nopagebreak

The momentum removed per second from the incident beam as seen
by the particle is ($S' A / c$) $\hat{\vec{S}}'$, where
$S'$ $=$ $S$ (1 $-$ 2 $\dot{r}/c$) ~, \\
if the accuracy to the first order in $\dot{r}/c$ is used
(see also Sec. 2.5 in Kla\v{c}ka 1992a, or, Eqs. 20-25 in
Kla\v{c}ka 2004, or, Kla\v{c}ka 2008b). \\
\noindent\\
3. \\
Burns {\it et al.} (1979, p. 5) write about the perfectly absorbing
particle: \\
"The absorbed energy flux $S' A$ is continuously reradiated from the
particle. Since the reradiation is nearly isotropic (small particles
being effectively isothermal), there is no net force exerted thereby
on the particle in its own frame. However, the reradiation is {\it equivalent}
to a mass loss rate of $S' A / c^{2}$ from the moving particle -- which has
velocity $\vec{v}$ as seen from the inertial frame of the Sun -- and,
as measured in the solar frame, this gives rise to a momentum flux
from the particle $-$ ($S' A / c^{2}$) $\vec{v}$, schematically shown in
Fig. 2." \\

Physics: \\
\nopagebreak

The force exerted on the particle, in its own frame, due to the outgoing
radiation, is: \\
$(d p^{\mu}_{out} / d \tau)_{particle}$ $=$ $-$
     ($w^{2} S A' \bar{Q'}_{ext} / c$)  $\left \{
     \bar{Q}' ~ u^{\mu} / c ~+~
     \left ( 1 - \bar{Q'} \right ) ~ b^{\mu} \right \}$ , \\
$\bar{Q}'$ $\equiv$ $\bar{Q'}_{pr} / \bar{Q'}_{ext}$ , \\
or, putting $\vec{v}$ $=$ 0 and
$\bar{Q'}_{pr}$ $=$ 1,  $\bar{Q'}_{ext}$ $=$ 2, \\
$(d\vec{p}'_{out} / d \tau)_{particle}$ $=$ $-$
     ($S' A' / c$)  $\vec{e}'$ ,   \\
according to Eq. (1).
Thus, the statement of Burns {\it et al.} (1979, p. 5) that
"there is no net force exerted thereby
on the particle in its own frame" is not true.

The following statement of Burns {\it et al.} (1979, p. 5) is:
"However, the reradiation is {\it equivalent}
to a mass loss rate of $S' A / c^{2}$ from the moving particle".
We have, in reality, according to Eq. (11), \\
$ ( d m / d \tau)_{out}$ $=$ $-$ 2 $w^{2} S A'/ c^{2}$ . \\
since $\bar{Q'}_{ext}$ $=$ 2 for the case discussed by the authors.

Again, the authors write: "this gives rise to a momentum flux
from the particle $-$ ($S' A / c^{2}$) $\vec{v}$, schematically shown in
Fig. 2." In reality, the
force exerted on the particle due to the outgoing
radiation is, according to Eq. (1): \\
$(d p^{\mu}_{out} / d \tau)_{particle}$ $=$ $-$
     ($w^{2} S A' \bar{Q'}_{ext} / c$)  $\left \{
     \bar{Q}' ~ u^{\mu} / c ~+~
     \left ( 1 - \bar{Q'} \right ) ~ b^{\mu} \right \}$ , \\
$\bar{Q}'$ $\equiv$ $\bar{Q'}_{pr} / \bar{Q'}_{ext}$ , \\
or, for the case treated by the authors, \\
$(d p^{\mu}_{out} / d \tau)_{particle}$ $=$ $-$
     ($w^{2} S A' / c$)  ($u^{\mu} / c ~+~
     b^{\mu} $) . \\
Approximation to terms of order $v/c$ yields: \\
$(d \vec{p}_{out} / d t)_{particle}$ $=$ $-$ ($w^{2} S A' / c$)
    ($\vec{v} / c ~+~ \vec{e} / w $) ~. \\
\noindent\\
4. \\
On the basis of the correct results summarized and discussed in the previous
three points (1., 2. and 3.) the reader immediately see that also Fig. 2
is physically incorrect. The outgoing radiation in the particle's frame
is not isotropic. Also the results $S'$ $=$ (1 $-$ 2 $\dot{r} / c$) $S$
are not taken into account. Also the result \\
$(d \vec{p}_{out} / d t)_{particle}$ $=$ $-$ ($w^{2} S A' / c$)
     ($\vec{v} / c ~+~ \vec{e} / w$) \\
is not considered. \\
\noindent\\
5. \\
Burns {\it et al.} (1979, p. 6) write: \\
"The net force on the particle is then the sum of the forces due to the
impulse exerted by the incident beam and the momentum density loss from the
particle or, for a particle of mass $m$, \\
$m$ $\dot{\vec{v}}$ $=$ ($S' A / c$) $\hat{\vec{S}}$ $-$
			($S' A / c^{2}$) $\vec{v}$ ~... ~(2)" \\

Physics: \\
\nopagebreak

The equation does not corresponds to correct physics.

At first, since $S'$ is the integrated flux density measured by the moving
particle (Burns {\it et al.} 1979, p. 5), the quantity ($S' A / c$)
cannot be used as the relevant quantity for the equation of motion
of the particle in the rest frame of reference of the source of light
(Sun). The second Newton's law does not admit such kind of mixing between
two different reference frames (simultaneous considering of
momenta measured in two different reference frames).

Secondly, the statement of Burns {\it et al.} (1979 -- Eq. 2) that \\
$d \vec{p}_{in} / d t$	$=$ ($S' A / c$) $\hat{\vec{S}}$ \\
$(d \vec{p}_{out} / d t)_{particle}$  $=$ $-$ ($S' A / c^{2}$) $\vec{v}$ \\
is incorrect.
The forces acting on the particle are given by Eqs. (1), or,
to the first order in $v/c$: \\
$d \vec{p}_{in} / d t$ $=$ (2 $w^{2} S A' / c$) $\vec{e} / w$ , \\
$(d \vec{p}_{out} / d t)_{particle}$ $=$ $-$ ($w^{2} S A' / c$)
     ($\vec{v} / c ~+~ \vec{e} / w $) , \\
$w$ $=$ 1 $-$ $\vec{v} \cdot \vec{e} / c$~. \\
\noindent\\
6. \\
Burns {\it et al.} (1979, p. 8) write: \\
" ... we have that \\
$\vec{F}_{rad}$ $=$ $m$ $\dot{\vec{v}}$ $=$ $Q_{pr}$ $M'$
($\vec{c}$ $-$ $\vec{v}$) , \\
where $M'$ is the "mass" of the photons which strike the particle or
($S' A / c^{2}$). In the above form, the radiation force can be
viewed as a drag force caused by the relative velocity $\vec{c}$ $-$ $\vec{v}$,
of the particle through a beam of photons. Finally, to terms of order $v/c$,
we can write the net force on the particle as \\
$m$ $\dot{\vec{v}}$ $=$ ($S A / c$) $Q_{pr}$ [ (1 $-$ $\dot{r}/c$)
			$\hat{\vec{S}}$ $-$
			$\vec{v} / c$ ]~. ~(5) \\
For a totally absorbing particle this reduces to (2) since then $Q_{pr}$ $=$ 1.
Equation (5) is identical to Robertson's expression except for being more
general by the inclusion of the important factor $Q_{pr}$. We note, however,
from the above derivation, that radiation forces -- including the
Poynting-Robertson drag" are {\it fundamentally classical} forces and are not
produced by relativistic effects as commonly believed." \\

Physics: \\
\nopagebreak

Our Eq. (8) yields that \\
$M'$ $\equiv$ $( d m / d \tau)_{in}$ $=$ $w^{2} S A' \bar{Q'}_{ext}/ c^{2}$. \\
Even if we take into account that $S'$ $=$ $w^{2} S$ (although we have already
discussed that Burns {\it et al.} use incorrect result given by their
Eq. 1), the correct result for $M'$ differs from the result presented by
the authors. Substituting the correct result for $M'$ into the equation
presented above Eq. (5) of Burns {\it et al.}, the authors would not
obtain Eq. (5).

We must admit that we have received the correct
result for $M'$ on the basis of relativity theory (Kla\v{c}ka 2008a, 2008b)
and not on the basis of {\it fundamentally classical} physics, as was
the access of the Burns {\it et al.}.

There is another problem with the access of the authors. They use
"the relative velocity $\vec{c}$ $-$ $\vec{v}$,
of the particle through a beam of photons".
There is a problem. We know that we can write $S'$ $=$ $n'$ $h$ $\nu '$ $c$
and, similarly, $S$ $=$ $n$ $h$ $\nu$ $c$, in the case of relativity
theory (Kla\v{c}ka 2008b -- Eqs. 20-23). Using transformations between
$n$ and $n'$, $\nu$ and $\nu '$, we obtain the correct relation
$S'$ $=$ $w^{2}$ $S$. One can say that the transformations between
$n$ and $n'$, $\nu$ and $\nu '$ can be obtained also on the basis of
{\it fundamentally classical} physics (see Sec. 2.4 in Kla\v{c}ka 1992a
and Doppler effect). And the result is consistent with the result
known from electromagnetism (see Sec 2.5 in Kla\v{c}ka 1992a).
But the access still assumes that $c'$ $=$ $c$, so there is no
{\it fundamentally classical} physics. How is it possible? The explanation
is evident: the result of electromagnetism is consistent with Maxwell's
equations and they are consistent with relativity (Maxwell's equations
motivated the creation of relativity).
If one would like to use also the transformation suggested by the
authors, "the relative velocity $\vec{c}$ $-$ $\vec{v}$", one would
obtain, on the basis of {\it fundamentally classical} physics, that
$S'$ $=$ $w^{3}$ $S$. And this result would be different from the correct
$S'$ $=$ $w^{2}$ $S$ and also from the result used by Burns {\it et al.}
(1979 -- Eq. 1) $S'$ $=$ $w$ $S$.

It seems to us that the statement "radiation forces are
{\it fundamentally classical} forces and are not produced by relativistic
effects as commonly believed" (Burns {\it et al.} 1979, p. 8) is
irrelevant. The complete correct results were obtained on the basis
of relativistic approach (Kla\v{c}ka 2008a, 2008b) and the results
are, at least in some partial steps, different from those presented
before the two papers (Kla\v{c}ka 2008a, 2008b). \\
\noindent\\
7.   \\
As for the relativistic approach presented by
Burns {\it et al.} (1979, Sec. IV), the authors consider \\
$\vec{p}_{o}$ $=$ (1 $-$ $Q_{pr}$) $E_{i} / c$ ...\\
(see Eqs. 10, 11, 14 in Burns {\it et al.} 1979).
Although the motivation for the value \\
$E_{i}$ $=$ $(1-\dot{r}/c)~S A$ \\
is not clear (see also Kla\v{c}ka 1992a -- Sec. 2.5), the important result
is that the usage of $\vec{p}_{o}$ $=$ (1 $-$ $Q_{pr}$) ... would produce
inconsistencies between relativity theory and Mie's solution of
Maxwell's equations (Kla\v{c}ka 2008a, 2008b).
The approach of Burns {\it et al.} (1979) leads to the inequality
0 $<$ $Q_{pr}$ $<$ 2, which is not consistent with the Mie's solution
of Maxwell's equations. Thus, even the
relativistic approach by Burns {\it et al.} (1979, Sec. IV) is not
physically correct. It is not consistent with relativity theory.
The correct equation is given by Eq. (16).
Moreover, the correct equation for $E_{i}$ is \\
$E_{i}$ $=$ $w S A' \bar{Q}_{ext}'$ \\
(Kla\v{c}ka 2008b -- Eqs. 11, 19, and 23)
and this is different from the result presented by Burns {\it et al.}
(1979 -- below Eq. 16). \\
\noindent\\
8. \\
We have discussed the result for $\left ( d m / d \tau \right )_{in}$
in Sec. 2.1. The approach of Burns {\it et al.} (1979) would yield
$\left ( d m / d \tau \right )_{in}$ $=$ $w S A' / c^{2}$ (but the
authors did not treat the problem  in a relativistically covariant
form, so they could not present the result for
$\left ( d m / d \tau \right )_{in}$).
Thus, the velocity dependence would be given by $w^{1}$ and not
by $w^{2}$. Moreover, optical properties of the particle would be
represented only by the particle geometrical cross-section $A'$ and not
by the extinction cross-section $A' \bar{Q'}_{ext}$. \\
\noindent\\
9. \\
Burns {\it et al.} (1979, Sec. V) discuss "solar wind corpuscular forces".
We want to say that the right-hand side of their Eq. (17*) must
be multiplied by a factor $x'$ depending on material properties
of dust particle (1 $<$ $x'$ $<$ 3, approximately,
see Eq. 29 in Kla\v{c}ka and Saniga 1993).
The force for the solar wind (if only radial velocity of the solar
wind is considered) is: \\
$\vec{F}_{sw}$ $=$ $F_{sw}$ [ (1 $-$
$\vec{v} \cdot \vec{e} /  v_{sw}$) $\vec{e}$ $-$
$x'$ $\vec{v} / v_{sw}$ ]~, \\
where $F_{sw}$ is the force on the particle and
$v_{sw}$ is the heliocentric solar-wind speed.

\subsection{Shive and Weber (1982), Dohnanyi (1978)}
Shive and Weber (1982, p. 221) present: "The particle absorbs light
and heat from the Sun and re-radiates photons uniformly in all
directions. However, as seen by an observer in a frame of reference
stationary with respect to the Sun, the photons emitted by the particle
in the forward direction of its orbital motion will be Doppler-shifted
towards higher frequencies than those emitted in the backward direction.
The energy density ahead of the particle will accordingly be greater
than that behind it, and the particle experiences a net tangential
radiation pressure which acts as a drag on its orbital motion. So it spirals
inward and is eventually engulfed by the Sun. Most of the interplanetary
dust originally present at the creation of the solar system has long ago
been swept out by this process."

Similar explanation can be found in Dohnanyi (1978, pp. 562-563):
"In the system of coordinates in which the particle is at rest, the Sun's
rays (and the force of radiation pressure as well) will not appear
to come from the centre of the Sun, but at an angle and in such a direction
as to oppose the particle's transverse component of motion. Analysis of
this effect, known as the aberration of light ... The opposing transverse
force is then [equal to the P-R effect].
An alternative explanation of this effect is obtained when we view the
particle's motion from the system of coordinates in which the Sun is at rest.
The direction of the Sun's rays and that of the force of radiation pressure
on the particle are now exactly in the radial direction. Because of the
particle transverse motion, the radiation reflected, diffracted, or emitted
by the particle is shifted in its wavelengths asymmetrically: radiation
leaving the particle in the direction of motion is blue-shifted and in the
other direction it is red-shifted. The net result is a drag force of the
same magnitude as equation [for the P-R effect]." \\

Physics of the phenomenon: \\
\nopagebreak
\noindent\\
1. \\
If the case treated by Poynting (1903) and Robertson (1937) is considered,
then $\bar{Q}'_{pr} / \bar{Q}'_{ext}$ $=$ 1/2 and Eq. (12), or Eq. (20),
yields that the outgoing radiation causes acceleration of the particle
and not its deceleration $(d\vec{v}/dt \propto +~\vec{v}/c)$. This is different from the statement done by
Shive and Weber (1982) and Dohnanyi (1982). If $\bar{Q}'_{pr}$ $/$
$\bar{Q}'_{ext}$ $=$ 1, then acceleration of the particle equals to zero
in both reference frames, i. e., in the frame of reference of the
particle and in the frame of reference stationary with respect to the Sun
(see Eqs. 12 or 20). The important phenomenon is that mass of the
particle decreases in the process of reradiation and this
leads to an increase of particle's acceleration: $d\vec{p}/dt$ $=$ ...,
$d\vec{v}/dt$ $=$ ($1/m$) [ $-$ ($dm/dt$) $\vec{v}$ $+$ ...] and the value
$dm/dt$ is given by Eq. (11) in the case of the P-R effect.

Maybe, there exists more simple illustration of the correct physics. If the
particle is not irradiated and only outgoing radiation due to the thermal
emission exists, then Eqs. (40)-(41) in Kla\v{c}ka (2008b) hold. As for
the spherical particle, one obtains
\begin{eqnarray}\label{29}
\left ( \frac{d p^{\mu}}{d \tau} \right )_{e} &=& -~
\frac{E'_{o}}{c} ~\frac{u^{\mu}}{c}  ~,
\nonumber \\
\left ( \frac{d u^{\mu}}{d \tau} \right )_{e} &=& 0 ~,
\nonumber \\
\left ( \frac{d m}{d \tau} \right )_{e} &=& -~
\frac{E'_{o}}{c^{2}} ~,
\end{eqnarray}
where $E'_{o}$ is thermal emission energy per unit time. The first of Eqs. (29)
really shows that a force acting against the motion of the particle exists,
but the second of Eqs. (29) states that no acceleration of the particle exists
and this holds in all reference frames.
The particle is decelerated if the following asymmetry exists, in the
proper frame of the particle: the total energy per unit time emitted
in the forward direction of particle's motion (with respect to the
surroundings) is greater than the corresponding energy
emitted in the backward direction (see Eqs. 40-41 in Kla\v{c}ka 2008b).

The flux density of radiation energy is influenced not only by the Doppler
effect, but also by the change of concentration of photons: $S'$ $=$
$w^{2}$ $S$. Moreover, the term $- u^{\mu} / c$ in Eq. (14) is generated by
the conservation of particle's mass (see Sec. 7.4.5 in Kla\v{c}ka 2008b)
and not by the shifts in wavelengths or frequencies. Similarly, the last two
equations of Eqs. (29) immediately lead to the first of Eqs. (29), if one
uses the fact that $p^{\mu}$ $=$ $m$ $u^{\mu}$. \\
\noindent\\
2. \\
As for the force acting on the particle due to the outgoing radiation for
the P-R effect, one has to use Eqs. (1) and (9): \\
$d p^{\mu} / d \tau$ $=$ $-$ ( $w^{2} S A' \bar{Q'}_{ext} / c$ ) [
$(\bar{Q'}_{pr}/\bar{Q'}_{ext}) ~ u^{\mu} / c ~+~
     ( 1 - \bar{Q'}_{pr} / \bar{Q'}_{ext} ) ~ b^{\mu}$ ] ~. \\
It is easily seen that this force is not of the type represented by Eq. (29),
if $\bar{Q'}_{pr} / \bar{Q'}_{ext}$ $\ne$ 1
($\bar{Q'}_{pr} / \bar{Q'}_{ext}$ $=$ 1/2 for the case treated by Poynting,
Robertson, Shive and Weber, Dohnanyi). Thus, the result does not correspond
to the idea of Shive and Weber (1982) and Dohnanyi (1978). \\
\noindent\\
3. \\
Let us consider the following simple thought experiment. Photons are
emitted from the two opposite rectangular faces of a cuboid
(rectangular prism). The total outgoing
energy per unit time is $E'_{o}$ $=$ 2 $S' A'$, where $A'$ is geometrical
cross-section of the rectangular faces. Then \\
($dp^{\mu}/d \tau$)$_{e}$ $=$ $-$ 2 ($S' A'/c$)
$u^{\mu}/c$ , \\
according to Eq. (29). This result can be obtained using the idea presented
by Dohnanyi (1978) and Shive and Weber (1982). However, also the change of
concentration of photons has to be taken into account, not only the Doppler
effect: $S'$ $=$ $w_{f}^{2} S_{f}$ $=$ $w_{b}^{2} S_{b}$, where the subscripts
"f", "b"  denote abbreviations of "forward" and "backward",
$w_{f}$ $=$ $\gamma (1 - v/c)$, $w_{b}$ $=$ $\gamma (1 + v/c)$.
According to Eqs. (18)-(20) and (24) in Kla\v{c}ka (2008b), we can write \\
$( dp^{\mu} / d\tau )_{e}$ $=$ $-$ [ ($w_{f}^{2} S_{f} A'/c$)
$b^{\mu}_{f}$ $+$ ($w_{b}^{2} S_{b} A'/c$) $b^{\mu}_{b}$ ] ~, \\
where $b^{\mu}_{f}$ $=$ ($1/w_{f}$ ; $\hat{\vec{v}} /w_{f}$),
$b^{\mu}_{b}$ $=$ ($1/w_{b}$ ; $- ~\hat{\vec{v}} /w_{b}$);
$~\hat{\vec{v}}$ is unit vector in the forward direction and orientation of the
emitted photons.
As a consequence, deceleration of the cuboid equals to zero:
$d u^{\mu} / d \tau$ $=$ 0.

Comment: \\
The four-vectors $b^{\mu}_{f}$, $b^{\mu}_{b}$ are important in the
previous equation. We can easily show it as follows. The forward and
backward directions would yield (to the first order in $v/c$)
$S_{f}$ $=$ $S'$ (1 $+$ 2 $v/c$),
$S_{b}$ $=$ $S'$ (1 $-$ 2 $v/c$).
But the  form
$dp/dt$ $=$ $-$ ($S_{f} A'/c$ $-$ $S_{b} A'/c$)
	      $=$ $-$ 4 ($S' A'/c$) $v/c$
would not correspond to the physical result. Surprisingly, the idea of
Burns {\it et al.} (1979) $S_{f}(B)$ $=$ $S'$ (1 $+$ $v/c$),
$S_{b}(B)$ $=$ $S'$ (1 $-$ $v/c$) -- results of classical mechanics --
would yield the correct final result! Perhaps, Shive and Weber (1982)
were influenced by the statement of Burns {\it et al.} (1979). Maybe,
classical access would be of the form
$dp/dt$ $=$ $-$  [ $S_{f} A'/$($c$ $+$ $v$) $-$ $S_{b}
A'/(c$ $-$ $v$)]
	      $=$ $-$ 2 ($S' A'/c$) $v/c$.
However, understanding nature requires much more effort than simple
collection and combination of some {\it ad hoc} formulae. \\
\noindent\\
4. Aberration of light? \\
As for the heuristic explanation of the term containing $- \vec{v}/c$ in
the P-R effect as a consequence of the aberration of light, we refer
the reader to Sec. 7.4.5 in Kla\v{c}ka (2008b).
Moreover, this can be easily seen from
Eqs. (1)-(2), or from the form to the first order in $\vec{v}/c$: \\
$\vec{F}_{in}$ $=$ ($w_{1}^{2} S A' \bar{Q}'_{ext} / c$)
(1 $+$ $\vec{v} \cdot \vec{e} / c$) $\vec{e}$ ~. \\
The form of the force $\vec{F}_{in}$ does not contain the aberrational term \\
(1 $+$ $\vec{v} \cdot \vec{e} / c$) $\vec{e}$ $-$ $\vec{v} / c$ ~. \\
Perhaps, the following idea can appear: the aberrational term exists
in the acceleration $\vec{a}_{in}$ $\equiv$ $\dot{\vec{v}}_{in}$. Really, something like this is produced by the covariant equation Eq. (8): \\
$d \vec{v}_{in} / d t$ $=$ [$w_{1}^{2} S A' \bar{Q}'_{ext} / (m c) $]
[ (1 $+$ $\vec{v} \cdot \vec{e} / c$) $\vec{e}$ $-$ $\vec{v} / c$ ]  \\
and this term is even almost consistent with the P-R effect!
However, there is the term $\bar{Q}'_{ext}$
instead of the term $\bar{Q}'_{pr}$, and, the relevant four-vector
$b^{\mu}$ $-$ $u^{\mu}/c$ is not aberration of light. And, for the special
case treated by Poynting (1903), Robertson (1937) and others,
$\bar{Q}'_{ext}$ $=$ 2, $\bar{Q}'_{pr}$ $=$ 1 (Kla\v{c}ka 2008a, 2008b).
Thus, aberration of light does not explain the term $-$ $\vec{v}/c$
in the P-R effect.

\subsection{Mignard (1982)}
Mignard (1982, p. 349) writes: "$Q$ [$\bar{Q'}_{pr}$] is a global
coefficient which expresses the particle surface properties related
to the incoming light." \\

Physics: \\
\nopagebreak

The dimensionless efficiency factor
for radiation pressure (integrated over stellar spectrum and calculated
for radial direction) considers volume properties of the particle, in general.
It results from the solution of Maxwell's equations (Mie 1908).

\subsection{Kapi\v{s}insk\'{y} (1984)}
Kapi\v{s}insk\'{y} (1984) discusses disturbing and dissipative
(destructive and disruptive) effects acting on interplanetary dust
particles. The author explains the action of the solar electromagnetic
radiation (Kapi\v{s}insk\'{y} 1984, disturbing forces -- p. 104):

"Effect of solar electromagnetic radiation (direct light pressure): \\
This effect is closely related to the Poynting-Robertson effect. Among
the non-gravitational effects, these two are of the greatest importance
for studying the dynamics of small particles in the Solar System.
The outdirected solar pressure always acts on the particle against the
gravitational force of the Sun."

Continuation on the same page is: \\
"Poynting-Robertson effect: Apart from the outdirected solar pressure acting
on the particle, the absorption of solar energy by the particle and its
isotropic emission causes a small force to be generated along the tangent
to the trajectory which decelerates the particle in the orbit. This tangential
force decreases the kinetic energy and the orbital angular momentum of the
particle. Consequently, the particle is forced to move to an orbit closer
to the Sun." \\

Physics: \\
\nopagebreak

The "effect of solar electromagnetic radiation" on spherical dust particle
is equivalent to the Poynting-Robertson effect, not to the
"direct light pressure". Direct light pressure is a velocity independent
part of the P-R effect. Thus, it is not possible to classify
"direct light pressure" and the P-R effect as two different disturbing effects
influencing motion of the particle.

Simlarly, we have the Poynting-Robertson effect and not three effects:
"direct light pressure", "Poynting-Robertson drag" and the
"Poynting-Robertson effect". We have only the
"Poynting-Robertson effect" and the other two "effects" are indispensable parts
of the "Poynting-Robertson effect". The two effects cannot be physically
separated from the P-R effect. Relativistically covariant equation of
motion, represented by Eq. (14), describes the P-R effect and it contains
both the "direct light pressure" and the "Poynting-Robertson drag".

Let us consider perfectly absorbing spherical particle, within geometrical
optics approximation. The "isotropic emission" is, in reality, described
by the value (see Eq. 16) \\
$\bar{Q}'$ $\equiv$ $\bar{Q'}_{pr} / \bar{Q'}_{ext}$ $=$ 1/2 , \\
and by the condition \\
$\vec{p'}_{outgoing}$ $=$ ( 1/2 ) $\vec{p'}_{incoming}$ . \\
Subsequently, the force corresponding to the incoming and outgoing radiation
contains also a component tangent to the trajectory
which decelerates the particle in the orbit. This tangential component
of the force decreases the total (kinetic plus potential) energy of the particle. The tangential component of the force increases kinetic
energy of the particle.

\subsection{Leinert and Gr\"{u}n (1990)}
Leinert and Gr\"{u}n (1990, p. 226) write: \\
Seen from the moving dust particle, the incident solar radiation
is displaced by an angle $v_{tan}/c$ from the radial direction,
where $v_{tan}$ is the tangential component of the orbital velocity.
Seen from outside, the particle motion causes scattering and thermal
emission to have a forward component. The result is the same: a braking force,
on the average over one orbit directed opposite to the direction of motion,
equal in first order to $v/c$ times the radiation pressure force for
a particle in a circular orbit."

Later on, the authors state: \\
"The impact of solar wind ions also exerts an outward pressure on orbiting
dust particles. It is more than three orders of magnitude weaker than radiation
pressure. But the aberration angle $v_{tan}/v_{sw}$, where $v_{sw}$ is
the velocity of the solar wind, is much larger,making the resulting
"ion impact" drag comparable to the Poynting-Robertson effect."
(Leinert and Gr\"{u}n 1990, pp. 226-227). \\

Physics: \\
\nopagebreak
\noindent\\
1. \\
At first, Leinert and Gr\"{u}n (1990, Secs. 5.4.2 and 5.4.3)
discuss "radiation pressure" and the "Poynting-Robertson effect"
as two different phenomena. However, the P-R effect is the effect of
radiation pressure on a moving body. \\
\noindent\\
2. \\
As for the forces, the mathematical forms for the P-R effect seem to be
consistent with the statements of the authors. However, the
force for the solar wind (under the assumption that
only radial velocity of the solar wind is considered) is: \\
$\vec{F}_{sw}$ $=$ $F_{sw}$ [ ( 1 $-$
$\vec{v} \cdot \vec{e} /  v_{sw}$ ) $\vec{e}$ $-$
$x'$ $\vec{v} / v_{sw}$ ]~, \\
where $F_{sw}$ is the force on the particle,
$v_{sw}$ is the heliocentric solar-wind speed, and,
the value of $x'$, 1 $<$ $x'$ $<$ 3 (approximately), depends on material
properties of the particle. This force does not correspond to the
statement of the authors. The presence of the term $x'$ shows
that any idea about importance of the aberration effect is not acceptable
(see also Sec. 7.4.5 in Kla\v{c}ka 2008b, as for light pressure).

Moreover, one has to take into account that Eqs. (1) hold.
Even if the authors want to take into account the simplest case
treated by Poynting (1903) and Robertson (1937), the explanations
of the authors will not produce the consistent results for the incoming
and outgoing radiation: it is sufficient to consider that
$\bar{Q}'_{ext}$ $=$ 2, $\bar{Q}'_{pr}$ $=$ 1 (see Kla\v{c}ka 2008a, 2008b).
The tacitly assumed values
$\bar{Q}'_{ext}$ $=$ 1, $\bar{Q}'_{pr}$ $=$ 1, considered by
Poynting (1903), Robertson (1937), Burns {\it et al.} (1979) and others,
are not correct. \\
\noindent\\
3. \\
Let us look if the substitution "force" $\rightarrow$ "acceleration"
would make the statements of the authors correct.

Eqs. (8) and (20) (see also Eqs. 106-107 of Kla\v{c}ka 2008b) yield \\
$(d \vec{v} / dt)_{in}$ $=$ $a_{ph}$ [ $(1 - \vec{v} \cdot e_{R} / c )$
$\vec{e}_{R}$ $-$ $\vec{v} / c$ ] ~, \\
and Eqs. (12) and (20) (see also Eqs. 109-110 of Kla\v{c}ka 2008b) yield \\
$(d \vec{v} / dt)_{out}$ $=$ $-$ $a_{ph}$ ( 1 $-$
$\bar{Q}'_{pr} / \bar{Q}'_{ext}$ ) [ $(1 - \vec{v} \cdot e_{R} / c )$
$\vec{e}_{R}$ $-$ $\vec{v} / c$ ] ~; \\
$a_{ph}$ $=$ $S A' \bar{Q}'_{ext} / ( m c)$. If
$\bar{Q}'_{pr} / \bar{Q}'_{ext}$ $<$ 1, then "scattering and thermal emission"
cause acceleration of the particle [the case treated by Poynting (1903),
Robertson (1937), and others, corresponds to
$\bar{Q}'_{pr} / \bar{Q}'_{ext}$ $=$ 1/2],
$(d \vec{v} / dt)_{out}$ $\propto$ $+$ $\vec{v} / c$. This result differs
from the statement of Leinert and Gr\"{u}n. The incoming
radiation causes deceleration of the particle,
$(d \vec{v} / dt)_{in}$ $\propto$ $-$ $\vec{v} / c$.

\subsection{Banaszkiewicz {\it et al.} (1994)}
The authors do not try to explain the P-R effect, but they use
the electromagnetic Poynting-Robertson force $\vec{F}_{e}$ in the form
(Banaszkiewicz {\it et al.} 1994 -- Eq. 10)\\
$\vec{F}_{e}$ $=$ [$S_{0} A r_{0}^{2}$ / ($ c r^{2}$) ] $Q_{pr}$
 [( 1 $-$ 2 $\dot{r}$ / $c$ ) $\vec{r}$ $-$ ( $r / c$) $\vec{v}$ ] ~,\\
where ... \\
The same relevant terms are used in Banaszkiewicz ({\it et al.} 1994 --
Eq. 22). \\

Physics: \\
\nopagebreak

Besides misprinting in the exponent of $r$ in the denominator,
the numerical factor 2 at the term $\dot{r}$ / $c$ is not correct and
the correct factor equals 1.

Maybe, Banaszkiewicz {\it et al.} (1994) were aware that
the flux of the incoming radiation, as seen by the particle, is
$S'$ $=$ $S$ ( 1 $-$ 2 $\dot{r}$ / $c$ )
and not by the relation used by Burns {\it et al.} (1979, p. 5 -- Eq. 1).
However, correct equation of motion is \\
$\vec{F}_{e}$ $=$ [$S_{0} A r_{0}^{2}$ / ($ c r^{3}$) ] $Q_{pr}$
 [( 1 $-$ $\dot{r}$ / $c$ ) $\vec{r}$ $-$ ( $r / c$) $\vec{v}$ ] ~, \\
see also Eq. (22) in this paper.

\subsection{Gustafson (1994)}
Gustafson (1994, p. 566) writes: "Poynting-Robertson drag can be thought
of as arising from an aberration of the sunlight as seen from the particle
and a Doppler-shift induced change in momentum. To the first order in
$v/c$, the radiation force acting on a spherical particle is \\
$|F_{g}|$ $\beta$ $\left[\left(1-2\dot{r}/c\right) \hat{\vec{r}}~-~
\left(r \dot{\Theta}/c\right) \hat{\vec{\Theta}} \right]$~, ~~~ (6-G) \\
where the unit vector $\hat{\vec{\Theta}}$ is normal to
$\hat{\vec{r}}$ in the orbital plane..." ($\hat{\vec{r}}$ is the unit heliocentric radius vector). The author continues: "The second term along
$\hat{\vec{r}}$ is due to Doppler shift. With the transverse last term,
this velocity-dependent
part of Equation (6) [Eq. 6-G in our notation] is the Poynting-Robertson
(PR) drag. Doppler shift enters twice: once due to the rate at which
energy is received and a second time due to the reradiated and scattered
radiation." \\

Physics: \\
\nopagebreak

As it was written in Sec. 7.4.5 in Kla\v{c}ka (2008b),
the third term does not originate from aberration
of light. The second term is a result of combination of the
change of concentration of photons, the Doppler shift of incident
radiation, the term $1/w$ present in $b^{\mu}$ (see Eqs. 2 and 14)
and the conservation of particle's mass. The Gustafson's statement,
that this term is due to the Doppler shift of the absorbed and re-radiated
radiation, is not correct.

\subsection{Considine (1995)}
Considine (1995, p. 2534) writes: \\
"{\bf Poynting-Robertson effect.}  The effect upon the motion of
a micrometeoroid or other very small particle due to the absorption and
emission of radiation. The particle absorbs radiation from the sun only in
one direction, but reradiates energy in all directions. This effect
produces a drag upon the particle that is directly tangential to the orbit
of the particle about the sun, and thus decreases the orbital angular
momentum. Since this angular momentum varies as the square root of the
orbital radius, so that the particle follows a spiral path directing
steadily closer to the sun. Although the solar radiation pressure upon
the particle opposes this effect, it is not sufficient to offset it
completely." \\

Physics: \\
\nopagebreak

The effect is characterized by an equation of motion and it holds for
(non-rotating) spherical body, irrespective of its size. In general,
it is not relevant that the process of absorption occurs, since the effect
holds also for total specular reflection (see Sec. 7.3.3 in
Kla\v{c}ka 2008b). May be, the explanation presented
above concentrates to the case treated by Poynting and Robertson, when
perfectly absorbing (within geometrical optics approximation) spherical
particle is considered. However, even in this special case
the radiation is absorbed only partially and the effect of diffraction
plays equally important role. In general, the fact that the particle
"reradiates energy in all directions" does not determine the behavior
corresponding to the Poynting-Robertson effect (see experimental results
of Krauss and Wurm 2004 and Eqs. 9-11, 38-39 in Kla\v{c}ka 2008b).
Eq. (16) must be fulfilled for the Poynting-Robertson effect.

The effect produces a drag upon the particle that consists both from radial
term and the term "directly tangential to the orbit
of the particle about the sun".

The solar radiation pressure upon the particle does not oppose the effect,
since it is an inevitable part of the Poynting-Robertson effect. More
correctly, the Poynting-Robertson effect is the effect of radiation
pressure force for a moving spherical body.

\subsection{Murray and Dermott (1999)}
Frequent idea about the essence of the P-R effect is that
"the term $- \vec{v}/{c}$ in the equation of motion exists
under the hypothesis of isotropic reemission of radiation".
However, this idea is in contradiction with the following statement:
''P-R drag is caused by the nonuniform reemission of the sunlight
that a particle absorbs.'' (Murray and Dermott 1999, p. 121). \\

Physics: \\
\nopagebreak

In order to be able to correctly understand physics of the P-R effect,
we have to realize that the fundamental condition for the validity
of the P-R effect exists: see Sec. 3 and Eq. (16).

The essence of the P-R effect is that nonradial components of the
particle's radiation pressure and components of the thermal force are
zero (in particle's frame of reference).
Eq. (16) states that the P-R effect holds if the total momentum (per unit time)
of the outgoing radiation, integrated over the whole space angle, is in the
direction of the incident radiation, in the particle's frame of reference.

If the spherical particle does not absorb, e. g., in the case of specular
reflection, no thermal emission exists and particle orbital evolution
corresponds to the P-R effect -- Eq. (16) is fulfilled. This result also
illustrates that the formulation "isotropic reemission of radiation is essential
for the P-R effect" is not correct. If at least part of the incident
radiation is absorbed, then, according to Eq. (16), thermal emission
force equals zero and reemission of the absorbed radiation is isotropic.
Thus, the statement "P-R drag is caused by the nonuniform reemission
of the sunlight that a particle absorbs" is not correct
(in the reference frame of the particle).

In general, the P-R effect admits both isotropic condition for
the outgoing radiation
\begin{eqnarray}\label{30}
\vec{p'}_{outgoing} &=& 0 ~,
\nonumber\\
\bar{Q}' &\equiv& \bar{Q}'_{pr} / \bar{Q}'_{ext} = 1~,
\end{eqnarray}
and anisotropic
\begin{eqnarray}\label{31}
\vec{p'}_{outgoing} &=& ( 1 ~-~ \bar{Q}' ) ~\vec{p'}_{incoming} ~,
\nonumber\\
\bar{Q}' &\equiv& \bar{Q}'_{pr} / \bar{Q}'_{ext} \ne 1 ~,
\end{eqnarray}
also. The form of anisotropy is exactly given by Eq. (31).

Conclusion:\\
Any statement on thermal emission is not sufficient for the existence of the
P-R effect, in general. Real conditions for the validity of the P-R effect
are presented by Eq. (16).

\subsection{Srikanth (1999)}

1. \\
Srikanth (1999) introduces "the two parameter-model". However, this model
does not respect the physics followed from Maxwell's equations. Correct
physics is given by Eq. (1) (Eqs. 75-76 in Kla\v{c}ka 2008b). \\
\noindent\\
2. \\
Srikanth (1999, p. 231) states: \\
"The principal conclusion is that the motion of a dust in circumsolar
orbit is governed only by solar radiation absorption and not by the
asymmetry of reemission, even when viewed in the rest-frame of the Sun." \\

Physics: \\
\nopagebreak

Motion of a body is governed by an equation of motion. As for the force
generated by the incoming radiation, we have, on the basis of Eq. (1)
(see also Eq. 75 in Kla\v{c}ka 2008b) \\
$d p^{\mu}_{in} / d \tau$ $=$ (
	$w^{2} S A' \bar{Q'}_{ext} / c$ ) $b^{\mu}$ ~, \\
or, to the first order in $\vec{v} / c$, \\
$d \vec{p}_{in} / d t$ $=$ ( $w^{2} S A' \bar{Q'}_{ext} / c$ )
	(1 $+$ $\vec{v} \cdot \vec{e} / c$ ) $\vec{e}$ ~, \\
where $\bar{Q'}_{ext}$ $=$ 2 for the case treated by Poynting (1903),
Robertson (1937), or, Srikanth (1999). The above presented equations
do not generate the P-R effect. Thus, the outgoing radiation plays
equally important role in the P-R effect: \\
$( d p^{\mu}_{out} / d \tau)_{particle}$ $=$ $-$ (
     $w^{2} S A' \bar{Q'}_{ext} / c$ )  $\left \{
     \bar{Q}' ~ u^{\mu} / c ~+~
     ( 1 - \bar{Q'} ) ~ b^{\mu} \right \}$ ~, \\
$\bar{Q}'$ $\equiv$ $\bar{Q}_{pr}' / \bar{Q}_{ext}'$, \\
or, to the first order in $\vec{v}/c$, \\
$( d \vec{p}_{out} / dt )_{particle}$ $=$ ( $S A'\bar{Q'}_{ext} / c$ )
$\left \{ \bar{Q'} \vec{v} / c ~+~ ( 1 ~-~ \bar{Q'} )
( 1 ~-~ \vec{v} \cdot \vec{e} / c ) ~\vec{e} \right \}$ ~, \\
$\bar{Q}'$ $\equiv$ $\bar{Q}_{pr}' / \bar{Q}_{ext}'$  \\
(see Eqs. 1, 2, 19).
The accelerations are given by Eqs. (8), (12), (20)
(see also Eqs. 106-111 in Kla\v{c}ka 2008b). \\
\noindent\\
3.  \\
Srikhant (1999, p. 233) discusses the origin of the term
1 $-$ $\vec{v} \cdot \vec{e} / c$ $\equiv$ 1 $-$ $\dot{r} / c$,
in the P-R effect. He states that the term "is a flux-modification factor
and not red-shift factor". \\

Physics: \\
\nopagebreak

Flux-modification factor is
( 1 $-$ 2 $\dot{r}$ / $c$ ), as it follows from the relation \\
$S'$ $=$ $w^{2} S$, $w$ $\doteq$  1 $-$ $\vec{v} \cdot \vec{e}$ / $c$  ~.  \\
This is discussed in Kla\v{c}ka (1992a -- Sec. 2.5; 1993b; 1994; 2000a;
2004 -- Sec. 3.1, Eqs. 20-23; 2008b -- Sec. 3.1, Eqs. 20-23). The change
of concentration of photons and red-shift (Doppler effect)
are equally important. The origin of the term
1 $-$ $\vec{v} \cdot \vec{e}$ / $c$ is evident from Eqs. (1) and (2)
(see also Sec. 10 in Kla\v{c}ka 2008b).

\subsection{Woolfson (2000)}
Woolfson (2000, p. 401) considers perfectly absorbing spherical particle
(within geometrical optics approximation)
of radius $a$ and density $\rho$ at a distance $R$ from the Sun. This particle
absorbs energy.
Woolfson writes: "The total energy absorbed per unit time is \\
$P~=~L_{\odot} ~ a^{2} / ( 4 R^{2} )$ ~. ~~~ (V.1-W) \\
The energy comes to the particle radially but then is
re-emitted by the particle moving in its orbit at speed $v$. The radiated
energy thus possesses momentum which must be taken from the particle.
Considering the mass equivalent of the radiated energy the rate of change
of momentum of the particle, or the tangential force on it, is\\
$F~=~L_{\odot} ~a^{2} v / ( 4 R^{2} c^{2} )$ ~. ~~~ (V.2-W) \\
This force exerts a torque that changes the angular momentum of the particle
at the rate (the sign in front of $L_{\odot}$ is corrected) \\
$dh / dt ~=~-~F R~=~-~L_{\odot}~ a^{2} v / ( 4 R c^{2} )$ ~. ~~~ (V.3-W) \\
If the mass of the particle is $m$ then its angular momentum is
$h~=~m~\sqrt{GM_{\odot}R}$ so that \\
$dh / dt ~=~ (1 / 2) ~m ~\sqrt{GM_{\odot} / R}	~dR / dt =
	     ( 1 / 2 )~m ~v~dR / dt$ ~.~~~ (V.4-W) \\
From (V.3-W) and (V.4-W)  and expressing $m$ in terms of
$a$ and $\rho$ we find \\
$dR / dt~=~-~3L_{\odot} / ( 8 \pi R a c^{2} \rho)$~. ~~~ (V.5-W) " \\

Physics -- access No. 1: \\
\nopagebreak

Woolfson's argument that re-radiated energy possesses momentum which must
be taken from the particle is irrelevant because the incident (absorbed)
energy also possesses momentum and it is given to the particle.
The author, maybe, tacitly assumes that the radially incoming radiation
does not change the particle's orbital angular momentum and he does not
take it into account.

However, according to Woolfson, the decrease of the particle's
angular momentum due to the re-emitted radiation would be responsible for
spiralling of the particle onto the Sun. This idea is wrong. The correct
rate of change of momentum of the particle, due to the outgoing radiation,
is given by Eq. (19) and not by Eq. (V.2-W).

Let us look at the rate of change of the particle angular momentum due
to the incoming and the outgoing radiation. For the outgoing radiation
we get to the first order in $\vec{v}/c$
\begin{eqnarray}\label{32}
\left( \frac{d \vec{h}}{dt} \right)_{out} &=& \frac{d}{dt}~
\left(m \vec{r} \times \vec{v} \right)~=~
\left(\frac{dm}{dt}\right)_{out} \vec{r} \times \vec{v}~+~
m \vec{r} \times \left( \frac{d\vec{v}}{dt} \right)_{out}~=~
\nonumber\\
&=& -\frac{SA'\bar{Q'}_{ext}}{c^{2}}~\vec{r} \times \vec{v}~+~
\frac{SA'\bar{Q'}_{ext}}{c^{2}}~\left(1-\bar{Q'}\right)~
\vec{r} \times \vec{v}~=~
\nonumber\\
&=& -~\frac{SA'\bar{Q'}_{ext}}{c^{2}}~\bar{Q'}~\vec{r} \times \vec{v}~,
\end{eqnarray}
where we used Eqs. (11) and (20). Similarly, for the incoming radiation
we have (see Eqs. 7 and 20)
\begin{eqnarray}\label{33}
\left( \frac{d \vec{h}}{dt} \right)_{in} &=&
\left(\frac{dm}{dt}\right)_{in} \vec{r} \times \vec{v}~+~
m \vec{r} \times \left( \frac{d\vec{v}}{dt} \right)_{in}~=~
\nonumber \\
&=& \frac{SA'\bar{Q'}_{ext}}{c^{2}}~\vec{r} \times \vec{v}~-~
\frac{SA'\bar{Q'}_{ext}}{c^{2}}~\vec{r} \times \vec{v}~=~\vec{0}~.
\end{eqnarray}
We see that the outgoing radiation really decreases the angular momentum
of the particle. But acceleration is the quantity determining
the particle's motion. Eq. (20) yields that $(d \vec{v} / dt)_{out}$
$\propto$ $\vec{v}$, for $\bar{Q'}$ $=$ 1/2 corresponding
to perfectly absorbing spherical  particle within geometrical optics approximation. Thus, the particle would accelerate and its
heliocentric distance would increase if the outgoing radiation
would be the only relevant process. Deceleration of the particle
is caused by the incident radiation, see Eq. (20).
We can say, from this viewpoint,
that the incident radiation is significant in spiralling of the particle
into the Sun. Eq. (32) shows that the decrease of the particle angular
momentum, in the case of the outgoing radiation,
is caused only by the decrease of the particle's mass
(re-emission of radiation). Thus, Woolfson's calculation
(Eqs. V.3-W, V.4-W, V.5-W) is not correct, since he assumes that
the change of angular momentum is connected only with the change of orbital
radius. On the other side, the incident radiation does not change
particle's angular momentum (see Eq. 33). \\

Physics -- access No. 2: \\
\nopagebreak

Formulae to the first order in $v/c$ will be considered.

The relation between Eqs. (V.1-W) and (V.2-W) is based on the relativistic
relation
\begin{equation}\label{34}
\vec{p} = \frac{E}{c^{2}} ~ \vec{v} ~.
\end{equation}
This relation yields
\begin{equation}\label{35}
\frac{d\vec{p}}{dt} = \frac{\dot{E}}{c^{2}} ~ \vec{v} ~+~
		      \frac{E}{c^{2}} ~ \frac{d\vec{v}}{dt}  ~.
\end{equation}

Woolfson access corresponds to $\dot{E}_{in}$ $=$ $\dot{E}_{out}$
and he uses $- \dot{E}_{out}$ instead of $\dot{E}$. Woolfson neglects
the second term on the right-hand side of Eq. (35). He obtains \\
$\dot{\vec{p}}_{W} = - ( \dot{E}_{out} / c^{2} ) ~ \vec{v}$ ~. \\

If we would use the correct result represented by Eq. (19), then
\begin{equation}\label{36}
\dot{\vec{p}}_{W} = - \frac{S ~A' ~ \bar{Q'}_{ext}}{c^{2}} ~ \vec{v} ~.
\end{equation}
This equation should correspond to Eq. (V.2-W).

However, one can immediately see that even the result in Eq. (36) does not
correspond to the result presented in Eq. (V.2-W). The difference can be
easily explained: Woolfson's considerations are consistent with those of
Poynting (1903), Robertson (1937), Wyatt and Whipple (1950),
Burns {\it et al.} (1979) and others -- all of them state that
$\bar{Q'}_{ext}$ $=$ 1, but the correct result is $\bar{Q'}_{ext}$ $=$ 2.

Moreover, Eq. (36) is not correct, in reality. It differes from Eq. (21).
The reason is that the second term in Eq. (35) cannot be neglected!
The first term in Eq. (35) is negligible, in reality! We will show it, now.

We have, on the basis of Eqs. (18)-(19),
\begin{equation}\label{37}
\frac{dE}{dt} = \frac{dE_{in}}{dt} ~-~ \frac{dE_{out}}{dt} =
		S ~A' ~ \bar{Q'}_{ext} ~ \bar{Q'} ~
	       \frac{\vec{v} \cdot \vec{e}}{c}	~.
\end{equation}
Using $E = m~c^{2}$, Eqs. (35) and (37) yield
\begin{equation}\label{38}
\frac{d\vec{p}}{dt} = m ~ \frac{d\vec{v}}{dt}  ~.
\end{equation}
Using $\dot{\vec{v}}$ $=$ $( d \vec{v} / dt )_{in}$ $+$
$( d \vec{v} / dt )_{out}$, Eqs. (20) yield the result consistent with Eq.
(21). But Eq. (21) differs from Eq. (36)!

Surprisingly, vector products $\vec{r} \times (d\vec{p} / dt)$ and
$\vec{r} \times \dot{\vec{p}}_{W}$, using Eqs. (21) and (36), yield
the same results if
$\bar{Q'}_{ext}$ $=$ 1 (incorrect) is assumed!

Let us continue. Woolfson writes $h$ $=$
$m ~\sqrt{G~M_{\odot} ~R}$. If we consider total energy $E$, as we have just
discussed, then $m$ is a constant. But, at first, the form of the
relation for angular momentum $h$, corresponding to Woolfson's idea,
is $m ~\sqrt{G~M_{\odot}~ ( 1 - \beta ) ~R}$
(see Secs. 6.1 and 6.2 in Kla\v{c}ka 2004). Secondly, the relation for $h$
assumes a circular orbit. But, in this case, Eq. (37) yields $dE/dt$ $=$ 0
and the Woolfson's access yields $\dot{\vec{p}}$ $=$ 0 and, thus,
$dR/dt$ $=$ 0. But it is O.K., since $R$ does not change when a particle
moves in a circular orbit. Of course, this is a trivial result
and one does not need the complicated "physics" presented by Woolfson.

Eq. (21), or Eq. (22), is the correct equation of motion (gravity of the Sun
is also added, this is tacitly assumed). The size of the
angular momentum $h$ is $m ~\sqrt{G~M_{\odot}~ ( 1 - \beta ) ~p_{\beta}}$,
where $p_{\beta}$ is semi-latus rectum. Eqs. (21)-(22) yield
\begin{eqnarray}\label{39}
\frac{dh}{dt} &=& -~\frac{S ~A' ~ \bar{Q'}_{ext}}{m~c^{2}} ~ \bar{Q'} ~ h
	       = -~\beta ~\frac{G~M_{\odot}}{c ~R^{2}} ~ h ~,
\nonumber \\
h &=& m ~\sqrt{G~M_{\odot}~ ( 1 - \beta ) ~p_{\beta}} ~.
\end{eqnarray}
Averaging over one orbital period (see Sec. 6.1 in Kla\v{c}ka 2004)
yields
\begin{eqnarray}\label{40}
\frac{dh}{dt} &=& -~\beta ~\frac{G~M_{\odot}}{c ~
a_{\beta}^{2} \sqrt{1 - e_{\beta}^{2}}} ~ h ~,
\end{eqnarray}
where $a_{\beta}$ is semimajor axis and $e_{\beta}$ is eccentricity
of the elliptical orbit.
We can use $p_{\beta}$ $=$ $a_{\beta}$ ($1 - e_{\beta}^{2}$) and Eq. (40) yields
\begin{eqnarray}\label{41}
\frac{dp_{\beta}}{dt} &=& -~2~\beta ~\frac{G~M_{\odot}}{c ~a_{\beta}^{2}
\sqrt{1 - e_{\beta}^{2}}} ~ p_{\beta} ~.
\end{eqnarray}
Eq. (41) is correct equation for secular evolution of semi-latus rectum.
If $e_{\beta}^{2}$ can be neglected in comparison with 1, then (but see
also Kla\v{c}ka and Kaufmannov\'{a} 1992, 1993, Breiter and Jackson 1998,
Kla\v{c}ka 2001a) $p_{\beta}$ equals approximately $a_{\beta}$ 
and Eq. (41) yields
\begin{eqnarray}\label{42}
a_{\beta} &=& \sqrt{a_{\beta ~0}^{2}~-~4~ \beta ~\frac{G~M_{\odot}}{c} ~
      \left ( t ~-~ t_{0} \right )} ~,
\end{eqnarray}
if $a_{\beta ~0}$ is the value of semimajor axis at the time $t_{0}$; this results
is consistent with the Eq. (V.5-W).

\subsection{McDonnell {\it et al.} (2001)}
McDonnell {\it et al.} (2001, p. 164) write that "small particulates
are subject to forces which significantly alter or dominate their
dynamics" and also the P-R effect is included: \\
"the Poynting-Robertson effect (non-radial radiation pressure due to
aberration of sunlight) resulting in a gradual decrease in eccentricity
and semimajor axis". \\

Physics: \\
\nopagebreak

If we would consider arbitrarily shaped dust particle, then also non-radial
components of radiation pressure may exist. Spherical particle exhibits
only radial component of the radiation pressure. This is exact physics
and it is defined in the proper frame of reference of the particle.
In the reference frame of the Sun, there is also a non-radial term coming from
$- \vec{v}$ $/$ $c$. But this term does not come from the aberration of light,
see Sec. 7.4.5 in Kla\v{c}ka (2008b), see also Sec. 6.2,
and it is a consequence of the
constancy of particle's mass. Moreover, other physical effects are important
in the P-R effect (Doppler effect, change of concentration of the incident
photons) and the P-R effect cannot be reduced only to the term
$- \vec{v}$ $/$ $c$. This term would not yield the correct result for
the secular decrease of eccentricity and semimajor axis. One has to bear
in mind that the P-R effect is the effect of radiation pressure on moving
spherical particles. Moreover, if the particles are small, then they
can be expelled from the Solar System.

\subsection{Gr\"{u}n {\it et al.} (2001a)}
Gr\"{u}n {\it et al.} (2001a, p. 335) write:
"Besides the direct effect of radiation pressure on trajectories
of small dust grains there is also a more subtle effect: the
Poynting-Robertson effect. This is caused by radiation pressure
force, that is not perfectly radial on moving dust particle but
has small component opposite to the particle motion. The strength
is order of $v_{t}/c$ ($v_{t}$ is the tangential particle
velocity) and it leads to loss of angular momentum and orbital
energy of orbiting particle. The effect is strongest when the
particle speed is highest, i.e. close to the Sun at its perihelion.
Thefore particle orbits slowly circularized while they spiral
toward the Sun." \\

Physics: \\
\nopagebreak

The Poynting-Robertson effect is the effect of radiation pressure on moving
spherical particles. The "direct effect of radiation pressure" is
an indispensable part of the P-R effect. There are not two different effects,
"the direct effect of radiation pressure" and
"a more subtle effect: the Poynting-Robertson effect", as the authors state.

The P-R effect is caused by radiation pressure force which is perfectly
radial. And this holds in the proper frame of reference of the particle.
Yes, the effect contains also a velocity component $v_{t}/c$, which leads to
the loss of angular momentum and orbital energy of the orbiting particle.
But $v_{t}$ is the transversal and not the tangential particle
velocity. The velocity vector is always tangential to the orbit
of the particle. The velocity vector can be decomposed into radial and
transversal components.

\subsection{Dermott {\it et al.} (2001)}
Dermott {\it et al.} (2001, pp. 574-575) write:
"The component of the radiation force tangential to a particle's
orbit is called the P-R drag force. This force is also proportional
to $\beta$. It results in an evolutionary decrease in both the
the semi-major axis and the osculating eccentricity of the
particle's orbit."  \\

Physics: \\
\nopagebreak

According to the conventional definition (see Sec. 5), the P-R drag
is the part of equation of motion which contains velocity terms.
So, not only the term containing $-$ $\vec{v}/c$, which is tangential
to the particle's orbit. However, relativistic covariant formulation
does not admit to decompose the P-R effect into various parts.

The osculating eccentricity does not exhibit only decrease: $de/dt$ can be
also positive (see Eqs. 55 and 94, 96 in Kla\v{c}ka 2004), but secular value
of $de/dt$ is negative (if analytical time averaging over one period is
admitted). An interesting result is presented in Figs. 5-6 in
Kla\v{c}ka {\it et al.} (2007).

\subsection{Gr\"{u}n {\it et al.} (2001b)}
Gr\"{u}n {\it et al.} (2001b, p. 791 - glossary) explain:
"Poynting-Robertson drag:
Drag on a moving particle that is due to the asymmetrical
momentum distribution of scattered and absorbed light; the drag
primarily affects small grains, causing them to spiral toward
the central object that they orbit, whether the Sun or planet." \\

Physics: \\
\nopagebreak

There is a preferred axis. It is given by the unit vector of the incoming
radiation $\hat{\vec{p}}'_{incoming}$. Total outgoing radiation preserves
this symmetry, since the condition given by Eqs. (16) is fulfilled: \\
$\vec{p}'_{outgoing} = ( 1 - \bar{Q}'_{pr} / \bar{Q}'_{ext} ) ~
		       \vec{p'}_{incoming}$ ~. \\
As a consequence, the equation of motion of the spherical particle \\
$d \vec{p'} / d \tau = ( \bar{Q}'_{pr} / \bar{Q}'_{ext} ) ~
		       \vec{p'}_{incoming}$  \\
also fulfills the symmetry.

The drag $- \vec{v} / c$ is generated by the first condition in Eqs. (16): \\
$E'_{outgoing} = E'_{incoming}$ ~. \\
The condition corresponds to the conservation of particle's mass.

Motion of spherical totally reflecting (within geometrical optics
approximation) body is also described by the P-R effect. Thus,
absorption of the incoming light is not an important condition for the
P-R effect.

\subsection{Matzner (2001)}
In the book of Matzner (2001, p. 524) we can find the following explanation of
the Poynting-Robertson effect. It is "a drag force arising on particles
orbiting the Sun, because solar radiation striking the leading surface is
blueshifted compared to that striking the following surface. Thus, the particles
receive a component of momentum from the radiation pressure which is opposite
to the direction of motion." \\

Physics: \\
\nopagebreak

If we consider a homogeneous flow of radiation from star at particle's location
(it is a good approximation, in practice), then this radiation impacts on the
particle in one direction. There is no difference between
Doppler shifts of photons striking different parts of particle's surface.
The origin of drag force component is in the conservation of particle's
mass. The role of the Doppler effect is in the relation
$S'$ $w^{2}$ $S$, where $S$ is the flux of radiation energy
(see Eqs. 1 and 2).

\subsection{Murdin (2001)}
Murdin (2001) presents that Poynting-Robertson effect is "a net force acting on
small particles which causes their orbits to decay, also known as
Poynting-Robertson drag". His explanation of this effect is following:
\\
"When photons of light strike a dust particle in orbit around the Sun,
they impinge not directly 'side-on' but slightly on its leading side
(the 'forward' side in terms of its orbital motion). Their energy is
absorbed, and subsequently re-radiated, but it is re-radiated
isotropically - in all directions. The transfer of energy from photon to
particle can be thought of as the application of a force to the particle.
There is a component of this force that acts in the opposite direction
to the tangential component of the particle's orbital motion which is not
compensated for when the absorbed energy is re-radiated. As a result the
particle loses kinetic energy, which reduces its velocity, so it spirals
inward, toward the Sun. ... The effect can be counteracted by radiation
pressure, more so with decreasing particle size below about 1 $\mu$m."
\\

Physics: \\
\nopagebreak

The first mistake is that Murdin distinguishes radiation pressure and
P-R drag which denotes as the P-R effect. In reality, only the P-R effect
exists. The P-R effect is the effect of radiation pressure on
a moving spherically symmetric body (see also Sec. 6. 2).

It is evident that Murdin considers perfectly absorbing spherical
particle. He writes that the incident radiation causes force with
component tangential to the particle's orbit due to the aberration.
This is not true in the stationary reference frame connected with the Sun.
As we can see from Eqs. (18), the force originating from incident
radiation has radial direction. Mentioned tangential force originates
just from the outgoing radiation.
Moreover, this tangential force decelerates the particle in the orbit and
decreases the total (kinetic plus potential) energy of the particle.
But increases the kinetic energy.

\subsection{Bottke {\it et al.} (2002a), Lauretta and McSween (2006)}
Bottke {\it et al.} (2002a, p. 769), Lauretta and McSween (2006, p. 914) write:
"Poynting-Robertson effect -- an effect of radiation on a small particle
orbiting the Sun that causes it to spiral slowly toward the Sun. It occurs
because the orbiting particle absorbs energy and momentum streaming
radially outward from the Sun, but reradiates energy isotropically in its
own frame of reference." \\

Physics: \\
\nopagebreak

The P-R effect holds for any spherically symmetric dust particle, not only for
"perfectly absorbing" particle which "isotropically reemits/reradiates
the absorbed radiation, in the reference frame of the particle".
E. g., also for totally reflecting sphere (within the geometrical
optics approximation), although no absorption
of the incoming radiation occurs. Moreover, even within the geometrical optics
approximation, one must be careful in stating that the radiation is perfectly
absorbed and isotropically reradiated in the particle's own frame of reference.
Although these ideas come back to Poynting (1903), and Robertson (1937), and
although they are conventionally used (e. g., Wyatt and Whipple 1950,
Burns {\it et al.} 1979), they do not contain the whole physics.
In reality, the particle accepts momentum per unit time given by Eq. (1), or, \\
$d \vec{p}'_{in} / d \tau$ $=$ ($S' A' \bar{Q'}_{ext} / c$) $\vec{e}'$ ~, \\
in the frame of reference of the particle, and, the outgoing radiation
is described by Eq. (16). The case treated by Poynting, Robertson and others
corresponds to $\bar{Q'}_{pr}$ $=$ 1, $\bar{Q'}_{ext}$ $=$ 2, in reality: \\
$\vec{p}'_{outgoing}$ $=$ (1$/$2) $\vec{p}'_{incoming}$ \\
and not $\vec{p}'_{outgoing}$ $=$ 0 (Poynting 1903, Robertson 1937,
Wyatt and Whipple 1950, Burns {\it et al.} 1979).

The P-R effect is the effect of radiation pressure.
The pressure can cause escape of small dust particles from the Solar System.
Only larger particles (of radius greater than 0.1 micron, approximately --
this depends on optical properties of the particles) spiral slowly towards
the Sun, due to the P-R effect.

\subsection{Bottke {\it et al.} (2002b), Bottke {\it et al.} (2002c)}
Bottke {\it et al.} (2002b, p. 11), Bottke {\it et al.} (2002c, p. 396) write:
"Poynting-Robertson drag -- a radiation effect that causes small objects
to spiral inward as they absorb energy and momentum streaming radially
outward from the Sun and then reradiate this energy isotropically in their
own reference frame." \\

Physics: \\
\nopagebreak

The P-R drag is not a physical effect, which can be treated as an
independent phenomenon. It is an indispensable part of the P-R effect. It
holds for any spherically symmetric dust particle. The
"perfect absorption" is not the essence of the P-R effect, as it was
discussed in the previous Sec. 6.18.

\subsection{Danby (2003)}
Danby (2003, p. 366) writes: ''We shall examine the
{\it Poynting-Robertson effect}. A small particle moves in the solar
system, subject to the gravitational attraction of the Sun and to
radiation pressure from the Sun.'' \\

Danby continues: " ... aberration of light ... leads to a transverse
drag component of the radiation force, proportional to the transverse
component of velocity, divided by the speed of light." And then:
"Next, consider radial motion away from the Sun with
speed $\dot{r}$. Because of this, the particle will absorb less radiative
energy in unit time than it would if it were at rest. Also, the radiation
that it receives is diluted by the Doppler effect. So, relative to the
'unperturbed' model, we have a radial drag force proportional to
2 $\dot{r}/r^{2}$." \\

Physical comments: \\
\nopagebreak
\noindent\\
1. \\
It must be said that spherical particle with spherically distributed mass
is considered. Moreover, equations presented by the author assume that
optical properties of the particle do not change. It may be wise to be stressed,
too. \\
\noindent\\
2. \\
As for the explanation of the term containing $- \vec{v}/c$ in the
P-R effect as a consequence of the aberration of light, we refer the reader to
Sec. 7.4.5 in Kla\v{c}ka (2008b), and, also to Sec. 6.2.
The term is not a consequence of the aberration of light. \\
\noindent\\
3. \\
Danby states, that the term $-$ 2 $\dot{r}/c$ comes from a decrease of
radiative energy and Doppler effect. This does not correspond to physics.
The access of Danby corresponds to the procedure presented in Burns {\it et al.}
(1979) which was commented in Sec. 2.5 in Kla\v{c}ka (1992a), see
also Sec. 6.1: this would
require violation of the second Newton's law, if one would like to obtain the
correct formula for the Poynting-Robertson effect. \\
\noindent\\
4. \\
According to physics, the energy flux density measured by the particle
generates the term $-$ 2 $\dot{r}/c$ (see Eqs. 113-114 in Kla\v{c}ka
1992a, Eq. 23 in Kla\v{c}ka 2004) and the term
is generated by the decrease of concentration of photons and frequency,
i.e. Doppler effect (see Eqs. 32 and 36 in Kla\v{c}ka 1992a, Eqs. 7-8 in
Kla\v{c}ka 1993b, Eqs. 17 and 22 in Kla\v{c}ka 2004, Eqs. 17, 20, 22
and 23 in Kla\v{c}ka 2008b). \\
\noindent\\
5. \\
The radial drag force proportional to
2 $\dot{r}/r^{2}$ comes from the transformation
(if the first order in $v/c$ is taken into account) \\
$S'$ $=$ $w^{2} ~S$ $=$ (1 $-$ 2 $\dot{r}/c$) $S$~, \\
then from the term \\
$\vec{b}$ $=$ $\vec{e}/w$ $=$ (1 $+$ $\dot{r}/c$) $\vec{e}$ ~, \\
and, finally, from the radial part of the term \\
$-$ ($\vec{v}/c$)$_{rad}$ $=$ $-$ $\dot{r}/c$ ~. \\
The term $1/r^{2}$ comes from the radiation energy flux density.

\subsection{Parker (2003), Moore (2002), K\"{o}hler and Mann (2002)}
Dictionary of Astronomy (Parker 2003, p. 99) explains the
Poynting-Robertson effect in the following way:
"The gradual decrease in orbital velocity of a small particle such as a
micrometeorite in orbit about the sun due to the absorption and
reemission of radiant energy by the particle." \\

Similar explanation is presented in Moore (2002): \\
"Poynting-Robertson effect. Non-gravitational force produced by the
action of solar radiation on small particles in the Solar system; it
causes the particles to spiral inwards towards the Sun. When particles
in orbit around the Sun absorb energy from the Sun and re-radiate it,
they lose kinetic energy and their orbital radius shrinks slightly.
The effect is most marked for particles of a few micrometers in size." \\

Similarly, K\"{o}hler and Mann (2002) explain: \\
"The Poynting-Robertson effect reduces the kinetic energy of the
particles. The semimajor axis and the eccentricity of the orbits
decrease so that finally the particles fall into the stars." \\

Physical explanation is based on the two-body problem: \\
At first, we immediately formulate the correct statement: the P-R effect
reduces the total (kinetic plus potential) energy of the particles. \\

The Keplerian orbital velocity of a body orbiting the Sun is characterized
by magnitude
\begin{equation}\label{43}
v~=~\sqrt{G \left ( M_{sun} + m \right ) \left ( \frac{2}{r} ~-~ \frac{1}{a}
\right)}~,
\end{equation}
where $a$ and $r$ are semi-major axis and heliocentric distance of the body
of mass $m$. Time averaging over one period yields
\begin{equation}\label{44}
\langle v^{2} \rangle~=~\frac{G\left(M_{sun}+m \right)}{a}~.
\end{equation}

Let us consider the P-R effect and let us assume that $\beta$ is a constant.
Then, $M_{sun}+m$ $\rightarrow$ $M_{sun}$ for a micrometeoroid.
One possibility is to use as a central Keplerian acceleration
$-~GM_{sun} ~\vec{r}/r^{3}$ and the corresponding
secular value of semi-major axis is $a$.
If the non-velocity radiation term is included into the central 
acceleration, then the central Keplerian acceleration is
$-~GM_{sun} (1 - \beta)~\vec{r}/r^{3}$ and the corresponding
secular value of semi-major axis is $a_{\beta}$. We can write
\begin{eqnarray}\label{45} 
\langle v^{2} \rangle &=&
\frac{GM_{sun} \left ( 1 - \beta \right )}{a_{\beta}}
\end{eqnarray}
for a micrometeoroid.
Since semi-major axis is a decreasing function of time
for the P-R effect, also
\begin{equation}\label{46}
\frac{d}{dt}~\langle v^{2} \rangle ~>~0~.
\end{equation}

\vspace*{0.5cm}

Qualitative physical explanation:\\
\nopagebreak

The above cited Parker's statement would be correct under the assumption that
potential energy is constant. If a particle is decelerated, then its total
energy is a decreasing function of time. The time averaged values of
the total, potential and the kinetic energies (per unit mass) of the
particle are
\begin{eqnarray}\label{47}
\langle E \rangle &=& -~
\frac{G M_{sun} \left ( 1 - \beta \right )}{2~a_{\beta}} ~,
\nonumber \\
\langle U \rangle &=& 2~\langle E \rangle~,
\nonumber\\
\langle K \rangle &=& -~\langle E \rangle~.
\end{eqnarray}
We want to stress that Eqs. (47) are results corresponding to secular
evolution. The last two equations of Eqs. (47) are similar to the virial theorem, where energy $E$ is constant.
Eqs. (47) immediately yield
\begin{eqnarray}\label{48}
\frac{d}{dt} ~ \langle E \rangle &=& 
\frac{GM_{sun}\left(1-\beta\right)}{
	     2~a_{\beta}^{2}} ~\frac{da_{\beta}}{dt}~< ~0~,
\nonumber \\
\frac{d}{dt} ~ \langle U \rangle &=& 
\frac{GM_{sun} \left(1-\beta\right)}{
	     a_{\beta}^{2}}~\frac{da_{\beta}}{dt}~ < ~0~,
\nonumber\\
\frac{d}{dt} ~ \langle K \rangle &=& 
-~\frac{GM_{sun}\left(1-\beta\right)}{
	     2~a_{\beta}^{2}}~\frac{da_{\beta}}{dt}~> ~0~.
\end{eqnarray}
The particle's potential energy decreases more rapidly than the
total energy, and, thus, the particle's kinetic energy must increase in order
to reduce the decrease of potential energy.
The increase of $\langle K \rangle$ $=$ $\langle \vec{v}^{2} \rangle /2$
occurs. So, the particle's velocity is an increasing function of  time. \\

Quantitative physical explanation: \\
\nopagebreak

Squared orbital velocity, averaged over one period, is
\begin{equation}\label{49}
\langle v^{2} \rangle ~=~ \frac{G M_{sun} \left(1-\beta\right)}{a_{\beta ~in}}~
\frac{e_{\beta~in}^{4/5}}{1-e_{\beta~in}^{2}}~\frac{1-e_{\beta}^{2}}{
e_{\beta}^{4/5}}~,
\end{equation}
where $a_{\beta ~in}$ and $e_{\beta ~in}$ are initial semi-major axis
and eccentricity
of the particle. The value of the eccentricity for a time
$t$ is given by the following differential equation:
\begin{eqnarray}\label{50}
\frac{de_{\beta}}{dx}&=&-~\frac{\left(1-e_{\beta}^{2}\right)^{3/2}}{e_{\beta}^{3/5}}~
\int_{0}^{e_{\beta ~in}} \frac{z^{3/5}}{\left(1-z^{2}\right)^{3/2}}~dz~,
\nonumber\\
t &=& x ~ T~,~~0 \leq x \leq 1
\end{eqnarray}
and $T$ is the time of spiralling of the particle into the
Sun:
\begin{equation}\label{51}
T~=~\frac{2}{5}~\left(\beta~\frac{G M_{sun}}{c}\right)^{-1}~
\frac{a_{\beta ~in}^{2} \left(1-e_{\beta ~in}^{2} \right)^{2}}{
e_{\beta ~in}^{8/5}}~
\int_{0}^{e_{\beta ~in}} \frac{z^{3/5}}{\left(1-z^{2}\right)^{3/2}}~dz~.
\end{equation}
Eq. (50) shows that eccentricity is a decreasing function of time.
As a consequence, $\langle v^{2} \rangle$ is an increasing function of time,
according to Eq. (49).

Finally, we have to stress that the P-R effect holds also for
nonabsorbing particles (totally reflecting sphere). Although the
case treated by Poynting (1903) and Robertson (1937) holds for
perfectly absorbing nonrotating spherical particle, the condition
for absorption is not relevant (and, moreover, Eq. 16 is valid).

\subsection{Encrenaz {\it et al.} (2004)}
Encrenaz {\it et al.} (2004, pp. 189-190) write: \\
"The particles in the rings are subject to two effects: \\
-- solar radiation pressure, which pushes the particles towards the outer
regions of the Solar System. Because it is proportional to the surface
area of a grain, while gravitational attraction is proportional to the
volume, radiation pressure is particularly effective on small-sized particles,
about one micrometre across; \\
-- the Poynting-Robertson effect results from the fact that a particle
orbiting the Sun receives solar radiation proprotional to its cross-section,
but radiates it away isotropically. The result is preferentially radiation
in the forward direction (the frequency of the photons is increased because
of the velocity of the particle) and this translates to a loss energy by the
particle. This form of braking, which continuously decreases the eccentricity
of the grain's elliptical path, tends to turn the latter into a circle,
and finally into a spiral. It is calculated that a grain one micrometre
across, subject to the Poynting-Robertson effect alone would fall into the Sun
in a few thousand years. In practice, grains of micrometre size are more
sensitive to the radiation pressure that tends to push them towards the
outer edge of the Solar System. The Poynting-Robertson effect is thus
particularly important for particles in the centimetre size range. \\

Physics: \\
\nopagebreak

Radiation pressure is an indispensable part of the P-R effect.

The incoming radiation generates the force acting on the spherical
particle. The force is proportional to the extinction cross-section
of the particle (see Eq. 1).
If the particle is perfectly absorbing, within geometrical
optics approximation, then
$\vec{p'}_{outgoing}$ $=$ (1/2) $\vec{p'}_{incoming}$ and not
$\vec{p'}_{outgoing}$ $=$ 0 (isotropic outgoing radiation), as it was
pointed out by Kla\v{c}ka (2008a, 2008b).

Momentum (per unit time) of the outgoing radiation is characterized by
Eq. (1), or, Eq. (19), where $\bar{Q}'$ $=$ 1/2 for the case treated by
Encrenaz {\it et al.} (2004): \\
$d \vec{p}_{out} / dt$ $=$ ($S A' / c$)
[ $\vec{v}/c$ $+$ ($1 ~-~ \vec{v}\cdot\vec{e}/c$) $\vec{e}$ ] . \\
A loss of energy by the particle, due to the outgoing radiation, follows
from Eq. (19): \\
$dE / dt$ $=$ $- 2~S ~A'$ [ $1 ~-~ (3/2) ~
	      \vec{v} \cdot \vec{e} / c$ ] . \\

\subsection{Encrenaz {\it et al.} (2004)}
Encrenaz {\it et al.} (2004, pp. 446) write: \\
"The Poynting-Robertson effect must be taken into account. As explained
earlier ... this effect results from the fact that the -- radial -- force
produced by the photons, which propagate with the velocity of light $c$,
is exerted on a body moving at a velocity $v$, which is not co-linear with $c$.
The transfer of momentum creates a braking force, which leads to a path that
spirals in towards the Sun." \\

Physics: \\
\nopagebreak

The Poynting-Robertson effect results from the fact that a particle
moves with respect to the source of radiation. The final equation of motion
-- represented by Eqs. (14), (21), or, (22) -- holds for any relative
direction and orientation of $\vec{v}$ and $\vec{e}$, including their
co-linearity.

\subsection{Quinn (2005)}
Quinn (2005, pp. 194-195) distinguishes the radiation pressure and the
Poynting-Robertson drag. He writes: "... as radiation is absorbed and
re-radiated by the particle, the re-radiated radiation is anisotropic
in the rest frame of the sun. This results in a drag force opposite
the direction of motion of the particle referred to as Poynting-Robertson
Drag. The total radiation force including this effect to the first order
in $v/c$ is:
\\
$\vec{F}_{rad}~=~(L_{\odot} Q_{pr} A)/(4 \pi c r^{2})~
\left[~\left(1-2 v_{r}/c\right) \hat{\vec{r}}~-~
(v_{\theta}/c)~\hat{\vec{\theta}}~\right]$
\\
where $L_{\odot}$ is the solar luminosity, $r$ is the radius of the particle,
and $v_{r}$ and $v_{\theta}$ are the radial and tangential component
respectively of the particle velocity in the frame of the Sun. The first
term in this expression is the radiation pressure, and the second and third
terms are the Poynting-Robertson drag. There is a factor of 2 in the
second radial term because of the combination of Doppler shift in
absorbing radiation and Doppler shift in emission." \\

Physics: \\
\nopagebreak

The factor 2 in the second radial term is a result of combination of the
change of concentration of photons, the Doppler shift of the incident
radiation, the term $1/w$ present in $b^{\mu}$ (see Eqs. 2 and 14)
and the conservation of particle's mass. The Quinn's statement that the
second radial term is due to the Doppler shift of the absorbed and re-radiated
radiation, is not correct.

Quinn's statements are in practice the same as Gustafson's ones (Gustafson
1994). Thus, the correct explanation of the origin of individual terms in
expression for drag force is also the same. See also Sec. 6.7.

\subsection{Harwit (2006)}
Harwit (2006, p. 174-175; see also Harwit 1988, p. 176-177,
Harwit 1973, p. 176-177)
belongs to the authors who try to explain the
Poynting-Robertson drag through a decrease of the angular momentum.

Harwit writes: \\
"Consider a grain of dust in interplanetary space. As it orbits the Sun
it absorbs sunlight, and re-emits this energy isotropically. We can view
this process from two different viewpoints.

(a) Seen from the Sun, a grain with mass $m$ absorbs light coming radially
from the Sun and re-emits it isotropically in its own rest frame. A
re-emitted photon carries off angular momentum proportional: (i) to its
equivalent mass $h \nu/c^{2}$, (ii) to the velocity of the grain
$R \dot{\theta}$; and (iii) to the grain's distance from the Sun $R$.
Considering only terms linear in $V/c$, and neglecting any higher terms
we see that the grain loses orbital angular momentum $L$ about the Sun
at a rate
\\
$dL~=~(h \nu/c^{2})~\dot{\theta} R^{2}~,~~(1/L)~dL~=~
(h \nu)/(m c^{2})~,~~~~~~(5-45-H)$
\\
for each photon whose energy is absorbed and re-emitted, or isotropically
scattered in grain's rest frame.

(b) Seen from the grain, radiation from the Sun arrives at an
aberrated angle $\theta '$ from the direction of motion, instead of
at $\theta$ $=$ $270^{\circ}$. Hence,
\\
$\cos{\theta}~=~[\cos{\theta '}+\left( V/c \right)]/
[1+\left( V/c \right)~\cos{\theta '}]~=~0~,~~\cos{\theta '}~=~-V/c~.
~~~~~~(5-46-H)$
\\
Here $V$ is $\dot{\theta} R$, the grain's orbital velocity, and the photon
imparts an angular momentum $p R \cos{\theta '}$ $=$ $-~(h \nu / c^{2})
R^{2} \dot{\theta}$ to the grain.

For a grain with cross-section $\sigma_{g}$
\\
$d L/d t~=~-~(L_{sun} \sigma_{g})/(4 \pi R^{2}m c^{2})~L~,~~~~~~(5-47-H)$
\\
where $L_{sun}$ is the solar luminosity.

Either way, the grain's velocity decreases on just absorbing sunlight.
From the first viewpoint, this happens because the grain gains mass,
which it then loses on re-emission; from the second, it is because
the grain is slowed down by the transfer of angular momentum." \\

Physics: \\
\nopagebreak

We have already explained in Sec. 6.11 that
although the orbital
angular momentum of the perfectly absorbing particle decreases
due to the outgoing radiation
(Eqs. 32, 39), the process increases acceleration of the particle
(see Eq. 12, $\bar{Q'}$ $=$ 1/2). The outgoing radiation
acts on the particle by the force containing also the term $-\vec{v}/c$
(see Eqs. 19, where $\bar{Q'}$ $=$ 1/2). Thus, the outgoing radiation
participates on the Poynting-Robertson drag force, but the corresponding
acceleration is positive (see Eqs. 20)!

Harwit's considerations are strictly based on the idea of Poynting (1903),
Robertson (1937) and others, when the authors assume that the outgoing
radiation corresponds to the isotropically re-emitted radiation.
We know that this is not true.

Moreover, even if the outgoing radiation would be equivalent
to the re-emitted radiation, the considerations of Harwit would be
incorrect. The Harwit's statement that each re-emitted photon carries
off the same angular momentum is not true, in a stationary reference
frame. One photon, emitted in a direction of unit vector $\vec{n'}$,
carries a momentum $d \vec{p'}$ $=$ $(h \nu ' / c) \vec{n'}$, in the
grain's proper reference frame. Each photon emitted in an arbitrary
direction is characterized by the frequency $\nu '$. Thus,
in the grain's reference frame, the photons are emitted isotropically
and the total momentum carried off by the photons per unit time is zero ($\vec{F}'_{e}$ $=$ 0). The momentum $d \vec{p'}$ transforms to the
stationary reference frame as
\begin{equation}\label{52}
d \vec{p}~=~\frac{h \nu}{c}~\vec{n}~=~
\frac{h\nu '}{c}~\left\{\vec{n'}~+~\left[\left(\gamma-1\right)~
\frac{\vec{v}\cdot\vec{n'}}{v^{2}}~+~\frac{\gamma}{c}\right]
\vec{v}\right\}~,
\end{equation}
due to the Doppler effect and the aberration of light.
Thus, the angular momentum carried off by the given photon is
\begin{equation}\label{53}
d \vec{L}~=~\vec{R}\times d\vec{p}~=~ \frac{h \nu '}{c}~R~
\left( \vec{e} \times \vec{n'}~+~\vec{e} \times \frac{\vec{v}}{c}\right)~,
\end{equation}
if the first order in $\vec{v}/c$ is considered.

In his second explanation (from the view of the grain), Harwit mixes terms
from stationary and proper reference frames. The incoming radiation really
hits the grain in an aberrated direction and the radiation acts on the grain
by a force (see Eqs. 18 and 20). But particle's velocity equals zero at every
moment, in the particle's proper frame of reference.
Thus, the particle has no orbital angular
momentum which could be decreased in the proper frame.

\subsection{de Pater and Lissauer (2006)}
de Pater and Lissauer (2006, pp. 35) state: \\
"Sections ... above describe the gravitational interactions between the Sun,
planets and moons. Solar radiation, which provides an important force
for small ($\leq$ 1 m) particles in the
Solar System, has been ignored. Three effects can be distinguished:
\begin{itemize}
\item {\it Radiation pressure}, which pushes particles (primarily
      micrometer-sized dust) outwards from the Sun.
\item {\it Poynting-Robertson drag}, which causes centimeter-sized particles to
      spiral inward towards the Sun.
\item The {\it Yarkovski effect}, which changes the orbits of meter to
      kilometer-sized objects due to uneven temperature distributions at their
      surfaces.
\end{itemize}

The solar wind produces a corpuscular drag similar in form to the
Poynting-Robertson drag;
corpuscular drag is more important for submicrometer particles." \\

The authors continue with a subsection
"Radiation Force (micrometer-sized particles)": \\

"The Sun's radiation exerts a repulsive force, $\vec{F}_{rad}$, on all bodies
in our	Solar System. This force is given by: \\
$\vec{F}_{rad}$ $\approx$ [ $L_{\odot} A / ( 4 \pi c r^{2}_{\odot} )$ ]
$Q_{pr}$ $\hat{\vec{r}}$~,~~~~ (2.45-P-L) \\
where $A$ is the particle's geometrical cross-section, $L_{\odot}$ the solar
luminosity, $r_{\odot}$ the heliocentric distance, $c$ is the speed of light,
and $Q_{pr}$ the {\it radiation pressure coefficient}. The radiation pressure
coefficient accounts for both absorption and scattering and is equal to
unity for a perfectly absorbing particle. Relativistic effects produced by the
Doppler shift between the frame of the Sun and that of the particle are
generally small, and have been omitted from equation (2.45-P-L), but they will
be considered in Sec. 2.7.2-P-L. The parameter $\beta$ is defined as the
ratio between the forces due to the radiation pressure
and the Sun's gravity: \\
$\beta$ $\equiv$ $|$ $F_{rad}$ / $F_{g}$ $|$ $=$ 5.7 $\times$ $10^{-5}$
$Q_{pr}$ / $(\rho R)$~,~~~~ (2.46-P-L) \\
with the particle's radius, $R$, in cm and its density, $\rho$, in $g$ $cm^{-3}$.
Note that $\beta$ is independent of heliocentric distance and that the solar
radiation force is only important for micrometer and submicrometer-sized particles.
Extremely small particles are not strongly affected by radiation pressure,
because $Q_{pr}$ decreases as the particle radius drops below the (visible
wavelength) peak in the solar spectrum (Fig. 2.12-P-L). The magnitude of the
Sun's effective gravitational attraction is given by: \\
$F_{g, eff}$ $=$ $-$ (1 $-$ $\beta$)~$G m M_{\odot}$ $/$ $r^{2}_{\odot}$
~.~~~~ (2.47-P-L) \\
It is clear that small particles with $\beta > 1$ are repelled by the Sun's
radiation, and thus quickly escape the Solar System, unless they are
gravitationally bound to one of the planets. Dust released at a keplerian
velocity from bodies on circular orbits is ejected from the Solar System if
$\beta > 0.5$; critical values of $\beta$ for dust released
from bodies on eccentric orbits are calculated in Problem 2.32-P-L.

The importance of solar radiation pressure can, for example, be seen in comets
(Sec. 10.4.1.-P-L): Cometary tails always point in the anti-solar direction
due to the Sun's radiation pressure. The dust tails are curved rather
than straight as a result of the continuous ejection of dust
grains from the comet, which itself is on an elliptical orbit around the Sun." \\

de Pater and Lissauer (2006, pp. 36-37) continue with a new subsection
"Poynting-Robertson Drag (centimeter-sized grains)": \\
"A particle in orbit around the Sun absorbs solar radiation and reradiates
the energy isotropically in its own frame. The particle thereby
preferentially radiates (and loses momentum)
in the forward direction in the inertial frame of the Sun (Fig. 2.13-P-L).
This leads to a decrease in the particle's energy and angular
momentum and causes dust in bound orbits
to spiral sunward. This effect is called {\it Poynting-Robertson drag}.

Let us consider a perfectly absorbing, rapidly rotating, dust grain.
The flux of solar radiation absorbed by the grain is equal to \\
$[$ $L_{\odot}$ $A$ $/$ (4 $\pi$ $r^{2}_{\odot}$) $]$ (1 $-$ $v_{r}$ $/$
$c$)~,~~~~ (2.48a-P-L) \\
where $v_{r}$ $=$ $\vec{v} \cdot \hat{\vec{r}}$
is the radial component of the particle's
velocity (i.e., the component which is parallel to the incident beam of light).
The second term in expression (2.48a-P-L) accounts for the Doppler shift between
the Sun's rest frame and that of the particle; the transverse Doppler shift
is of order $(v_{\theta} / c)^{2}$ $\ll$ 1 and will be ignored here.
The absorbed flux is reradiated isotropically and can be written as a mass
loss rate in the particle's frame of motion (using $E$ $=$ $m c^{2}$): \\
$[$ $L_{\odot}$ $A$ $/$ (4 $\pi$ $c^{2}$ $r^{2}_{\odot}$) $]$
(1 $-$ $v_{r}$ $/$ $c$)~.~~~~ (2.48b-P-L) \\
As the particle moves relative to the Sun with velocity $\vec{v}$,
there is a momentum flux from the particle as seen in the rest frame
of the Sun, since the particle emits more momentum in the
forward direction than in the backward direction (Fig. 2.13-P-L).
The momentum flux is equal to \\
$[$ $-$ $L_{\odot}$ $A$ $/$ (4 $\pi$ $c^{2}$ $r^{2}_{\odot}$) $]$
(1 $-$ $v_{r}$ $/$ $c$) $\vec{v}$~,~~~~ (2.48c-P-L) \\
which can be generalized to the case in which the particle reflects and/or
scatters some of the radiation impingent upon it via multiplication
by $Q_{pr}$. The net force on the particle in
this more general case is given by: \\
$\vec{F}_{rad}$ $=$ $[$ $L_{\odot}$ $Q_{pr}$ $A$ $/$
		   (4 $\pi$ $c$ $r^{2}_{\odot}$) $]$ (1 $-$ $v_{r}$ $/$ $c$)
		   $\hat{\vec{r}}$ \\
\hspace*{1.1cm} $-$ $[$ $L_{\odot}$ $Q_{pr}$ $A$ $v$ $/$
(4 $\pi$ $c^{2}$ $r^{2}_{\odot}$) $]$
(1 $-$ $v_{r}$ $/$ $c$) $\hat{\vec{v}}$ ~~~~ (2.49a-P-L) \\
\hspace*{0.8cm} $\approx$ $[$ $L_{\odot}$ $Q_{pr}$ $A$ $/$ (4$\pi$ $c$ $r^{2}_{\odot}$) $]$
$[$ (1 $-$ 2 $v_{r}$ $/$ $c$ ) $\hat{\vec{r}}$ $-$ ($v_{\theta}$ $/$ $c$)
$\hat{\theta}$ $]$~.~~~~ (2.49b-P-L) \\
The first term in equation (2.49b-P-L) is that due to radiation pressure and
the second and third terms (those involving
the velocity of the particle) represent the Poynting-Robertson drag.

From the above discussion, it is clear that small dust grains in the
interplanetary medium disappear, with (sub-)micrometer-sized grains
being blown out of the Solar System, while centimeter-sized particles
spiral inward towards the Sun."  \\

Physics: \\
\nopagebreak

The authors distinguish the two effects: the
{\it radiation pressure} and the {\it Poynting-Robertson drag}.
However, physics states that the effect of radiation pressure
on a moving spherical particle is equivalent to the Poynting-Robertson effect.
The Poynting-Robertson effect is the relevant physical effect. The
mentioned two "effects", the {\it radiation pressure} and the
{\it Poynting-Robertson drag} cannot be separated from the P-R effect,
as it follows from the relativistic formulation of the equation of motion.
Similarly, the two "effects" are relevant for the same particles: it is not
correct to say that one part of the P-R effect is relevant for the
micrometer-sized dust, while the rest part is relevant for the
centimeter-sized particles. Moreover, the P-R effect is not the
dominant effect for centimeter-sized interplanetary particles -- collisions
among particles, impact erosion are important. Moreover, the P-R effect
is dominant in the inner part of the Solar System (say, for heliocentric
distances less than 10-20 AU -- solar wind bombardment, effect of the
interstellar hydrogen are relevant for larger heliocentric distances)
for micrometer-sized (and tens of micrometer-sized) dust.
Small particles can be blown out from the Solar System, due to the
radiation pressure.

As for the solar wind effect ("corpuscular drag"), it is
important for the same sizes of the particles as the P-R effect.
As for submicrometer-sized particles, the Lorentz force, due to the
presence of interplanetary magnetic field, is relevant.
Small particles can be blown out from the Solar System, due to the
(electromagnetic) radiation pressure. \\

As for the Sec. 2.7.2-P-L:
"A particle in orbit around the Sun absorbs solar radiation and reradiates
the energy isotropically in its own frame. The particle thereby
preferentially radiates (and loses momentum)
in the forward direction in the inertial frame of the Sun (Fig. 2.13-P-L).
This leads to a decrease in the particle's energy and angular
momentum and causes dust in bound orbits
to spiral sunward. This effect is called {\it Poynting-Robertson drag}." \\

As the authors state, the derivations presented in the textbook is based
on the derivation presented by Burns {\it et al.} (1979). Unfortunately,
the derivation is physically incorrect (Kla\v{c}ka 1992a - see mainly Sec. 2.5).
We will present the relevant comments, now. \\
\noindent\\
1. \\
"Let us consider a perfectly absorbing, rapidly rotating, dust grain.
The flux of solar radiation absorbed by the grain is equal to \\
$[$ $L_{\odot}$ $A$ $/$ (4 $\pi$ $r^{2}_{\odot}$) $]$ (1 $-$ $v_{r}$ $/$
$c$)~,~~~~ (2.48a-P-L)" \\

Comment: \\
If one would like to consider "rapidly rotating" particle, the result
would be more complicated. So, we will consider that the particle does
not rotate. The correct result for the flux of the incoming solar
radiation interacting with the grain equals to \\
$\bar{Q}'_{ext}$ $[$ $L_{\odot}$ $A$ $/$ (4 $\pi$ $r^{2}_{\odot}$) $]$
(1 $-$ 2 $v_{r}$ $/$ $c$)~,~~~~ (2.48a-P-L-correct) ~. \\
$\bar{Q}'_{ext}$ $=$ 2 for the special case treated by the authors.
The second term in expression (2.48a-P-L-correct) accounts for the
Doppler shift and the change of concentration of photons between
the Sun's rest frame and that of the particle. \\
The difference between Eqs. (2.48a-P-L) and (2.48a-P-L-correct) is evident.
Incorrect form of Eq. (2.48a-P-L) can be found also in, e. g., Mukai and
Yamamoto (1982). \\
\noindent\\
2. \\
"The absorbed flux is reradiated isotropically and can be written as a mass
loss rate in the particle's frame of motion (using $E$ $=$ $m c^{2}$): \\
$[$ $L_{\odot}$ $A$ $/$ (4 $\pi$ $c^{2}$ $r^{2}_{\odot}$) $]$
(1 $-$ $v_{r}$ $/$ $c$)~.~~~~ (2.48b-P-L)" \\

Comment: \\
In general, the absorbed flux can be reradiated in an arbitrary way, not
only isotropically. The mass loss rate, corresponding to the outgoing
radiation in the particle's frame of motion, is: \\
$\bar{Q}'_{ext}$ $[$ $L_{\odot}$ $A$ $/$
(4 $\pi$ $c^{2}$ $r^{2}_{\odot}$) $]$
(1 $-$ 2 $v_{r}$ $/$ $c$)~.~~~~ (2.48b-P-L-correct) \\
\noindent\\
3. \\
"As the particle moves relative to the Sun with velocity $\vec{v}$,
there is a momentum flux from the particle as seen in the rest frame
of the Sun, since the particle emits more momentum in the
forward direction than in the backward direction (Fig. 2.13-P-L).
The momentum flux is equal to \\
$[$ $-$ $L_{\odot}$ $A$ $/$ (4 $\pi$ $c^{2}$ $r^{2}_{\odot}$) $]$
(1 $-$ $v_{r}$ $/$ $c$) $\vec{v}$~,~~~~ (2.48c-P-L) \\
which can be generalized to the case in which the particle reflects and/or
scatters some of the radiation impingent upon it via multiplication by
$Q_{pr}$." \\

Comment: \\
As the particle moves relative to the Sun with velocity $\vec{v}$,
there is a momentum flux from the particle as seen in the rest frame
of the Sun, since the particle emits more momentum in the
forward direction than in the backward direction (Fig. 2.13-P-L).
The momentum flux is equal to, for the case treated by Poynting (1903) and
Robertson (1937), \\
$[$ $-$ 2 $L_{\odot}$ $A$ $/$ (4 $\pi$ $c$ $r^{2}_{\odot}$) $]$
$[$ (1 $-$ $v_{r}$ $/$ $c$) $\hat{\vec{r}}$ $-$ (1 / 2) $\times$ \\
((1 $-$ $v_{r}$ $/$ $c$) $\hat{\vec{r}}$ $-$
(1 $-$ 2 $v_{r}$ $/$ $c$) $\vec{v}$ / $c$) $]$~,~~~~ (2.48c-P-L-correct)\\
see Eq. (76) in Kla\v{c}ka (2008b). The last result can be
generalized to the case in which the particle reflects and/or
scatters some of the radiation impingent upon it. However, this cannot be
done in the simple way as stated by the authors (via multiplication by
$Q_{pr}$). The correct result is given by Eq. (76) in Kla\v{c}ka (2008b). \\
\noindent\\
4. \\
"The net force on the particle in
this more general case is given by: \\
$\vec{F}_{rad}$ $=$ $[$ $L_{\odot}$ $Q_{pr}$ $A$ $/$
		   (4 $\pi$ $c$ $r^{2}_{\odot}$) $]$ (1 $-$ $v_{r}$ $/$ $c$)
		   $\hat{\vec{r}}$ \\
\hspace*{1.1cm} $-$ $[$ $L_{\odot}$ $Q_{pr}$ $A$ $v$ $/$
(4 $\pi$ $c^{2}$ $r^{2}_{\odot}$) $]$
(1 $-$ $v_{r}$ $/$ $c$) $\hat{\vec{v}}$ ~~~~ (2.49a-P-L) \\
\hspace*{0.8cm} $\approx$ $[$ $L_{\odot}$ $Q_{pr}$ $A$ $/$
(4$\pi$ $c$ $r^{2}_{\odot}$) $]$
$[$ (1 $-$ 2 $v_{r}$ $/$ $c$) $\hat{\vec{r}}$ $-$ ($v_{\theta}$ $/$ $c$)
$\hat{\theta}$ $]$~.~~~~ (2.49b-P-L) \\
The first term in equation (2.49b-P-L) is that due to radiation pressure and
the second and third terms (those involving
the velocity of the particle) represent the Poynting-Robertson drag." \\

Comment: \\
The net force on the particle in
this more general case is given by: \\
$\vec{F}_{rad}$ $=$ $[$ $L_{\odot}$ $Q_{ext}$ $A$ $/$
		   (4 $\pi$ $c$ $r^{2}_{\odot}$) $]$
(1 $-$ $v_{r}$ $/$ $c$) $\hat{\vec{r}}$ \\
\hspace*{1.1cm} $-$ $[$ $L_{\odot}$ $Q_{ext}$
$A$ $/$ (4 $\pi$ $c$ $r^{2}_{\odot}$) $]$ $[$
(1 $-$ $v_{r}$ $/$ $c$) $\hat{\vec{r}}$ $-$ ($Q_{pr}$ / $Q_{ext}$) $\times$ \\
\hspace*{0.8cm} ((1 $-$ $v_{r}$ $/$ $c$) $\hat{\vec{r}}$ $-$
(1 $-$ 2 $v_{r}$ $/$ $c$) $\vec{v}$ / $c$) $]$ ~~~~ (2.49a-P-L-correct) \\
\hspace*{1.1cm} $\approx$ $[$ $L_{\odot}$ $Q_{pr}$ $A$ $/$
(4$\pi$ $c$ $r^{2}_{\odot}$) $]$
$[$ (1 $-$ 2 $v_{r}$ $/$ $c$) $\hat{\vec{r}}$ $-$ ($v_{\theta}$ $/$ $c$)
$\hat{\theta}$ $]$ ~.
\\
The last force is the force due to the radiation pressure.

\subsection{Carroll and Ostlie (2007)}
We find the following definition in Carroll and Ostlie (2007,
p. 806): \\
"The Poynting-Robertson effect (a consequence of the headlight effect
discussed in example 4.3.3) can cause ring particles to spiral in toward
the planet. When particles in the rings absorb sunlight, they must
re-radiate that energy again if they are to remain in thermal equilibrium.
The original light was emitted from the Sun isotropically, but in the
Sun's rest frame the re-radiated light is concentrated in the direction
of motion of the particle. Since the re-radiated light carries away
momentum as well as energy, the particle slows down and its orbit decays."
\\

Physics: \\
\nopagebreak

The statement introduced above is one of the wide-spread ideas about the
essence of P-R effect.
\\
Let us consider spherical particle.
Re-emitted thermal radiation carries off energy and momentum
per unit time (in the proper frame of particle)
\begin{eqnarray}\label{54}
E'_{T} &=& C'_{abs} S'~,
\nonumber\\
\vec{p'}_{T} &=& -~\vec{F'}_{e} ~=~\vec{0}~,
\end{eqnarray}
where $C'_{abs}$ is an absorption cross-section of the particle.
I.e., the four-momentum of thermal radiation per unit time is
\begin{equation}\label{55}
p'^{\mu}_{T}~=~\frac{1}{c}~C'_{abs}~S'~\left(1~;~\vec{0}\right)~.
\end{equation}
Due to the emission of the energy, the rest mass of the particle will
decrease:
\begin{equation}\label{56}
\left(\frac{dm}{d\tau}\right)_{T}~=~- ~\frac{E'_{T}}{c^{2}}~~=~
-~ \frac{C'_{abs}~S'}{c^{2}}~.
\end{equation}
In the stationary frame, the four-momentum of thermal radiation per unit
time will be
\begin{equation}\label{57}
p^{\mu}_{T} = \frac{1}{c}~C'_{abs}~S'~\frac{u^{\mu}}{c}~,
\end{equation}
where $u^{\mu}$ is the particle's four-velocity.
Thus, thermal radiation acts on the perfectly absorbing particle by
four-force
\begin{equation}\label{58}
\left(\frac{dp^{\mu}}{d\tau}\right)_{T}~=~-~p^{\mu}_{T}~=~-~
\frac{1}{c}~C'_{abs}~S'~\frac{u^{\mu}}{c}~.
\end{equation}
Using the fact that
\begin{equation}\label{59}
\left(\frac{dp^{\mu}}{d\tau}\right)_{T}~=~\left(\frac{dm}{d\tau}
\right)_{T}~u^{\mu}~+~m~\left(\frac{du^{\mu}}{d\tau}\right)_{T}
\end{equation}
and also Eqs. (56) and (58), the four-acceleration of the particle is
\begin{equation}\label{60}
\left ( \frac{du^{\mu}}{d\tau} \right )_{T}~=~0~,
\end{equation}
due to the thermal emission.
Thus, the re-emitted thermal radiation produces no acceleration of the
spherical particle. The particle is accelerated by incoming and by
outgoing scattered radiation. But although the acceleration has the
component in opposite direction to the particle's velocity $\vec{v}$,
the particle does not slow down, but its orbital velocity increases
(see Sec. 6.21).

\subsection{Meyer {\it et al.} (2007)}
Meyer {\it et al.} (2007, p. 580) write: \\
"The radial component of the radiation force is known as radiation pressure."
... "The tangential component of the radiation force is known as the
Poynting-Robertson (P-R) drag. This acts on all grains and causes their orbits
to decay into the star (where the grains evaporate) at a rate $\dot{a}$ $=$
$-$ 2 $\alpha$ $/$ $a$, where ..." \\

Physics: \\
\nopagebreak

In general, the radiation pressure force has three components, for 
arbitrarily shaped body. One radial and two nonradial. Equation of motion
does not correspond to the P-R effect, in this general case. 
The difference in motion of spherical and nonspherical dust grains
was confirmed experimentally (Krauss and Wurm 2004).
Only spherical grains behave in accordance with the P-R effect.

The Poynting-Robertson effect is the effect of radiation pressure upon 
the motion of spherical particle. Nonradial components of the radiation
pressure force equal zero, in this case. As for the P-R drag definition,
much confusion exists in the literature, see Sec. 5. In any case,
relativistically covariant equation of motion yields the P-R effect
and its decomposition into several parts (classified as "effects")
is nonphysical.

\subsection{Sykes (2007)}
Sykes (2007, p. 685) writes: "Comets were long thought to be the origin
of the zodiacal cloud. However, estimates of dust production by
short-period comets fell far short of that needed to maintain the cloud
in steady state against losses from particles spiralling into the Sun. This mechanism, where the absorption and reemission of solar radiation
continually decrease particle velocity, is called Poynting-Robertson drag." \\

Physics: \\
\nopagebreak

What does physics say, in reality? Particle speed of the spiralling toward the
Sun is an increasing function of time, according to Eqs. (46)-(48).
The same states Eq. (45), since particle's semi-major axis $a$ is a
decreasing function of time and $\beta$ is considered to be a constant.
The same result follows from Eqs. (49)-(50): Eq. (50) yields decrease
of eccentricity $e$ and Eq. (49) produces an increase of the speed of
motion for a decreasing eccentricity. Thus, the statement of the Sykes
that "the absorption and reemission of solar radiation continually
decreases particle velocity" does not correspond to reality.
Moreover, totally reflecting particle is also described by the P-R effect
-- no absorption/reemission exists in this case.

\subsection{Kr\"{u}gel (2008)}
Kr\"{u}gel (2008, p. 173-174) offers very similar explanation as
Woolfson (2000), see also Sec. 6.9.
He considers "a grain of mass $m$, radius
$a$ and geometrical cross-section $\sigma_{geo}$ $=$ $\pi a^{2}$ circling
a star at distance $r$ with frequency $\omega$ and velocity $v$ $=$
$\omega r$. The angular momentum of the grain is $l$ $=$ $m r^{2} \omega$
$=$ $m r v$". Further he states:" Let $M_{\star}$ and $L_{\star}$ be
the mass and luminosity of the star. Because there is mostly optical
radiation, the absorption coefficient of the grain is $C^{abs}$ $\simeq$ $\sigma_{geo}$ and the particle absorbs per unit time the energy
\\
$\Delta E~=~L_{\star} \sigma_{geo}/(4 \pi r^{2})~.$
\\
In thermal balance, the same amount is reemitted. Seen from a non-rotating
rest frame, the stellar photons that are absorbed travel in radial
direction and carry no angular momentum, whereas the emitted photons do
because they partake in the circular motion of the grain around the star.
If we associate with the absorbed energy $\Delta E$ a mass $m_{phot}$
$=$ $\Delta E/c^{2}$, the angular momentum of the grain decreases per
unit time through emission by
\\
$dl/dt~=~-~m_{phot}~r v~=~-~(\Delta E/c^{2})~r v~=~
-~L_{\star} \sigma_{geo} l/(4 \pi c^{2} r^{2} m)~.~$"
\\
Next procedure is in principle the same as the Woolfson's one. Kr\"{u}gel
calculates a decrease of the particle's distance from the star based on the
decrease of the orbital angular momentum of the grain.

Kr\"{u}gel also tries to explain the P-R effect from the view of the observer
in a frame corotating with the particle (i.e. particle's proper frame).
He states: "In such a reference frame, the stellar photons approach the
grain not exactly along the radius vector from the star, but hit it slightly
head-on in view of the aberration of light.... The photons thus decrease
the angular momentum of the grain and force it to spiral into the star."
\\

Physics: \\
\nopagebreak

From the view of the Sun.\\
Calculation in Sec. 6.27 shows that the re-emitted thermal radiation from
spherical grain acts on it by nonzero force.
Thus, the re-emitted radiation changes the grain's orbital angular momentum.
But this action is caused only by the decrease of the grain's mass due to
the re-emission of the radiation and the corresponding acceleration equals
zero. Thus, this phenomenon is not an essence of the Poynting-Robertson
effect in the sense that electromagnetic radiation gives some acceleration
to the particle.
Moreover, thermal radiation does not represent the whole outgoing
radiation which decreases particle's angular momentum. See Secs. 6.11 and
6.25 for more detail explanation. \\

From the view of the particle.\\
Kr\"{u}gel's statement that the force originating from the incident
radiation decreases (orbital) angular momentum of the particle is,
similarly to Harwit's one (see Sec. 6.25), senseless. The velocity of
the particle equals zero in its proper reference frame and thus the
particle has no orbital angular momentum in this frame.

\subsection{Liou and Kaufmann (2008)}
1. \\
Liou and Kaufmann (2008, p. 426) write about "forces due to solar radiation
and solar wind, including solar radiation pressure, PR drag, and
solar wind drag." \\

Comment: \\
There are two physical forces: solar radiation force and solar wind force.
These forces cannot be physically decomposed into various parts, as it is
presented by the authors. This holds also for explanations
presented in Liou and Kaufmann (2008, p. 426-427).
Similarly, physically correct solar wind force
should automatically contain possibility of nonradial velocity of the
solar wind and time dependent properties of the solar wind, but no
separation of these "effects" is physically acceptable. \\
\noindent\\
2. \\
Liou and Kaufmann (2008, p. 426) write about the drag forces (PR drag and
solar wind drag): "Although these drag forces are weak compared to the
velocity-independent radial pressure force component, they dissipate
energy and momentum and thereby cause the particles to eventually
spiral into the Sun." \\

Comment: \\
The velocity-dependent terms of the radiation pressure and solar wind pressure
forces dissipate total energy. But the momentum is not dissipated, since the
speed of the particle increases during the process of spiralling toward
the Sun. The magnitude of the momentum increases. The magnitude of the
angular momentum decreases. Moreover, nonradial velocity component of the solar
wind can increase particle's total energy and angular momentum
(Kla\v{c}ka {\it et al.} 2008). \\
\noindent\\
3. \\
Liou and Kaufmann (2008, pp. 426-427) write: "The
PR drag can be thought of as arising from an aberration of the
sunlight as ssen from the particle and a Doppler-shift induced change
in momentum." \\

Comment: \\
As it is discussed in Sec. 7.4.5 in Kla\v{c}ka (2008b), the
velocity-dependent term $-$ $\vec{v}$/$c$ ($\vec{v}$ -- heliocentric velocity
of the particle, $c$ is the speed of light in vacuum) in the P-R effect is not
produced by the aberration of light. The same holds for the effect
of solar wind. Moreover, the force for the solar wind effect does not
contain the term $-$ $\vec{v}$/$u$ where $u$ is the speed of solar
wind particles (see Eq. 29 in Kla\v{c}ka and Saniga 1993).
The effect of aberration is discussed also in Sec. 6.2.

In reality, not only the Doppler effect causes the P-R effect.
Also change of concentration of photons plays an equally important role
(see also Kla\v{c}ka 1992a: Eqs. 31, 32, 36, 37, or, Secs. 2.4 and 2.5;
see also Kla\v{c}ka 1993b: Eqs. 2-5). \\
\noindent\\
4. \\
Liou and Kaufmann (2008, p. 427) write: "Thus, forces due to collisions
with solar wind particles are analogous to forces due to radiation." \\

Comment: \\
Yes, we can say that the force for the solar wind and solar
electromagnetic radiation effects are analogous. However, one has to bear
in mind that there are important differences.
The force for the solar wind effect on dust particle contains decrease of
mass of the dust particle (Kla\v{c}ka and Saniga 1993, Kla\v{c}ka 1993a,
Kocifaj and Kla\v{c}ka 2008 -- this physical difference is often not considered,
although the change of mass of the particle is admitted;
see, e.g., Dohnanyi 1978 -- Secs. 2.1.3 and 2.2.6,
Mukai and Yamamoto 1982, Mann 2009 -- Eqs. 7-10, 7-11),
nonradial velocity of the solar wind
(Kla\v{c}ka 1994, Kla\v{c}ka {\it et al.} 2008), time dependence
of the solar wind (Svalgaard 1977). \\
\noindent\\
5. \\
Liou and Kaufmann (2008, p. 427) write:
"The contribution of the solar wind to the drag forces on dust particles
is typically parametrized as a constant fraction $sw$ of the PR drag." \\

Comment: \\
The statement of the authors is used in their equations: Eqs. (5-LK),
(7-LK), (10-LK), (11-LK), (17-LK), (18-LK), (21-LK), (23-LK), ..., i.e.,
the equations contain the term $\beta ( 1 + sw )$.
The same access can be found, e.g., in Liou and Zook (1997),
Holmes {\it et al.} (2003).
However, this is not correct. The correct form is (except for the forces)
$\beta ( 1 + sw / \bar{Q}'_{pr} )$ as it is presented in
Kla\v{c}ka (1994, Eq. 24 -- although the value of $sw$ is different
from those used by Liou and Kaufmann), Kla\v{c}ka (2004: Eqs. 176, 177,
182, 184, 186, ...), Kla\v{c}ka {\it et al.} (2008: Eqs. 4, 5, 15, 16).
Moreover, the correct form for the force is even more complicated: \\
$\vec{F}_{drag}$ $=$ $-$ [ $S_{0} A Q_{pr} / ( r_{0}^{2} c )$ ]
[ $( 1 + sw / Q_{pr} )$ ($\vec{v} \cdot \hat{\vec{r}}_{0} / c$)
$\hat{\vec{r}}_{0}$ $+$
$( 1 + x ' sw / Q_{pr} )$ $\vec{v} / c$ ] ~, \\
where the value of $x'$ ($>$ 1) depends on material properties of dust
particle. The last equation is the correct equation, not Eq. (5) in
Liou and Kaufmann (2008, p. 427).

\subsection{http:$//$en.wikipedia.org$/$wiki$/$Poynting-Robertson\_effect
(last modification August 23, 2008)}
In this very popular webpage we can find the following definition of
Poynting-Robertson effect:\\
"The Poynting-Robertson effect, also known as Poynting-Robertson drag, ...,
is a proces by which solar radiation causes a dust grain in the solar
system to slowly spiral inward. The drag is essentially a component of
radiation pressure tangential to the grain's motion." \\

There is also an explanation of the effect:\\
"From the perspective of the grain of dust circling the Sun ..., the
Sun's radiation appears to be coming from a slightly forward direction
(aberration of light). Therefore the absorption of this radiation
leads to a force with a component against the direction of movement. ...
From the perspective of the solar system as a whole (...), the dust grain
absorbs sunlight entirely in a radial direction, thus the grain's angular
momentum remains unchanged. However, in absorbing photons, the dust
acquires added mass via mass-energy equivalence. In order to conserve
angular momentum (which is proportional to mass), the dust grain must
drop into a lower orbit.\\
Note that the re-emission of photons, which is isotropic in the frame of
the grain, does not affect the dust particle's orbital motion. However,
in the frame of the solar system, the emission is beamed anisotropically,
and hence the photons carry away angular momentum from the dust grain.
It is somewhat counter-intuitive that angular momentum is lost while the
orbital motion of the grain is unchanged, but this is an immediate
consequence of the dust grain shedding mass during emission and that
angular momentum is proportional to mass.\\
The Poynting-Robertson drag can be understood as an effective force
opposite the direction of the dust grain's orbital motion, leading to
a drop in the grain's angular momentum. It should be mentioned that while
the dust grain thus spirals slowly into the Sun, its orbital speed
increases continuously.\\
The Poynting-Robertson force is equal to:\\
$F_{PR}~=~Wv/c^{2}~=~ [ r^{2}/(4c^{2}) ] \sqrt{GM_{s}L_{s}^{2}/R^{5}}$
\\
where $W$ is the power of the incoming radiation, $v$ is the grain's
velocity, $c$ is the speed of light, $r$ the object's radius, $G$ is the
universal gravitational constant, $M_{s}$ the Sun's mass, $L_{s}$ is
the solar luminosity and $R$ the object's orbital radius."
\\

Comments: \\
\nopagebreak

At first, we must  point out that the Poynting-Robertson drag is
not the same as the P-R effect, according to the standard approach
used in literature (see also Sec. 5). The radial radiative pressure
is also included into this effect. Therefore also the introduced
equation for the P-R force is
not correct, it corresponds only to the component proportional to
$-\vec{v}/c$, moreover not precisely (see Eqs. 21).

We have to stress that the P-R effect is the effect describing motion of
a spherical body (with spherically symmetric mass distribution)
under the action of radiation pressure. It has no sense to give a special
names to individual parts of the total force. At first, the confusion
exists in scientific literature, and, moreover, the readers
are utterly confused by such nonphysical steps. \\

It has been already mentioned, for several times, that the aberration
of the incident light does not lead to a force with component against the
direction of particle's movement. For an explanation see Sec. 7.4.5 in
Kla\v{c}ka (2008b) and also Sec. 6.2.
\\

From the author's explanation it follows that dynamics of the particle
is influenced (its orbit is changed) only by the incident (absorbed;
he does not consider complete interaction including also scattering)
radiation because of conservation of particle's orbital angular momentum
at absorption of this radiation. The outgoing (thermal re-emitted)
radiation does not influence the particle's dynamics.

The incident radiation really does not change particle's angular momentum
(see Eqs. 8 and 18 and also Sec. 6.11) because the action of the
increase of particle's mass is compensated by the action of deceleration
of the particle. Thus, the author is partially correct in this sense.
But his argumentation implies that the total acceleration of the particle
should be proportional to $\bar{Q'}_{abs}$ (particle's mass increase is proportional to this one) and not to $\bar{Q'}_{pr}$, as it actually is
(see Eq. 22). Moreover, the statement of the author is not consistent
with the fundamental condition represented by Eq. (16). Decisive
role plays incoming and outgoing radiation. In reality, the incoming
radiation generates particle's acceleration proportional to
$\bar{Q}'_{ext}$ $=$ 2, for the case treated by the author (see Eq. 20).\\

Equation of motion is the relevant physical result. The
equation of motion contains (linear)
momentum vector of the particle, not its angular momentum vector.
If we have equation of motion in disposal, it is not problem to obtain
exact equations for the angular momentum of the particle.
Really, we can consider angular momentum of the particle \\
$\vec{L}_{in}$ $=$ $\vec{r}$ $\times$ $\vec{p}_{in}$ $\equiv$
$m$ $\vec{H}_{in}$ ~, \\
where \\
$\vec{H}_{in}$ $\equiv$ $\vec{r}$
$\times$ $\vec{v}_{in}$ \\
is angular momentum per unit mass of the particle, and, similarly, \\
$\vec{L}_{out}$ $=$ $\vec{r}$ $\times$ $\vec{p}_{out}$ $\equiv$
$m$ $\vec{H}_{out}$~, \\
where \\
$\vec{H}_{out}$ $\equiv$ $\vec{r}$
$\times$ $\vec{v}_{out}$ ~. \\
The results for these quantities can be summarized as follows: \\
$d \vec{L}_{in} / dt$ $=$ 0 ~,	\\
$d \vec{H}_{in} / dt$ $=$ $-$ ($| dm/dt |/m$) $\vec{H}_{in}$ ~,  \\
$d \vec{L}_{out} / dt$ $=$ $-$ ($| dm/dt |/m$)
($\bar{Q}'_{pr} / \bar{Q}'_{ext}$) $\vec{L}_{out}$ ~,  \\
$d \vec{H}_{out} / dt$ $=$ $+$ ($| dm/dt |/m$) (1 $-$
$\bar{Q}'_{pr} / \bar{Q}'_{ext}$) $\vec{H}_{out}$ ~,  \\
where $| dm/dt |$ $=$ $w^{2} S A' \bar{Q}'_{ext} / c^{2}$
(of course, $\vec{v}_{in}$ $=$ $\vec{v}_{out}$ $=$ $\vec{v}$). These
results immediately yield, e.g., that the magnitude of the vector
$\vec{H}_{in}$ decreases (particle spirals toward the central star)
and the magnitude of the vector $\vec{H}_{out}$ increases for the cases
$\bar{Q}'_{pr} / \bar{Q}'_{ext}$ $<$ 1 (the particle would spiral outward
from the central star, if only "out" process would be important) --
everything is consistent with the results obtained from linear momenta
vectors (compare with Sec. 7.4.1 in Kla\v{c}ka 2008b).

\subsection{Zimmermann and G\"{u}rtler (2008)}
Zimmermann and G\"{u}rtler (2008, p. 307) write:
"P-R-Effekt, das Ph\"{a}nomen, womach kleine sich um die Sonne bewegende
Teilchen infolge der Absorption von Sonnenstrahlung abgebremst werden,
was zur allm\"{a}hlichen Ann\"{a}herung an die Sonne f\"{u}hrt.
Die von der Sonne emittierten Lichtquanten trager einen zu ihre.
Energie proportionalen Impuls, der auf ein aborbierendes Teilchen
\"{u}bertragen wird. Infolge der Bahngeschwindigkeit der umlaufenden
Partikeln und der endlichen Lichtgeschwindigkeit ergibt sich bei der
Absorption ein Aberrationseffekt. Der auf ein Teilchen \"{u}bertragene Impuls
hat auBer einer von der Sonne wegge richteten radialen Komponente eine der
Bewegung des Teilchens entgegenge richtete tangentiale, abbrem sendwirkende
Komponente. Je kleiner und masse\"{a}rmer eine Partikel ist, umso gr\"{o}Ber
ist des Verh\"{a}ltnis des tangentialen Impulses gegen\"{u}ber dem Umlaufimpuls
des Tielchens, umso schneller erfolgt die Abbremsung. Andererseits w\"{a}chst
mit atnehmendem Tielchenradius das Verh\"{a}ltnis des radialen Strahlungsdrucks
zur gravitativen Anziehurg durch die Sonne. Bei Tielchen gr\"{o}Ber als etwa
0.1 $\mu$m \"{u}berwiegt die Abbreusung, die Teilchen spieralen in die Sonne.
Bei kleineren Partiklen \"{u}berwiegt der Strahlungsdruck, sie werden von der
Sonne weggetrieben." \\

Physics: \\
\nopagebreak
\noindent\\
1. \\
The P-R effect holds for spherically symmetric bodies (see also
laboratory measurements obtained by Krauss and Wurm, 2004). Solar
electromagnetic radiation and gravity cause secular spiralling
of spherical grains toward the Sun (if their radii are larger than
$\approx$ 0.1 $\mu$ m).\\
\noindent\\
2. \\
The essence of the P-R effect is not an absorption of the incoming light.
Even motion of the particle characterized with mirror-like spherical
surface (specular reflection) is characterized by the P-R effect
(see Sec. 7.3.3 in Kla\v{c}ka 2008b).\\
\noindent\\
3. \\
The aberration of the incoming light does not lead to a force
with component against the direction of particle's movement.
For an explanation see Sec. 6.2 (or Sec. 7.4.5 in Kla\v{c}ka 2008b).\\
\noindent\\
4. \\
The P-R effect is the effect of electromagnetic radiation
pressure on moving spherically symmetric bodies.

\subsection{Mann (2009)}
1. \\
Mann discusses "radiation pressure force" (Mann 2009, Sec. 7.3.1).
Moreover, she discusses the "Poynting-Robertson effect", independently
(Mann 2009, Sec. 7.3.3). Finally, she writes: "Particles in bound orbit
about the star for which radiation pressure force is smaller than the
stellar gravity force migrate toward the star due to the azimuthal
component of the radiation pressure force and this is called the Poynting
Robertson effect." (Mann 2009, Sec. 7.3.3).  \\

Mann (2009) makes difference between the radiation pressure force and
the P-R effect. As we have already mentioned in Sec. 5, this access
does not correspond to physics. The P-R effect is the radiation pressure
force acting on moving (nonrotating) body with spherically symmetric mass
distribution. Moreover, the P-R effect contains not only the
transversal velocity term, but also other terms, including radial
velocity term. In reality, nonradial components of radiation pressure
force are zero for spherical bodies. The nonradial components of
radiation pressure force may exist for nonspherical particles
(Kla\v{c}ka 2004, 2008b) -- proper frame of reference of the particle
is the relevant system for defining radial and nonradial components.   \\
\noindent\\
2. \\
Mann (2009, p. 201) states: \\
"The stellar wind force in the frame of the moving particles depends on the
dust velocity $\vec{v}$ as: \\
$\vec{F}_{sw}$ $=$ $F_{sw}$ [ (1 $-$ $\vec{v} \cdot \vec{r} / (r v_{sw})$)
$\vec{r} / r$ $-$ $\vec{v} / v_{sw}$ ]~, ~~~ (7.10-M) \\
where $F_{sw}$ is the force on the dust for $\vec{v}$ $=$ 0 and
$\vec{v}_{sw}$ is the  bulk velocity of the wind. The non-radial term in (7.10)
is referred to as plasma or pseudo Poynting-Robertson drag force."
(We have corrected the evident errorneous term in the original equation,
Eq. (7-10-M) is correct.)  \\

Physics: \\
\nopagebreak
\noindent\\
i)\\
The solar-wind force acting on a body is not of the form
presented by Mann (2009 --  see Eq. 7.10-M).
The correct form is given by Kla\v{c}ka and Saniga (1993 -- Eq. 29): \\
$\vec{F}_{sw}$ $=$ $F_{sw}$ [ (1 $-$
$\vec{v} \cdot \vec{r} / (r v_{sw})$) $\vec{r} / r$ $-$
$x'$ $\vec{v} / v_{sw}$ ]~, ~~~~(29-K-S) \\
where $x'$ denotes the part of the incident energy which is lost
from the grain during the interaction with the solar wind (measured in the
reference frame of the grain; 1 $<$ $x'$ $<$ 3, approximately,
depending on material properties of the particle).
The force is even more complicated if nonradial component of the solar
wind velocity is included. \\
\noindent\\
ii)\\
What is the sense of the words "plasma or pseudo Poynting-Robertson drag
force"? We have just shown that the last term  of the real  force
$\vec{F}_{SW}$ does not correspond to that presented by Mann (2009 -
Eq. 7.10). Thus, the term "plasma or..." is misleading.
And what about the rest velocity term? \\
\noindent\\
3. \\
Mann (2009, p. 202) states: \\
"The ratio of plasma P-R drags force to (photon) P-R drag force can be
written as \\
($F_{sw}/F_{rad}$) $c  / v_{sw}$ $\approx$ ...	(7.11-M) \\
The factor $c  / v_{sw}$ results from the difference of the aberration
angles for photons or solar wind particles." \\

Physics: \\
\nopagebreak

We are aware that this kind of explanation is presented also in
Burns {\it et al.} (1979, p. 12), or, Leinert and Gr\"{u}n (1990, pp. 226-227)
-- see also Sec. 6.5. However, the solar-wind force does not produce the term
$-$ $F_{sw}$ $\vec{v} / v_{sw}$ (see Eq. 29-K-S presented above). Thus,
explanation of the term as a consequence of the aberration is incorrect.
Even the light-term $-$ $F_{ph}$ $\vec{v} / c$
is not produced by the aberration of light, see Sec. 6.2 and also
Sec. 7.4.5 in Kla\v{c}ka (2008b).

\subsection{Minato {\it et al.} (2004)}
We have already mentioned an analogy between the solar electromagnetic and
corpuscular radiation in Secs. 6.30, 6.34. We will present detail comments
on the paper by Minato {\it et al.} (2004), now: the incorrect statements
on the solar wind effect may enable better understanding of the P-R effect.\\
\noindent\\
1. \\
Minato {\it et al.} (2004 -- Sec. 2.1) state that the contribution of
solar-wind forces to the motion of dust grain is described as \\
$F_{sw}$ [ (1 $-$ 2 $\dot{r} / v_{sw}$) $\vec{e}_{R}$ $-$
($r \dot{\theta} / v_{sw}$)  $\vec{e}_{\theta}$ ]~, ~~~ (1-M) \\
where $\vec{v}$ is the velocity of the grain $\vec{v}$ $=$
$\dot{r}$ $\vec{e}_{R}$ $+$ $r \dot{\theta}$ $\vec{e}_{\theta}$ and
$v_{sw}$ is the heliocentric solar-wind speed. \\

Physics: \\
\nopagebreak

The solar-wind force acting on a body is not of the form
presented by Minato {\it et al.} (2004 -- Sec. 2.1, see Eq. 1-M).
The correct form is given by Kla\v{c}ka and Saniga (1993 -- Eq. 29): \\
$F_{sw}$ [ (1 $-$ (1 $+$ $x'$) $\dot{r} / v_{sw}$) $\vec{e}_{R}$ $-$
($x'$ $r \dot{\theta} / v_{sw}$) $\vec{e}_{\theta}$ ]~, ~~~(29-K-S) \\
where $x'$ denotes the part of the incident energy which is lost
from the grain during the interaction with the solar wind (measured in the
reference frame of the grain; $x'$ $>$ 1).
The force is even more complicated if one
uses also the nonradial component of the solar wind velocity. \\
\noindent\\
2. \\
"The second term on the right-hand side of Eq.(1-M) expresses the
pseudo PR drag which causes dust grains to spiral into the Sun."
(Minato {\it et al.} 2004 -- Sec. 3.2) \\

Physics: \\
\nopagebreak
\noindent\\
i) \\
There does not exist the term $-$ $r \dot{\theta} / v_{sw}$ $\vec{e}_{\theta}$
in the solar wind force. The real term is multiplied by the value $x'$. This is
one of the differences between the P-R effect and the solar wind effect. Thus,
there does not exist a simple analogy to the "P-R drag". \\
ii)  \\
There is another inconsistency in defining the "P-R drag". We have already
discussed (see Sec. 5) that various definitions are used. Now,
Minato {\it et al.} (2004 -- Sec. 3.2) present another definition for the
"P-R drag". Thus, according to the authors, the "P-R drag" is neither
the term $-$ $F_{ph}$ [ 2 ($\dot{r} / c$) $\vec{e}_{R}$ $+$
($r \dot{\theta} / c$) $\vec{e}_{\theta}$ ], nor
the term $-$ $F_{ph}$ [ ($\dot{r} / c$) $\vec{e}_{R}$ $+$
($r \dot{\theta} / c$) $\vec{e}_{\theta}$ ], but the term
$-$ $F_{ph}$ ($r \dot{\theta} / c$) $\vec{e}_{\theta}$.
And what about the rest velocity terms in the P-R effect? How are they
called in the "physics" of Wyatt (2009), Mann (2009) and Minato {\it et al.} (2004)?

Really, the radiation pressure force for a body with spherically
distributed mass corresponds to the P-R effect. Relativistically covariant
formulation of the equation of motion shows that any separation of
the equation into various (ad-hoc) components is of no physical sense and
the term "P-R drag" should be omitted in future.

By the way, the term $-$ $F_{sw}$ ($r \dot{\theta} / v_{sw}$)
$\vec{e}_{\theta}$ may not cause that dust grain spirals into the Sun.
Also the outspiralling from the Sun may occur if the solar wind bombardment
significantly decreases mass of the grain. Moreover, non-radial solar
wind velocity may be more important than the term
$-$ $F_{sw}$ ($r \dot{\theta} / v_{sw}$) $\vec{e}_{\theta}$. \\
\noindent\\
3. \\
"The ratio of $\gamma$ of the pseudo P-R drag to the P-R drag is given by
$\gamma$ $=$ ($F_{sw}/F_{ph}$) $c/v_{sw}$, where the factor
$c/v_{sw}$ comes from the difference of the aberration angles for solar
wind and light."
(Minato {\it et al.} 2004 -- Sec. 3.2) \\

Physics: \\
\nopagebreak

The solar-wind force does not produce the term
$-$ $F_{sw}$ $r \dot{\theta} / v_{sw}$ $\vec{e}_{\theta}$. Thus,
explanation of it as a consequence of the aberration is incorrect.
Even the light-term $-$ $F_{ph}$ ($r \dot{\theta} / c$) $\vec{e}_{\theta}$
is not produced by the aberration of light, see Sec. 6.2 (Sec. 7.4.5 in Kla\v{c}ka 2008b).

\section{Simple explanation of the P-R effect}
Astronomers often want to present a simple explanation of a physical effect.
We have just presented many incorrect explanations of the P-R effect.
The correct explanation can be found in Kla\v{c}ka (2008b). However,
very simple explanation can be very useful for scientists and students.

\subsection{Simple explanation without equations}
Physical correct and simple explanation with no equation
can be as follows: \\
{\bf Poynting-Robertson effect} -- The effect of electromagnetic radiation
pressure on a moving spherical particle. The incident radiation can be represented by a
parallel beam of photons and the particle is a nonrotating body of spherically
symmetric mass distribution.

Let the Sun (or another star)
be the only source of radiative and gravitational forces
acting on the particle. The particle of radius smaller than about
0.1 micron (the value depends on the particle's optical properties
and the luminosity of the star)
is expelled from the Solar System (the star) while the larger particle
slowly spirals toward the Sun (the star).

Solar radiation pressure force acting on small dust particles
(less than 0.1 micron in radii) is larger than the solar gravitational
force. As for larger particles, the conservation of particle's mass,
during particle's interaction with the incoming solar radiation, is important.
The conservation of the mass produces the term proportional to
$-$ $\vec{v}/c$ in the radiation pressure force acting on the particle;
$\vec{v}$ is particle's heliocentric orbital velocity and $c$ is
the speed of light. The drag/frictional term proportional
to $-$ $\vec{v}/c$ causes decrease of particle's total energy (and
angular momentum) and the particle slowly spirals toward the Sun.
The magnitude of the momentum (and the speed of the particle) slowly increases.

\subsection{Mathematical description}
Homogeneous beam of parallel photons interacts with non-rotating spherical
body/particle. If the incoming momentum per unit time is $\vec{p'}_{incoming}$
in the proper frame of reference of the particle, then the total
outgoing momentum per unit time is \\
$\vec{p'}_{outgoing}$ $=$ ($1 ~-~C'_{pr} / C'_{ext}$)
$\vec{p'}_{incoming}$ . \\
$C'_{pr}$ and $C'_{ext}$ are pressure and extinction cross-sections.
The cross-sections are calculated from Mie's solution of Maxwell's equations
(Mie 1908). The following inequalities hold: $C'_{pr} >$ 0 and
0 $< C'_{pr} / C'_{ext} <$ 2. The case of "perfectly absorbing spherical
particle" (treated by Poynting 1903 and Robertson 1937) is, in reality,
characterized by $C'_{pr}$ $=$ $\pi R^{2}$,
$C'_{ext}$ $=$ $2 \pi R^{2}$, where $R$ is radius of the particle. The force
acting on the particle is \\
$d \vec{p'} / d \tau$ $=$ $\vec{p'}_{incoming}$ $-$
$\vec{p'}_{outgoing}$ $=$ ($C'_{pr} / C'_{ext}$) $\vec{p'}_{incoming}$ , \\
where $\tau$ is proper time measured in the reference frame of the particle.
Using the fact that \\
$\vec{p'}_{incoming}$ $=$ ($S' C'_{ext} / c$) $\vec{e'}$ , \\
where $S'$ is the flux density of radiation energy (energy flow through unit
area perpendicular to the ray per unit time), $c$ is the speed of light
and $\vec{e'}$ is unit vector describing direction and orientation of
the travelling photons with respect to the particle. The last two equations
yield \\
$d \vec{p'} / d \tau$ $=$ ($S' C'_{pr} / c$) $\vec{e'}$ . \\
This equation represents an electromagnetic (light) pressure acting on
the particle.
Moreover, the mass of the particle does not change and this corresponds to
the condition for the energy of the particle $E'$ \\
$d E' / d \tau$ $=$ 0 .

The left-hand sides of the last two equations are components of the
energy-momentum four-vector (four-momentum). This enables us to find the
equation of motion of the particle in any frame of reference, if we
use Lorentz transformation. If the particle moves with velocity $\vec{v}$
in a frame of reference, then the equation of motion of the particle is \\
$d p^{\mu} / d \tau$ $=$ ($w^{2} S C'_{pr} / c$)
($b^{\mu} ~-~ u^{\mu} / c$)~, ~~~~(*) \\
where $p^{\mu}$ $=$ ($E/c$ ; $\vec{p}$) is four-momentum of the particle,
$w$ $=$ $\gamma$ ($1 - \vec{v} \cdot \vec{e} / c$),
$b^{\mu}$ $=$ ($1/w$ ; $\vec{e} / w$),
$u^{\mu}$ $=$ ($\gamma c$ ; $\gamma \vec{v}$) is four-velocity of the particle,
$\vec{e}$ is unit vector describing direction and orientation of
the travelling photons with respect to the stationary frame of reference
(laboratory frame) in which the particle moves with velocity $\vec{v}$, and,
finally, $\gamma$ $=$ $1/\sqrt{1 - (\vec{v}/c)^{2}}$ is the Lorentz factor.

The equation of motion (*) is known as the Poynting-Robertson effect.
The Poynting-Robertson effect is the effect corresponding to the
electromagnetic (light) pressure.

To the first order in $\vec{v}/c$, Eq. (*) reduces to \\
$d \vec{v} / d t$ $=$ [$S C'_{pr} / (m c)$]
[ ($1 - \vec{v} \cdot \vec{e}$) $\vec{e}$ $-$ $\vec{v} / c$ ] ~. \\

Remark: \\
\nopagebreak

The process of interaction between the incident radiation and the
spherical particle can be treated as a two-step process. \\

The incoming radiation acts on the particle with a force \\
$( d p^{\mu} / d \tau )_{in}$ $=$
	($w^{2} S \bar{C'}_{ext} / c$) $b^{\mu}$ ~, \\
and the force causes the following acceleration of the particle \\
$( d u^{\mu} / d \tau ) _{in}$	$=$
	[ $w^{2} S \bar{C'}_{ext}/(m c)$ ]  ($b^{\mu}$ $-$
		   $u^{\mu} / c$) ~, \\
or, to the first order in $\vec{v}/c$: \\
$( d \vec{v} / d t )_{in}$  $=$
		 [ $S \bar{C'}_{ext}/(m c)$ ] $\times$
		 [ ($1 - \vec{v} \cdot \vec{e} / c$) $\vec{e}$ $-$
		 $\vec{v} / c$ ] ~. \\

The outgoing radiation acts on the particle with a force \\
$( d p^{\mu} / d \tau )_{out}$ $=$ $-$ ($w^{2} S \bar{C'}_{ext} / c$)
	 [ ($\bar{C'}_{pr} / \bar{C'}_{ext}$) $u^{\mu} / c$ $+$
	 (1 $-$ $\bar{C'}_{pr} / \bar{C'}_{ext}$) $b^{\mu}$ ]~, \\
and the force causes the following acceleration of the particle \\
$( d u^{\mu} / d \tau ) _{out}$  $=$ $-$
	[ $w^{2} S \bar{C'}_{ext}/(m c)$ ]
	(1 $-$ $\bar{C'}_{pr} / \bar{C'}_{ext}$) ($b^{\mu}$ $-$
	$u^{\mu} / c$) ~, \\
or, to the first order in $\vec{v}/c$: \\
$( d \vec{v} / d t )_{out}$  $=$ $-$
		 [ $S \bar{C'}_{ext}/(m c)$ ]
		 (1 $-$ $\bar{C'}_{pr} / \bar{C'}_{ext}$)
		 $\times$
		 [ ($1 - \vec{v} \cdot \vec{e} / c$) $\vec{e}$ $-$
		 $\vec{v} / c$ ] ~. \\

It can be easily seen that $(d \vec{v} / d t)_{out}$ $\propto$ $+$ $\vec{v}$
if $\bar{C'}_{pr} / \bar{C'}_{ext}$ $<$ 1. This holds also for the case
treated by Poynting and Robertson ($\bar{C'}_{pr} / \bar{C'}_{ext}$ $=$
1/2).

Often the dimensionless efficiency factors of extinction and pressure are used:
$\bar{Q'}_{ext}$ $=$ $\bar{C'}_{ext}/(\pi R^{2})$,
$\bar{Q'}_{pr}$ $=$ $\bar{C'}_{pr}/(\pi R^{2})$, where $R$ is radius
of the particle.

Orbital evolution of interplanetary spherical dust particle under the
action of solar electromagnetic radiation and the solar gravity is
thoroughly understood and explained in Kla\v{c}ka (2004) and
Kla\v{c}ka {\it et al.} (2007). Orbital evolution of spherical particle
under the action of solar gravity and electromagnetic radiation
leads to a secular decrease of semi-major axis and eccentricity.
Shift of perihelion may exist already in the case when the equation
of motion is considered to the first order in $\vec{v}/c$, if optical
properties of the particle change.

\section{Secular evolution of orbital elements}
Let us consider orbital evolution of a body under the action
of gravity and electromagnetic radiation of a central star (e. g., Sun).
The action of electromagnetic radiation is taken to be the P-R effect.
Accuracy to the first order in $\vec{v}/c$ is considered.

We have already discussed that physics does not admit to divide the P-R 
effect into partial components. Thus, we have to take into account that 
the P-R effect represents the disturbing force to the Keplerian motion
given by the gravity of the central star. We are interested in secular
evolution of orbital elements (short-term oscillations in orbital elements
are not important for us). We will proceed in accordance with Kla\v{c}ka
(1994) and Kla\v{c}ka (2004 -- Sec. 6).

If parameter $\beta$ changes, then only numerical calculations can
be realized. If equation of motion in the form
\begin{eqnarray}\label{61}
\dot{\vec{v}} &=& - ~ \frac{G M}{r^{3}} ~\vec{r}
\nonumber \\
& & +~\beta ~\frac{G M}{r^{2}} ~ \left \{  \left ( 1~-~
      \frac{\vec{v} \cdot \vec{e}}{c} \right ) ~ \vec{e}
      ~-~ \frac{\vec{v}}{c} \right \} ~,
\end{eqnarray}
is numerically solved, then
orbital elements can be calculated on the basis of Eqs. (47) in Kla\v{c}ka
(2004 -- the correct right-hand side of the last equation contains
$\vec{v} \cdot \vec{e}_{T}$/$(e~\sqrt{G M_{\odot} / p}~)$ $-$ 1/$e$,
if $M$ $=$ $M_{\odot}$, instead of ... $-$ 1).

In any case, an approximation of constant $\beta$ can be useful, at least
for comparison. This case can be treated in a fully analytical approach,
if secular evolution of orbital elements is searched. The results can
be summarized as follows, as for semi-major axis $a$ and eccentricity $e$:
\begin{eqnarray}\label{62}
a &=& a_{\beta} \left ( 1 - e_{\beta}^{2} \right ) ^{3/2} ~
     \frac{1}{2 \pi} ~ \int_{0}^{2 \pi}
    \frac{\left [ 1 + \beta \left ( 1 + e_{\beta}^{2} + 2 e_{\beta}
    \cos x \right ) / \left ( 1 - e_{\beta}^{2} \right ) \right ]^{-1}}
    {\left ( 1 + e_{\beta} \cos x \right )^{2}}
    ~dx ~,
\nonumber \\
e &=& \left ( 1 - e_{\beta}^{2} \right ) ^{3/2} ~
     \frac{1}{2 \pi} ~ \int_{0}^{2 \pi}
\frac{\sqrt{\left ( 1 - \beta \right ) ^{2}  e_{\beta}^{2} + \beta ^{2} -
    2 \beta \left ( 1 - \beta \right )	e_{\beta}
    \cos x}}{\left ( 1 + e_{\beta} \cos x \right )^{2}}
    ~dx ~,
\end{eqnarray}
where
\begin{eqnarray}\label{63}
\frac{d a_{\beta}}{d t} &=& -~ \beta ~ \frac{G M}{c} ~
       \frac{2 + 3 e_{\beta}^{2}}{a_{\beta} \left ( 1 - e_{\beta}^{2} \right )^{3/2}} ~,
\nonumber \\
\frac{d e_{\beta}}{d t} &=& -~ \frac{5}{2}~ \beta ~ \frac{G M}{c} ~
       \frac{e_{\beta}}{a_{\beta}^{2} \left ( 1 - e_{\beta}^{2} \right )^{1/2}} ~.
\end{eqnarray}
Orbital elements with subscript $\beta$ correspond to the central Keplerian
acceleration $-$ $GM (1-\beta) \vec{r} / r^{3}$  --
non-velocity radial term of acceleration caused by radiation force 
in Eq. (61) is added to the gravitational acceleration of the star 
(however, it is often stated that Eqs. 63 hold for central acceleration 
$-$ $GM \vec{r} / r^{3}$, see, e. g., Burns {\it et al.} 1979 -- Eqs. 30a, 
30b, 30c, 30d, 31a, 31b, 46a, 46b, 47, 48, Mignard 1992). The set of differential equations (Eqs. 63)
is the set presented by Robertson (1937), Wyatt and Whipple (1950).
Moreover, initial conditions
for Eqs. (63) are required. If the particle is released from the
parent body (quantities relating to the parent body are denoted by
subscript $P$) with velocity $\vec{\Delta}$
\begin{equation}\label{64}
\vec{\Delta} = \Delta v_{R} ~\vec{e}_{PR} ~+~  \Delta v_{T} ~\vec{e}_{PT} ~+~
       \Delta v_{N} ~\vec{e}_{PN} ~,
\end{equation}
while
\begin{eqnarray}\label{65}
\vec{v} &=& v_{R} ~ \vec{e}_{PR} ~+~ v_{T} ~ \vec{e}_{PT} ~,
\nonumber \\
v_{R} &=& \sqrt{G M ~p_{P}^{-1}} ~e_{P} ~
\sin \left ( \Theta_{P} - \omega_{P} \right ) ~,
\nonumber \\
v_{T} &=& \sqrt{G M ~p_{P}^{-1}} ~\left [ 1 ~+~ e_{P} ~
    \cos \left ( \Theta_{P} - \omega_{P} \right ) \right ] ~;
\end{eqnarray}
$p_{P} = a_{P}  \left ( 1 ~-~ e_{P}^{2} \right )$,
$p_{\beta} = a_{\beta}  \left ( 1 ~-~ e_{\beta}^{2} \right )$ and
$\vec{e}_{PN} = \vec{e}_{PR} \times~\vec{e}_{PT}$,
then (compare Gajdo\v{s}\'{\i}k and Kla\v{c}ka 1999; Kla\v{c}ka 2004 -- Eq. 59):
\begin{eqnarray}\label{66}
p_{\beta ~in} &=& \frac{p_{P}}{1 ~-~ \beta} ~
      \frac{\left ( v_{T} ~+~ \Delta v_{T} \right ) ^{2} ~+~
      \left ( \Delta v_{N} \right ) ^{2}}
      {v_{T}^{2}} ~,
\nonumber \\
e_{\beta ~in}^{2} &=& \left ( \frac{1 ~+~ e_{P} ~ \cos f_{P}}{1 ~-~ \beta} ~
  \frac{v_{TS}^{2}}{v_{T}^{2}} ~-~ 1 \right ) ^{2} ~+~
\nonumber \\
& & \left ( \frac{1 ~+~ e_{P} ~ \cos f_{P}}{1 ~-~ \beta} ~
  \frac{v_{TS}}{v_{T}}	\right ) ^{2}
  \left ( \frac{v_{R} ~+~ \Delta v_{R}}{v_{T}} \right ) ^{2} ~,
\nonumber \\
a_{\beta ~in} &=& \frac{p_{\beta ~in}}{1 - e_{\beta ~in}^{2}} ~,
\nonumber \\
\cos i_{\beta ~in} &=& \frac{v_{T} ~+~ \Delta v_{T}}{v_{TS}} \cos i_{P} ~-~
\frac{\Delta v_{N}}{v_{TS}} ~ \cos \Theta_{P} ~ \sin i_{P} ~, \nonumber \\
\sin i_{\beta ~in} ~\cos \Omega_{\beta ~in} &=&
\frac{v_{T} ~+~ \Delta v_{T}}{v_{TS}} \sin i_{P}
~ \cos \Omega_{P} ~+~ \nonumber \\
& & \frac{\Delta v_{N}}{v_{TS}} ~  \left ( \cos \Theta_{P} ~
\cos i_{P} ~ \cos \Omega_{P}  ~-~ \sin \Theta_{P} ~ \sin \Omega_{P} \right ) ~,
\nonumber \\
\sin i_{\beta ~in} ~\sin \Omega_{\beta ~in} &=& \frac{v_{T} ~+~
      \Delta v_{T}}{v_{TS}} \sin i_{P}
      ~ \sin \Omega_{P} ~+~ \nonumber \\
& & \frac{\Delta v_{N}}{v_{TS}} ~  \left ( \cos \Theta_{P} ~
\cos i_{P} ~ \sin \Omega_{P}  ~+~ \sin \Theta_{P} ~ \cos \Omega_{P} \right ) ~,
\nonumber \\
e_{\beta ~in} ~ \cos f_{\beta ~in} &=& \frac{p_{\beta ~in}}{p_{P}} ~
      \left ( 1 ~+~ e_{P} ~\cos f_{P} \right ) ~-~ 1 ~,
\nonumber \\
e_{\beta ~in} ~ \sin f_{\beta ~in} &=& \frac{1 ~+~ e_{P} ~\cos f_{P}}{1 ~-~ \beta} ~
    \frac{v_{TS} ~ \left ( v_{R} ~+~ \Delta v_{R} \right )}{v_{T}^{2}} ~,
\nonumber \\
\sin i_{\beta ~in} ~\cos \Theta_{\beta ~in} &=& \frac{v_{T} ~+~ \Delta v_{T}}{v_{TS}} ~\sin i_{P}
~ \cos \Theta_{P} ~+~ \nonumber \\
& & \frac{\Delta v_{N}}{v_{TS}} ~ \cos i_{P} ~,
\nonumber \\
\sin i_{\beta ~in} ~\sin \Theta_{\beta ~in} &=& \sin i_{P} ~ \sin \Theta_{P} ~,
\nonumber \\
v_{TS}^{2} &\equiv& ( v_{T} ~+~ \Delta v_{T} )^{2} ~+~ ( \Delta v_{N} )^{2} ~.
\end{eqnarray}

The set of equations represented by Eqs. (62)-(66) fully corresponds to
detailed numerical calculations of vectorial equation of motion, if we 
are interested in secular evolution of eccentricity and semi-major axis
for the case when central acceleration is defined by gravity alone.
We have to stress that
"$e_{\beta ~in} <$ 1" and "$e_{\beta}$ does not correspond to
pseudo-circular orbit" are tacitly assumed.

\begin{figure}[!h]
\begin{center}
\includegraphics{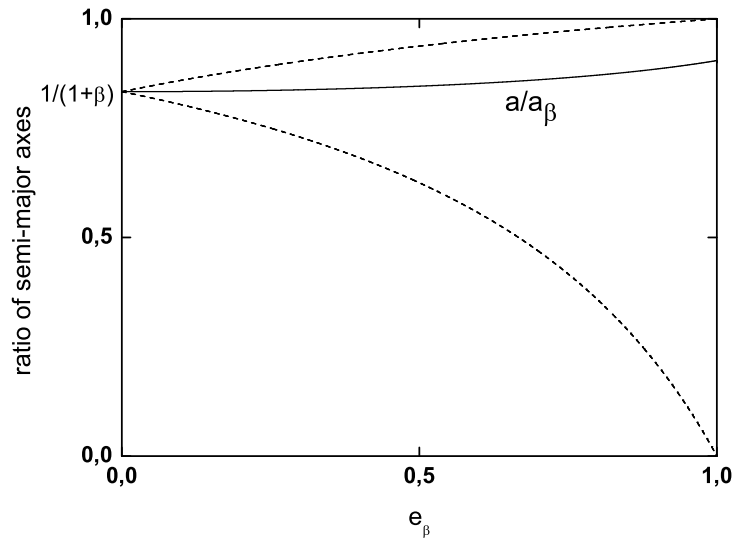}
\end{center}
\label{F1}
\caption{Dependence of $a / a_{\beta}$ as a function of
secular value of $e_{\beta}$. Dashed lines depict functions
$(1 - e_{\beta} )$ / $[1 - e_{\beta} + \beta (1 + e_{\beta} )]$
and 
$(1 + e_{\beta} )$ / $[1 + e_{\beta} + \beta (1 - e_{\beta} )]$.
The dashed lines confine the zone where instantaneous ratio of semi-major
axis (central acceleration $-~G M \vec{r} / r^{3}$)
and $a_{\beta}$ can occur.}
\end{figure}

\begin{figure}[!h]
\begin{center}
\includegraphics{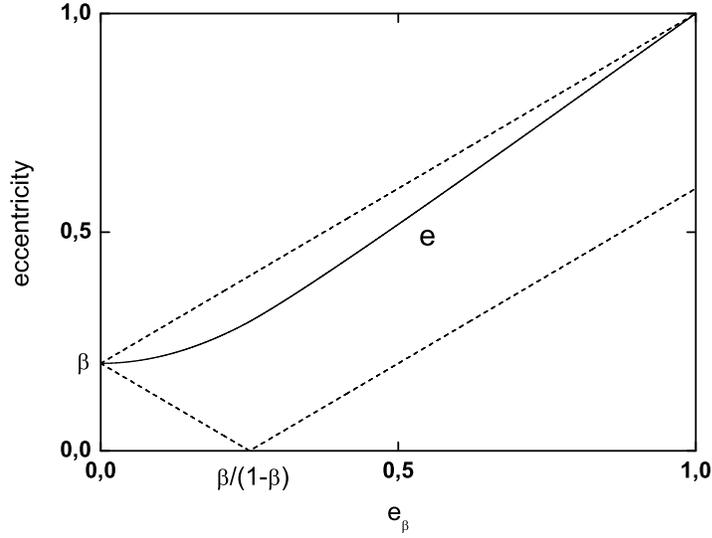}
\end{center}
\label{F2}
\caption{Dependence of $e$ as a function of
secular value of $e_{\beta}$. Dashed lines depict functions
$| (1 - \beta) e_{\beta} - \beta |$ and 
$(1 - \beta) e_{\beta} + \beta$.
The dashed lines confine the zone where instantaneous eccentricity 
(central acceleration $-~G M \vec{r} / r^{3}$) can occur.}
\end{figure}

\begin{figure}[h]
\begin{center}
\includegraphics{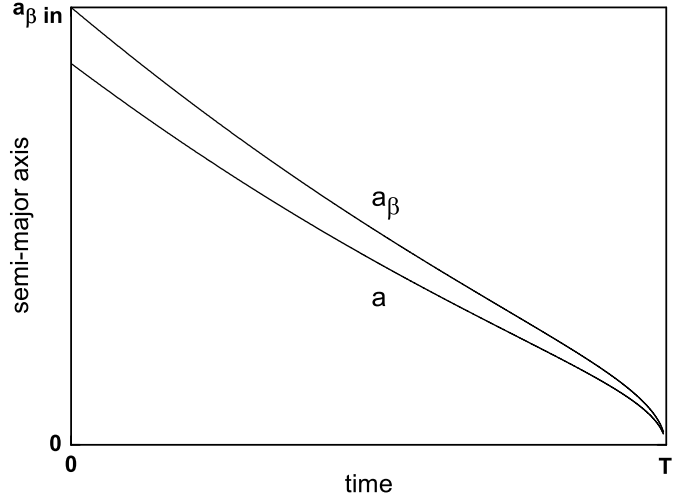}
\end{center}
\label{F3}
\caption{Time evolution of semi-major axes $a$ and $a_{\beta}$.
$T$ is the time of spiralling toward the star. The value of $a_{in}$
is given by the first of Eqs. (62).
The following time limits hold: $\lim_{t \rightarrow T} a$ $=$ 
$\lim_{t \rightarrow T} a_{\beta}$ $=$ 0.}
\end{figure}

\begin{figure}[h]
\begin{center}
\includegraphics{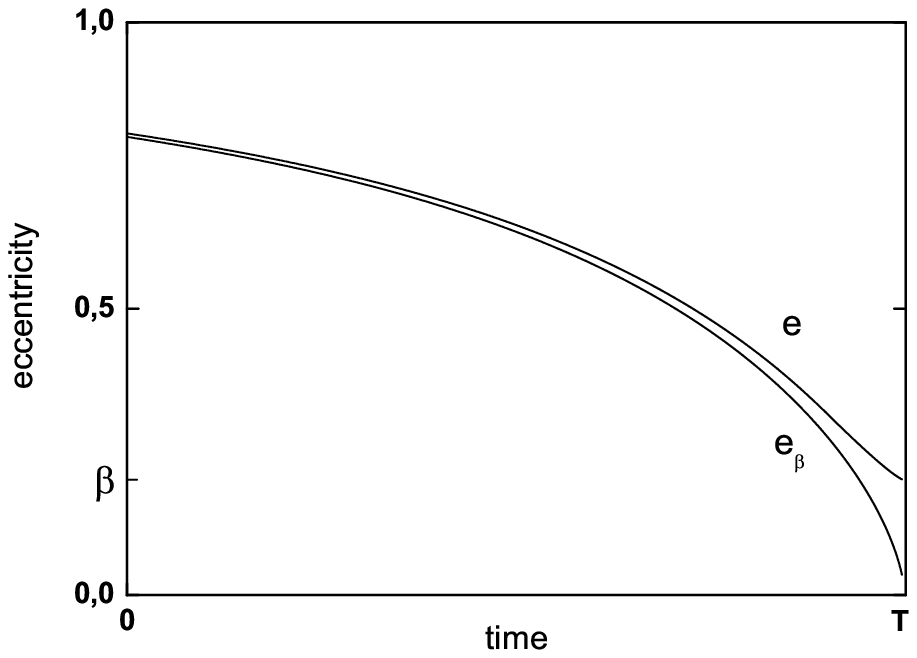}
\end{center}
\label{F4}
\caption{Time evolution of eccentricities $e$ and $e_{\beta}$. 
$T$ is the time of spiralling toward the star.
The following time limits hold: $\lim_{t \rightarrow T} e$ $=$ $\beta$,
$\lim_{t \rightarrow T} e_{\beta}$ $=$ 0.}
\end{figure}

It is worth mentioning that instantaneous time derivatives of semi-major 
axis and eccentricity defined by central acceleration 
$-$ $GM \vec{r} / r^{3}$ may be both positive and negative,
while secular evolution yields that semi-major axis $a$ and
eccentricity $e$ are decreasing functions of time.
As for the central acceleration $-$ $GM (1 - \beta) \vec{r} / r^{3}$,
the instantaneous time derivative of semi-major axis is always negative, 
while the instantaneous time derivative of eccentricity 
may be both negative (near apocentrum) and positive (secular
evolution yields that semi-major axis $a_{\beta}$ and eccentricity 
$e_{\beta}$ are decreasing functions of time -- see Eqs. 63).

For the special case $\vec{\Delta} =$ 0 Eqs. (66) reduce to
\begin{equation}\label{67}
a_{\beta ~in} = a_{P} \left ( 1 - \beta \right ) \left ( 1 - 2 \beta
       \frac{1 + e_{P} \cos f_{P}}{1~-~e_{P}^{2}}
       \right ) ^{-1} ~,
\end{equation}
\begin{equation}\label{68}
e_{\beta ~in} = \sqrt{1 - \frac{1 - e_{P}^{2} - 2 \beta \left (
      1 + e_{P} \cos f_{P} \right )}{
      \left ( 1 - \beta \right )^{2}}} ~,
\end{equation}
where $f_{P} \equiv \Theta_{P} - \omega_{P}$,
$\omega _{\beta ~in}$ has to be obtained from
\begin{eqnarray}\label{69}
e_{\beta~ in} ~ \cos \left ( \Theta _{P} - \omega _{\beta ~in} \right ) &=&
  \frac{\beta + e_{P} \cos f_{P}}{1 - \beta} ~,
\nonumber \\
e_{\beta ~in} ~ \sin \left ( \Theta _{P} - \omega _{\beta ~in} \right ) &=&
  \frac{e_{P} \sin f_{P}}{1 - \beta} ~,
\end{eqnarray}
\begin{equation}\label{70}
\Omega_{\beta~ in} = \Omega_{P} ~, ~~i_{\beta~ in} = i_{P} ~,~~
\Theta_{\beta ~in} = \Theta_{P} ~.
\end{equation}
Physics of Eqs. (67)-(70) is following: meteoroid escapes from
the Solar System when the orbital energy becomes positive and this
can happen when the energy due to the radial component of the
radiation force is included, without it being necessary for this
force to exceed the gravitational attraction (Harwit 1963). Some
figures may be found in Kla\v{c}ka (1992b, 1993c). As an example we may
mention that particle of $\beta = ( 1 - e_{P} ) / 2$ moves in
parabolic orbit if ejected at perihelion of the parent body, and,
in an orbit with eccentricity $e_{\beta ~in} = | 1 - 3 e_{P} | / (
1 + e_{P} )$ if released at aphelion; $e_{\beta ~in} =$ 0 for
$\beta = e_{P} =$ 1/3 and aphelion ejection.

We may mention that a little more simple procedure is obtained when
semi-major axis $a_{\beta}$ is replaced by 
$p_{\beta}$ $=$ $a_{\beta} (1 - e_{\beta}^{2})$
(a quantity referred to as semi-latus rectum). Then
$p_{\beta}$ $=$ $p_{\beta ~in}\left(e_{\beta}/e_{\beta~in}\right)^{4/5}$, and:
\begin{equation}\label{71}
\frac{d p_{\beta}}{dt} = -~2~\beta~ \frac{G M}{c} ~\frac{ \left [
1 - e_{\beta ~in}^{2} \left ( p_{\beta} / p_{\beta ~in} \right ) ^{5/2}
\right ] ^{3/2}}{p_{\beta}} ~,
\end{equation}
\begin{equation}\label{72}
\frac{d e_{\beta}}{dt} = -~ \frac{5}{2} ~\beta~ \frac{G M}{c} ~
	  \frac{e_{\beta ~in}^{8/5}}{p_{\beta ~in} ^{2}} ~
	  \frac{\left ( 1 - e_{\beta}^{2} \right ) ^{3/2}}{e_{\beta}^{3/5}} ~.
\end{equation}

Eqs. (62)-(63) show that secular evolutions of semi-major axes $a$,
$a_{\beta}$ and eccentricities $e$, $e_{\beta}$ are decreasing functions
of time. This means that eccentric orbits become more circular, during the particle's
orbital evolution (constant $\beta$ is assumed). This corresponds to
the fact that apocentric distances (aphelion distances) $Q$, $Q_{\beta}$
decrease more rapidly than the pericentric distances (perihelion distances)
$q$, $q_{\beta}$, as it is evident also from the equations
\begin{eqnarray}\label{73}
q &=& a_{\beta} \left ( 1 - e_{\beta}^{2} \right ) ^{5/2} ~
     \frac{1}{2 \pi} ~ \int_{0}^{2 \pi} Z_{q} ~dx ~,
\nonumber \\
Z_{q} &=& \frac{1 ~-~
    \sqrt{\left ( 1 - \beta \right ) ^{2}  e_{\beta}^{2} + \beta ^{2} -
    2 \beta \left ( 1 - \beta \right )	e_{\beta} \cos x}}{
    \left [ 1 - e_{\beta}^{2} + \beta \left ( 1 + e_{\beta}^{2} + 2 e_{\beta}
    \cos x \right ) \right ] \left (
    1 + e_{\beta} \cos x \right )^{2}} ~,
\end{eqnarray}
\begin{eqnarray}\label{74}
Q &=& a_{\beta} \left ( 1 - e_{\beta}^{2} \right ) ^{5/2} ~
     \frac{1}{2 \pi} ~ \int_{0}^{2 \pi} Z_{Q} ~dx ~,
\nonumber \\
Z_{Q} &=& \frac{1 ~+~
    \sqrt{\left ( 1 - \beta \right ) ^{2}  e_{\beta}^{2} + \beta ^{2} -
    2 \beta \left ( 1 - \beta \right )	e_{\beta} \cos x}}{
    \left [ 1 - e_{\beta}^{2} + \beta \left ( 1 + e_{\beta}^{2} + 2 e_{\beta}
    \cos x \right ) \right ] \left (
    1 + e_{\beta} \cos x \right )^{2}} ~,
\end{eqnarray}
where Eqs. (63)-(66) can be used (or, Eqs. 71-72, too), or
\begin{eqnarray}\label{75}
a_{\beta} &=& \frac{1}{2} ~\left ( q_{\beta} ~+~ Q_{\beta} \right ) ~,
\nonumber \\
e_{\beta} &=& \frac{1 ~-~  q_{\beta} / Q_{\beta}}{
	      1 ~+~  q_{\beta} / Q_{\beta}} ~,
\end{eqnarray}
and $q_{\beta}$, $Q_{\beta}$ are given by the following equations:
\begin{eqnarray}\label{76}
\frac{d q_{\beta}}{d t} &=& -~\beta~\frac{G M}{c}~\frac{q_{\beta}}
{2Q_{\beta}}~\frac{1}{\sqrt{q_{\beta}Q_{\beta}}}~
\frac{5q_{\beta}+3Q_{\beta}}{q_{\beta}+Q_{\beta}}~,
\nonumber \\
\frac{d Q_{\beta}}{d t} &=& -~\beta~\frac{G M}{c}~\frac{Q_{\beta}}
{2q_{\beta}}~\frac{1}{\sqrt{q_{\beta}Q_{\beta}}}~
\frac{3q_{\beta}+5Q_{\beta}}{q_{\beta}+Q_{\beta}}~,
\nonumber \\
\left(\frac{d q_{\beta}}{d t}\right)/\left(\frac{d Q_{\beta}}{d t}\right)
&=& \left(\frac{q_{\beta}}{Q_{\beta}}\right)^{2}
\left(1-2~\frac{Q_{\beta}-q_{\beta}}{3q_{\beta}+5Q_{\beta}}\right)~ < ~ 1 ~.
\end{eqnarray}
For completeness we have to introduce initial conditions
\begin{eqnarray}\label{77}
q_{\beta~in} &=& \frac{p_{\beta~in}}{1~+~e_{\beta~in}}~,
\nonumber\\
Q_{\beta~in} &=& \frac{p_{\beta~in}}{1~-~e_{\beta~in}}~,
\end{eqnarray}
where $p_{\beta~in}$ and $e_{\beta~in}$ are given by Eqs. (66).

It is worth mentioning that instantaneous time derivatives of pericentric 
and apocentric distances defined by central acceleration 
$-$ $GM \vec{r} / r^{3}$ may be both positive and negative, while 
secular evolution yields that $q$ and $Q$ are decreasing functions 
of time. As for the central acceleration 
$-$ $GM (1 - \beta) \vec{r} / r^{3}$,
the instantaneous time derivatives of pericentric and apocentric 
distances are both less than or equal zero (secular
evolution yields that $q_{\beta}$ and $Q_{\beta}$ are 
decreasing functions of time -- see Eqs. 76).

The time of spiralling of the particle into the star is
\begin{equation}\label{78}
T [ yrs ]~=~\frac{6.4 \times 10^{2}}{\beta}~\frac{M_{sun}}{M} ~
\frac{\left ( a_{\beta ~in} [ AU ] \right )^{2} 
\left(1-e_{\beta ~in}^{2} \right)^{2}}{e_{\beta ~in}^{8/5}}~
\int_{0}^{e_{\beta ~in}} \frac{z^{3/5}}{\left(1-z^{2}\right)^{3/2}}~dz~.
\end{equation}
(see also Eq. 51).

Figs. 1 and 2 depict the behaviour of Eqs.(62). The dashed lines
correspond to the dependences
$(1 - e_{\beta} )$ / $[1 - e_{\beta} + \beta (1 + e_{\beta} )]$
and 
$(1 + e_{\beta} )$ / $[1 + e_{\beta} + \beta (1 - e_{\beta} )]$
for the ratio of instantaneous values of semi-major axes and
$| (1 - \beta) e_{\beta} - \beta |$ and 
$(1 - \beta) e_{\beta} + \beta$ for instantaneous values of eccentricities
corresponding to the central acceleration $-~G M \vec{r} / r^{3}$
(see Eqs. 96-97 in Kla\v{c}ka 2004). Time evolution of semi-major
axes and eccentricities is presented in Figs. 3 and 4. The time of
spiralling toward the central star is given by Eq. (78). While the
values of $a$, $a_{\beta}$ and $e_{\beta}$ approach to zero, the value
of $e$ tends to the value $\beta$ (see Eqs. 62-63).

\section{Conclusion}
We have presented physical comments to "explanations" of the
Poynting-Robertson effect published during the past thirty years by almost
hundred of scientists working in the field. We have used mainly
scientific papers and monographs. Although we have only restricted access 
to literature, we hope that physics of other published presentations
can be covered by our paper. Moreover, we hope that
our paper, together with papers by Kla\v{c}ka (1992a, 1993b, 2004, 2008a,
2008b), can help in better understanding the dynamical interaction of
electromagnetic radiation with a moving body.

Moreover, we have added two simple explanations of the P-R effect (see Sec. 7).
They are not only simple, but they concentrate the physical essence
of the phenomenon. These simple explanations (or their modifications)
may be used in scientific literature, encyclopediae and textbooks.

The confusion in the scientific literature about the term called the
"Poynting-Robertson drag" (see, e.g., Dohnanyi 1978, Gustafson 1994,
Minato {\it et al.} 2004, Wyatt 2009, Mann 2009) can be easily removed 
if one takes into account that the relativistically covariant form of 
the P-R effect does not enable any division into partial (covariant) 
terms. The correct physics yields the P-R effect as a whole effect of 
the radiation force on a moving body and no "P-R drag" exists as a 
partial relativistically covariant equation of motion.

The difference between the effects of solar electromagnetic
and corpuscular (solar wind) radiation is stressed (see Secs. 6.34, 6.35).

Sec. 8 presents secular evolution of orbital elements of a spherical
particle if the action of gravity of a central star and it's 
electromagnetic radiation is considered. 

The physical understanding of the P-R effect enables correct derivation
of the action of stellar winds on evolution of dust grains. This can help
in physical explanation of the observed dust distribution at stars
with planetary systems.

\section*{Acknowledgement}
This work was supported by the Scientific Grant Agency VEGA, Slovakia,
grant No. 2/0016/09.

\end{document}